\documentclass[12pt]{article}
\usepackage{epsfig,amsmath,amsfonts,amssymb,amstext,afterpage}
\def\spose#1{\hbox to 0pt{#1\hss}}
\def\lsim{\mathrel{\spose{\lower 3pt\hbox{$\mathchar"218$}}
 \raise 2.0pt\hbox{$\mathchar"13C$}}}
\def\gsim{\mathrel{\spose{\lower 3pt\hbox{$\mathchar"218$}}
 \raise 2.0pt\hbox{$\mathchar"13E$}}}

\catcode`@=11
\def\@citex[#1]#2{%
  \if@filesw\immediate\write\@auxout{\string\citation{#2}}\fi
  \def\@citea{}\@cite{\@for\@citeb:=#2\do
    {\@citea\def\@citea{,\penalty\@m}\@ifundefined
      {b@\@citeb}{{\bf ?}\@warning
{Citation `\@citeb' on page \thepage \space undefined}}%
      \hbox{\csname b@\@citeb\endcsname}}}{#1}}
\def\citer{\@ifnextchar [{\@tempswatrue\@citexr}{\@tempswafalse\@citexr[]}}
  \def\@citexr[#1]#2{%
    \if@filesw\immediate\write\@auxout{\string\citation{#2}}\fi
    \def\@citea{}\@cite{\@for\@citeb:=#2\do
      {\@citea\def\@citea{--\penalty\@m}\@ifundefined
{b@\@citeb}{{\bf ?}\@warning
{Citation `\@citeb' on page \thepage \space undefined}}%
\hbox{\csname b@\@citeb\endcsname}}}{#1}}

\begin{document}

\begin{titlepage}

\begin{flushright}
{\small
LMU-ASC~43/08\\
September 2008
}
\end{flushright}

\vspace{0.5cm}
\begin{center}
{\Large\bf \boldmath
$B\to V_L V_L$ Decays\\
at Next-to-Leading Order in QCD 
\unboldmath}
\end{center}

\vspace{0.5cm}
\begin{center}
{\sc Matth\"aus Bartsch$^1$, Gerhard Buchalla$^1$ and Christina Kraus$^{1,2}$}
\end{center}

\vspace*{0.4cm}

\begin{center}
$^1$Ludwig-Maximilians-Universit\"at M\"unchen, Fakult\"at f\"ur Physik,\\
Arnold Sommerfeld Center for Theoretical Physics, 
D--80333 M\"unchen, Germany\\
$^2$Max-Planck-Institut f\"ur Quantenoptik,\\ 
Hans-Kopfermann-Str. 1, D--85748 Garching, Germany
\end{center}

\vspace{0.5cm}
\begin{abstract}
\vspace{0.2cm}\noindent
We compute the amplitudes for the two-body decay of $B$ mesons into
longitudinally polarized light vector mesons at next-to-leading
order in QCD. We give the explicit expressions in QCD factorization
for all 34 transitions of a heavy-light $B$ meson into a pair of
longitudinal vector mesons $\rho$, $\omega$, $\phi$, $K^*$ within
the Standard Model. Decay rates and CP asymmetries are discussed
in detail and compared with available data. Exploiting the fact that
QCD penguins are systematically smaller for vector mesons in comparison
to pseudoscalars in the final state, we investigate several methods
to achieve high-precision determinations of CKM parameters and
New Physics tests. We propose a method to use V-spin symmetry and
data on $\bar B_d\to\bar K^{*0}_L K^{*0}_L$ to constrain the
penguin contribution in $\bar B_d\to\rho^+_L\rho^-_L$. 
CP violation in the latter decay together with a measurement of
$\sin 2\beta$ determines the unitarity triangle with high accuracy.
We show that CP violation in $\bar B_d\to\rho^+_L\rho^-_L$ and
$\bar B_d\to\psi K_S$ alone implies $|V_{ub}|=(3.54\pm 0.17)\cdot 10^{-3}$,
presently the most accurate determination of this quantity.
\end{abstract}

\vspace*{3cm}

\vfil
\end{titlepage}

\section{Introduction}
\label{sec:intro}

Charmless hadronic $B$-meson decays provide us with important
information on the physics of flavour, within and beyond the
Standard Model. After the very successful first generation of 
$B$-factory experiments, BaBar and Belle, the interest in this
field is re-inforced by the upcoming start of the LHC with the
dedicated LHCb experiment \cite{Buchalla:2008jp}. In the longer
run the second generation $B$-factory projects 
SuperB \cite{Bona:2007qt} and SuperKEKB \cite{Kageyama:2006zd} 
promise excellent future opportunities 
\cite{Browder:2007gg,Browder:2008em}, 
as well as the LHCb upgrade options \cite{Buchalla:2008jp}.

In the present paper we investigate in detail the class of hadronic
two-body decays of $B^-$, $\bar B_d$ and $\bar B_s$ mesons with
longitudinally polarized light vector mesons
($\rho^\pm$, $\rho^0$, $\omega$, $\phi$, $K^{*\pm}$, $\bar K^{*0}$, $K^{*0}$) 
in the final state.
These can be computed systematically using QCD factorization
in the heavy-quark limit \cite{Beneke:2001ev,Beneke:1999br,Beneke:2000ry}.
In comparison, $B$ decays into vector mesons with transverse polarization
are suppressed by powers of $\Lambda_{QCD}/m_b$. Their amplitudes do not
factorize, as indicated by infrared singularities in convolution
integrals, and this introduces a model-dependence in the computation
of the related observables \cite{Kagan:2004uw,Kraus:2004,Beneke:2006hg}. 
For this reason we do not include final states with transversely
polarized vector mesons in our present analysis, although they
raise issues of interest in their own right \cite{Kagan:2004uw}. 
Rather, we will pursue the physics of $B\to V_L V_L$ decays, 
which comprise 28 calculable channels and 6 annihilation modes
within the Standard Model
and offer a rich phenomenology by themselves. The various decay modes
probe different types of amplitudes, such as tree, QCD and
electroweak penguin, or weak annihilation. The large number of channels
will allow us to test our understanding of QCD in hadronic $B$ decays
and at the same time to extract flavour parameters and search for
New Physics effects. 
An important motivation for the study of final states with two
vector mesons is the considerably smaller size of QCD penguin
contributions, as compared to the decay into pseudoscalars. 
This improves our control of QCD effects in several observables of
interest for flavour physics. Mixing-induced CP violation in
$\bar B_d\to\rho^+_L\rho^-_L$ is an important example.

Several studies of $B\to VV$ decays based on factorization in the 
heavy-quark limit have been made in the literature
\citer{Kagan:2004uw,Cheng:2008gx}.
So far the only comprehensive analysis, including all channels,
has been given in \cite{Beneke:2006hg}, where also transverse
polarization is discussed. The other articles have addressed various
aspects of $B\to VV$ decays, investigating particular channels, 
polarization effects and, in part, New Physics contributions.

In this paper we present a detailed analysis of the complete set
of $B\to V_LV_L$ decays within the Standard Model.
We work to next-to-leading order (NLO) in QCD, using factorization
in the heavy-quark limit. Power corrections to this limit, in particular
from weak annihilation, are estimated using a model description
\cite{Beneke:2001ev}. 
Our presentation includes some additional aspects and several new
applications of $B\to V_LV_L$ transitions. We give detailed analytical 
expressions for all 34 $B\to V_LV_L$ decay amplitudes at next-to-leading
order. We suggest an alternative treatment of long-distance electromagnetic
penguins, which contribute in channels with the neutral vector mesons
$\rho^0$, $\omega$ or $\phi$. The effects of $\omega$-$\phi$ mixing are
briefly discussed. We apply the results in new ways to the
phenomenology of the unitarity triangle: we use the formulation
of \cite{Buchalla:2003jr,Buchalla:2004tw} for the analysis of
CP violation in $\bar B_d\to\rho^+_L\rho^-_L$, we propose an approach
to independently constrain the $\bar B_d\to\rho^+_L\rho^-_L$-penguin
with $\bar B_d\to\bar K^{*0}_L K^{*0}_L$ and V-spin symmetry,  and
we extract accurate values of the CKM quantities
$\bar\rho$, $\bar\eta$, $\gamma$, $\alpha$ and $|V_{ub}|$.

The paper is organized as follows.
Section 2 collects basic ingredients of the calculation, in particular
the effective Hamiltonians, form factors, light-cone wave functions
and meson projectors. Section 3 presents the NLO results for the
$B\to V_L V_L$ decay amplitudes. Input parameters and experimental results
are summarized in section 4. Section 5 contains the phenomenological
analysis, with a discussion of branching fractions and CP asymmetries
and with precision determinations of the unitarity triangle.  
A brief comparison with the literature is given in section 6, before
we conclude in section 7. Our treatment of long-distance electromagnetic
penguins is discussed in the appendix.

\afterpage{\clearpage}

\section{Preliminaries}
\label{sec:prelim}

The effective weak Hamiltonian for charmless hadronic $B$ decays,
without change in strangeness ($\Delta S=0$), is given by
\cite{Buchalla:1995vs}
\begin{equation}\label{heff}
   {\cal H}^{\Delta S=0}_{\rm eff} = \frac{G_F}{\sqrt2} \sum_{p=u,c} \!
   \lambda_p \bigg( C_1\,Q_1^p + C_2\,Q_2^p
   + \!\sum_{i=3,\dots, 10}\! C_i\,Q_i + C_{7\gamma}\,Q_{7\gamma}
   + C_{8g}\,Q_{8g} \bigg) + \mbox{h.c.} 
\end{equation}
where the elements of the CKM quark-mixing matrix $V$ enter
as $\lambda_p=V_{pb} V_{pd}^*$, $C_i$ are Wilson coefficients,
and the operators $Q_i$ read 
\begin{equation}
\begin{aligned}
   Q_1^p &= (\bar p b)_{V-A} (\bar d p)_{V-A} \,, \\
   Q_3 &= (\bar d b)_{V-A} \sum{}_{\!q}\,(\bar q q)_{V-A} \,, \\
   Q_5 &= (\bar d b)_{V-A} \sum{}_{\!q}\,(\bar q q)_{V+A} \,, \\
   Q_7 &= (\bar d b)_{V-A} \sum{}_{\!q}\,{\textstyle\frac32} e_q  
    (\bar q q)_{V+A} \,,  \\
   Q_9 &= (\bar d b)_{V-A} \sum{}_{\!q}\,{\textstyle\frac32} e_q 
    (\bar q q)_{V-A} \,, \\
   Q_{7\gamma} &= \frac{e}{8\pi^2}\,m_b\, 
    \bar d\sigma_{\mu\nu}(1+\gamma_5) F^{\mu\nu} b \,,
\end{aligned}
\qquad\quad
\begin{aligned}
    Q^p_2 &= (\bar p_i b_j)_{V-A} (\bar d_j p_i)_{V-A} \,, \\
    Q_4 &= (\bar d_i b_j)_{V-A} \sum{}_{\!q}\,(\bar q_j q_i)_{V-A} \,, \\
    Q_6 &= (\bar d_i b_j)_{V-A} \sum{}_{\!q}\,(\bar q_j q_i)_{V+A} \,, \\
    Q_8 &= (\bar d_i b_j)_{V-A} \sum{}_{\!q}\,{\textstyle\frac32} e_q
    (\bar q_j q_i)_{V+A} \,, \\
    Q_{10} &= (\bar d_i b_j)_{V-A} \sum{}_{\!q}\,{\textstyle\frac32} e_q
    (\bar q_j q_i)_{V-A} \,, \\
   Q_{8g} &= \frac{g}{8\pi^2}\,m_b\, 
    \bar d\sigma_{\mu\nu}(1+\gamma_5) G^{\mu\nu} b 
\end{aligned}
\label{qqi}
\end{equation}
Here $i,j$ are colour indices, $e_q$ are the quark charges,
and the sums extend over $q=u,d,s,c,b$. 
The Wilson coefficients $C_i$ will be taken at 
next-to-leading order (NLO),
using the treatment of electroweak contributions described
in detail in \cite{Beneke:2001ev}.
The sign conventions for the electromagnetic and strong coupling
correspond to the covariant derivative
$D_\mu=\partial_\mu + ie Q_f A_\mu+ig T^a A^a_\mu$.
With these definitions the coefficients $C_{7\gamma}$,
$C_{8g}$ are negative in the Standard Model, which is the
convention usually adopted in the literature.

The effective Hamiltonian for charmless decays
of $B$ mesons with $\Delta S=1$ can be obtained from (\ref{heff})
by interchanging $d$- and $s$-quark labels. The CKM factors 
governing these transitions are then $\lambda_p'=V_{pb} V_{ps}^*$.

To obtain the amplitudes for $\bar B\to V_{1L}V_{2L}$ decays from the
Hamiltonian, the matrix elements of the operators $Q_i$ have to be
computed in QCD factorization 
\cite{Beneke:2001ev,Beneke:1999br,Beneke:2000ry}. 
To lowest order the matrix elements are expressed in terms of 
$\bar B\to V$ form factors and vector-meson decay constants.
The required form factors are defined by
(see e.g. \cite{Charles:1998dr})
\begin{eqnarray}
\langle V(p,\epsilon)|\bar q\gamma^\mu\gamma_5 b|\bar B(p_B)\rangle &=&
2m_V A_0(q^2)\frac{\epsilon\cdot q}{q^2}q^\mu +
(m_B+m_V)A_1(q^2)\left[\epsilon^\mu-\frac{\epsilon\cdot q}{q^2}q^\mu\right]
\nonumber\\
&& -A_2(q^2)\frac{\epsilon\cdot q}{m_B+m_V}
\left[(p_B+p)^\mu -\frac{m^2_B-m^2_V}{q^2}q^\mu\right]
\label{formfa}
\end{eqnarray}
\begin{equation}\label{formfv}
\langle V(p,\epsilon)|\bar q\gamma^\mu b|\bar B(p_B)\rangle =
-2 i\frac{V(q^2)}{m_B+m_V}\varepsilon^{\mu\nu\rho\sigma}
p_{B\nu}p_\rho \epsilon_\sigma
\end{equation}
where the momentum transfer is $q=p_B-p$ and the totally antisymmetric tensor
$\varepsilon^{\mu\nu\rho\sigma}$ is normalized by 
$\varepsilon^{0123}=-1$.

The vector-meson decay constant $f_V$ is given by
\begin{equation}\label{fvdef}
\langle V(q,\eta)|\bar q\gamma_\mu q'|0\rangle = -i f_V m_V \eta_\mu
\end{equation}
for a vector meson with flavour content $V=\bar q' q$.
The corresponding matrix element where $\gamma_\mu\to\gamma_\mu\gamma_5$
is zero.
For energetic vector mesons $V(p,\epsilon)$ with longitudinal polarization
\begin{equation}\label{epslong}
\epsilon^\mu = \frac{p^\mu}{m_V}
\end{equation}
up to corrections of second order in $m_V/m_B$. 

The factorized matrix element of a $(V-A)\otimes (V-A)$ operator then
reads to lowest order ($\alpha^0_s$)
\begin{equation}\label{melo}
\langle V_{1L}V_{2L}|(\bar q_1b)_{V-A}(\bar q_2 q'_2)_{V-A}|\bar B_q\rangle
= i m^2_B A^{B\to V_1}_0(m^2_{V_2}) f_{V_2}
\end{equation}
where $V_1=\bar qq_1$ and $V_2=\bar q'_2 q_2$ ($q_1\not= q_2$).

The corrections at higher order in $\alpha_s$ are expressed in terms of
calculable hard-scattering kernels and meson light-cone distribution
amplitudes. The latter quantities enter through the meson projectors
in momentum space. For the $B$ meson this projector is given by
\begin{equation}\label{bpro}
b\bar q=\frac{i f_B}{4}(\not\! p_B + m_B)\gamma_5
\left[\phi_{B1}(\xi) + \phi_{B2}(\xi) \not\! n\right]
\end{equation}
In our notation $(b\bar q)$ denotes a matrix in Dirac space displaying the flavour
composition of a $\bar B_q$ meson in the initial state.
The $4$-vector $n^\mu = (1,0,0,-1)$ is chosen to be in the direction of the
recoiling meson $V_1$. The parameter $\xi$ is the light-cone momentum fraction of
the spectator quark $\bar q$. The distribution amplitudes are
normalized as
\begin{equation}\label{phibnorm}
\int_0^1 d\xi\, \phi_{B1}(\xi)=1, \qquad\quad
\int_0^1 d\xi\, \phi_{B2}(\xi)=0
\end{equation}
In the present analysis $\phi_{B2}$ does not enter the results and
$\phi_{B1}$ appears only through the first inverse moment
\begin{equation}\label{lambdef}
\int_0^1 d\xi\, \frac{\phi_{B1}}{\xi} = \frac{m_B}{\lambda_B}
\end{equation}
which defines the hadronic parameter $\lambda_B={\cal O}(\Lambda_{QCD})$.
Colour indices have been suppressed in writing (\ref{bpro}). They are taken 
into account by replacing $b\bar q\to b_i\bar q_j$ and including a factor of 
$\delta_{ij}/N_c$ on the right-hand side.

For a longitudinally polarized vector meson in the final state with flavour 
content $\bar q_2 q_1$ and momentum $p$, the projector can be written as
\cite{Beneke:2003zv}
\begin{equation}\label{vpro}
q_2\bar q_1 =\frac{i f_V}{4}\not\! p\, \phi_{||}(x) -\frac{i f^\perp_V}{4}
m_V\frac{\not\! k_2 \not\! k_1}{k_2\cdot k_1}\, \Phi_v(x)
\end{equation}
Here $x$ is the momentum fraction of the final-state quark $q_1$ and
\begin{eqnarray}
k^\mu_1 &=& x p^\mu + k^\mu_\perp +\frac{\vec k^2_\perp}{2 x m_B} n^\mu \label{kk1}\\
k^\mu_2 &=& \bar x p^\mu - k^\mu_\perp +\frac{\vec k^2_\perp}{2 \bar x m_B} n^\mu \label{kk2}
\end{eqnarray}
(with $\bar x=1-x$) are the momenta of $q_1$ and $\bar q_2$, respectively.
$n^\mu$ is a light-like vector with spatial direction opposite to $p^\mu$:
If $p\sim n_+$ ($p\sim n_-$) then $n=n_-$ ($n=n_+$), where
$n^\mu_\pm = (1,0,0,\pm 1)$.

The function $\phi_{||}(x)$ is the light-cone distribution amplitude
of leading twist for a longitudinal vector meson.
The subleading-twist amplitude $\Phi_v(x)$ has been treated in (\ref{vpro})
in the Wandzura-Wilczek approximation. It gives rise to contributions
suppressed by one power of $\Lambda_{QCD}/m_B$.
We will nevertheless include it in order to estimate the impact of
this particular source of power corrections to factorization in
the heavy-quark limit.

The functions $\phi_{||}$ and $\Phi_v$ can be expanded in terms of
Gegenbauer and Legendre polynomials, respectively,
\begin{eqnarray}
\phi_{||}(x) &=& 6 x\bar x \sum_{n=0}^\infty \alpha_n C^{3/2}_n(2x-1)
\label{phipcn}\\
\Phi_v(x) &=& 3 \sum_{n=0}^\infty \alpha_{n\perp} P_{n+1}(2x-1)
\label{phivpn}
\end{eqnarray}
where $\alpha_0=\alpha_{0\perp}=1$.
In the Wandzura-Wilczek approximation $\Phi_v$ can be expressed in terms
of the twist-2 wave function of a transversely polarized vector meson,
$\phi_\perp$, as
\begin{equation}\label{phivt}
\Phi_v(x) = \int_0^x du\, \frac{\phi_\perp(u)}{\bar u} -
            \int_x^1 du\, \frac{\phi_\perp(u)}{u}
\end{equation}
$\phi_\perp$ has an expansion similar to (\ref{phipcn}) and this leads
to (\ref{phivpn}).

Note that (\ref{phivt}) implies
\begin{equation}\label{intphiv0}
\int_0^1 dx\, \Phi_v(x) = 0
\end{equation}
even though $\Phi_v(x)$ is not necessarily antisymmetric under
$x\leftrightarrow\bar x$ for general $\alpha_{n\perp}$.
The normalization of $\phi_{||,\perp}$ is 
$\int_0^1 dx\, \phi_{||,\perp}(x)=1$.

For phenomenological applications we shall truncate the expansions
of $\phi_{||}$ and $\Phi_v$ and use, for a particular meson $V$,
\begin{equation}\label{phip12}
\phi_{||}^V(x) = 6x\bar x\left[1+\alpha^V_1\, 3(2x-1) 
+\alpha^V_2\, 6(5x^2-5x+1)\right]
\end{equation}
\begin{equation}\label{phiv0}
\Phi_v^V(x) = 3(2x-1) 
\end{equation}

Taking the vacuum-to-meson matrix element of a local current,
the projector (\ref{vpro}) reproduces (\ref{fvdef}), (\ref{epslong})
\begin{equation}\label{fvpro}
\langle V(p)|\bar q_1\gamma_\mu q_2 |0\rangle =
-\int_0^1dx\,{\rm tr}\,\gamma_\mu q_2\bar q_1 =-i f_V p_\mu
\end{equation}
and
\begin{equation}\label{pro0}
\langle V(p)|\bar q_1\Gamma q_2 |0\rangle = 0
\end{equation}
for $\Gamma=1$, $\gamma_5$, $\gamma_\mu\gamma_5$.

\afterpage{\clearpage}

\section{\boldmath QCD factorization in $B\to V_LV_L$ decays\unboldmath}
\label{sec:qcdf}

The amplitudes for the $\Delta S=0$ decay of a $\bar B$ meson into a pair
of light vector mesons with longitudinal polarization
can be conveniently expressed as follows
(the case of $\Delta S=1$ is obtained by replacing $d\leftrightarrow s$):
\begin{equation}\label{hefftp}
   \langle V_{1L}V_{2L}|{\cal H}^{\Delta S=0}_{\rm eff}|\bar B\rangle
   = \frac{G_F}{\sqrt2} \sum_{p=u,c} \lambda_p\,
\langle V_{1L}V_{2L}|{\cal T}^d_p+{\cal T}_p^{{\rm ann},d}|\bar B\rangle 
\end{equation}
where
\begin{eqnarray}\label{tpdef}
   {\cal T}^d_p &=& a_1(V_1V_2)\,\delta_{pu}\,
    (\bar u b)_{V-A} \otimes (\bar d u)_{V-A} \nonumber\\
   &+& a_2(V_1V_2)\,\delta_{pu}\,
    (\bar d b)_{V-A} \otimes (\bar u u)_{V-A} \nonumber\\
   &+& a_3(V_1V_2) \sum{}_{\!q}\, (\bar d b)_{V-A} \otimes
    (\bar q q)_{V-A} \nonumber\\
   &+& a_4^p(V_1V_2) \sum{}_{\!q}\, (\bar q b)_{V-A} \otimes
    (\bar d q)_{V-A} \nonumber\\
   &+& a_5(V_1V_2) \sum{}_{\!q}\, (\bar d b)_{V-A} \otimes
    (\bar q q)_{V+A} \nonumber\\
   &+& a_7(V_1V_2) \sum{}_{\!q}\, (\bar d b)_{V-A} \otimes
    {\textstyle\frac32} e_q(\bar q q)_{V+A} \nonumber\\
   &+& a_9(V_1V_2)\sum{}_{\!q}\, (\bar d b)_{V-A} \otimes
    {\textstyle\frac32} e_q(\bar q q)_{V-A} \nonumber\\
   &+& a_{10}^p(V_1V_2) \sum{}_{\!q}\, (\bar q b)_{V-A} \otimes
    {\textstyle\frac32} e_q(\bar d q)_{V-A} 
\end{eqnarray}
Here the summation is over $q=u,d,s$. The symbol $\otimes$ indicates that 
the matrix elements of the operators in ${\cal T}^d_p$ are to be evaluated 
in factorized form \cite{Beneke:2001ev}. 
The factorization coefficients $a_i$ include 
hard QCD corrections to the $B$-decay matrix elements at NLO,
as well as electroweak effects in the systematic approximation
of \cite{Beneke:2001ev}. Note that structures with
scalar and pseudoscalar currents are absent in (\ref{tpdef}),
in contrast to the case of $B\to K\pi$ considered in \cite{Beneke:2001ev}.
Because (pseudo)scalar currents cannot create a vector meson from the 
vacuum, these structures can give no contribution to $B\to VV$ decays. 

The term ${\cal T}_p^{{\rm ann},d}$ in (\ref{hefftp}) 
describes the effects of weak annihilation.
These are power suppressed in the heavy-quark limit and
cannot be computed in QCD factorization. We shall use
model calculations to estimate this important class of
power corrections to the leading, factorizable
amplitudes. Weak annihilation will be discussed in
Section~\ref{subsec:annihilation}.

\subsection{\boldmath Results for the parameters $a_i$\unboldmath}
\label{subsec:airesults}

The factorization coefficients can be written as 
$a_i=a_{i,\rm I}+a_{i,\rm II}$. We find
\begin{eqnarray}\label{ai}
   a_{1,\rm I} &=& C_1 + \frac{C_2}{N_c} \left[ 1
    + \frac{C_F\alpha_s}{4\pi}\,V_{V} \right] \,,
    \nonumber\\[-1.25cm]
   &&\hspace{7.7cm}
    a_{1,\rm II} = \frac{C_2}{N_c}\,\frac{C_F\pi\alpha_s}{N_c}\,
    H_{V_1V_2} \,,
    \nonumber\\
   a_{2,\rm I} &=& C_2 + \frac{C_1}{N_c} \left[ 1
    + \frac{C_F\alpha_s}{4\pi}\,V_{V} \right] \,,
    \nonumber\\[-1.25cm]
   &&\hspace{7.7cm}
    a_{2,\rm II} = \frac{C_1}{N_c}\,\frac{C_F\pi\alpha_s}{N_c}\,
    H_{V_1V_2} \,,
    \nonumber\\
   a_{3,\rm I} &=& C_3 + \frac{C_4}{N_c} \left[ 1
    + \frac{C_F\alpha_s}{4\pi}\,V_{V} \right] \,,
    \nonumber\\[-1.25cm]
   &&\hspace{7.7cm} 
    a_{3,\rm II} = \frac{C_4}{N_c}\,\frac{C_F\pi\alpha_s}{N_c}\,
    H_{V_1V_2} \,,
    \nonumber\\
   a_{4,\rm I}^p &=& C_4 + \frac{C_3}{N_c} \left[ 1
    + \frac{C_F\alpha_s}{4\pi}\,V_{V} \right]
    - \frac{C_5}{N_c} \frac{C_F\alpha_s}{4\pi}\, r^{V}_\perp V^\perp_{V} 
    \nonumber\\
    &&+ \frac{C_F\alpha_s}{4\pi N_c}\,
       (P_{V ,2}^p -r^{V}_\perp P_{V ,3}^p) \,, 
    \nonumber\\[-1.25cm]
   &&\hspace{7.7cm}
    a_{4,\rm II} = \frac{C_3}{N_c}\,\frac{C_F\pi\alpha_s}{N_c}\,
    H_{V_1V_2} \,,
    \nonumber\\
   a_{5,\rm I} &=& C_5 + \frac{C_6}{N_c} \left[ 1
    + \frac{C_F\alpha_s}{4\pi}\,(-V_{V}') \right] \,,
    \nonumber\\[-1.25cm]
   &&\hspace{7.7cm} 
    a_{5,\rm II} = \frac{C_6}{N_c}\,\frac{C_F\pi\alpha_s}{N_c}\,
    (-H'_{V_1V_2})
    \,, \nonumber\\
   a_{7,\rm I}^p &=& C_7 + \frac{C_8}{N_c} \left[ 1
    + \frac{C_F\alpha_s}{4\pi}\,(-V_{V}') \right] 
    + \frac{\alpha}{9\pi} P_{V,n}^{p,{\rm EW}}   \,,
    \nonumber\\[-1.25cm]
   &&\hspace{7.7cm} 
    a_{7,\rm II} = \frac{C_8}{N_c}\,\frac{C_F\pi\alpha_s}{N_c}\,
    (-H'_{V_1V_2})
    \,, \nonumber\\
   a_{9,\rm I}^p &=& C_9 + \frac{C_{10}}{N_c} \left[ 1
    + \frac{C_F\alpha_s}{4\pi}\,V_{V} \right] 
    + \frac{\alpha}{9\pi} P_{V,n}^{p,{\rm EW}} \,,
    \nonumber\\[-1.25cm]
   &&\hspace{7.7cm} 
    a_{9,\rm II} = \frac{C_{10}}{N_c}\,\frac{C_F\pi\alpha_s}{N_c}\,
    H_{V_1V_2} \,,
    \nonumber\\
   a_{10,\rm I}^p &=& C_{10} + \frac{C_9}{N_c} \left[ 1
    + \frac{C_F\alpha_s}{4\pi}\,V_{V} \right]
   - \frac{C_7}{N_c} \frac{C_F\alpha_s}{4\pi}\, r^{V}_\perp V^\perp_{V}
   \nonumber\\ 
   &&+ \frac{\alpha}{9\pi N_c}\,
      (P_{V,2}^{p,{\rm EW}}-r^{V}_\perp P_{V,3}^{p,{\rm EW}}) \,,
    \nonumber\\[-1.25cm]
   &&\hspace{7.7cm} 
    a_{10,\rm II} = \frac{C_9}{N_c}\,\frac{C_F\pi\alpha_s}{N_c}\,
    H_{V_1V_2} 
\end{eqnarray}
where $C_i\equiv C_i(\mu)$, $\alpha_s\equiv\alpha_s(\mu)$, 
$C_F=(N_c^2-1)/(2N_c)$, and $N_c=3$. The hadronic
quantities $V_V^{(\prime)}$, 
$H_{V_1 V_2}^{(\prime)}$, $P_{V,2}^p$, $P_{V,3}^p$, $P_{V,2}^{p,{\rm EW}}$, 
$P_{V,3}^{p,{\rm EW}}$, and $P_{V,n}^{p,{\rm EW}}$ are given below. 
All indices ${}_V$ in $V_V$, $P_V$, $r^V$ are understood to refer to the
emitted meson $V_2$.  

Contributions suppressed by one power of $\Lambda_{QCD}/m_b$ that
arise from the twist-3 component of the vector-meson wave function
have been included in the above expressions.
They are related to the scalar penguin operator 
$(\bar qb)_{S-P}(\bar sq)_{S+P}$ and come with a factor
\begin{equation}\label{rperpdef}
r^V_\perp(\mu) = \frac{2 m_V f^\perp_V(\mu)}{{m}_b(\mu) f_V} =
\frac{2 m_V f^\perp_V(1\rm{GeV})}{{m}_b(m_b) f_V}
\left[\frac{\alpha_s(\mu)}{\alpha_s(m_b)}\right]^{-3 C_F/\beta_0}
\left[\frac{\alpha_s(\mu)}{\alpha_s(1\rm{GeV})}\right]^{C_F/\beta_0}
\end{equation}
Here ${m}_b(\mu)$ is the $\overline{MS}$-mass 
of the $b$ quark at scale $\mu$, and $\beta_0=23/3$ for $f=5$ flavours
of quarks.

\subsubsection{Vertex and penguin contributions}

\begin{figure}[t]
\begin{center}
\includegraphics[width=\linewidth]{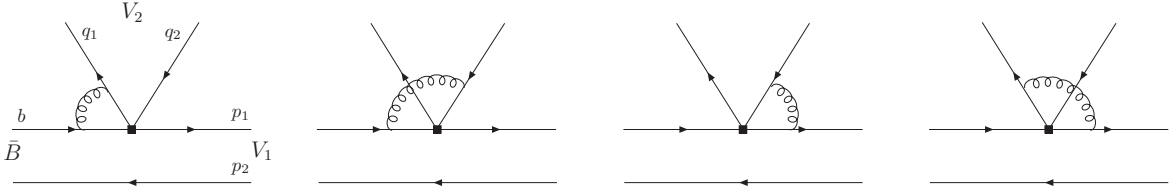}
\end{center}
{\caption{\label{fig:vertex} Vertex diagrams.}}
\end{figure}
The vertex corrections (Fig. \ref{fig:vertex}) are given by 
\begin{eqnarray}\label{vvg}
   V_V &=& 12\ln\frac{m_b}{\mu} - 18 + \int_0^1\! dx\,g(x)\,\phi^V_{||}(x)
    \,, \nonumber\\
   V_V' &=& 12\ln\frac{m_b}{\mu} - 6 + \int_0^1\! dx\,g(1-x)\,\phi^V_{||}(x)
    \,, \nonumber\\
   g(x) &=& 3\left( \frac{1-2x}{1-x}\ln x-i\pi \right) \nonumber\\
   &&\mbox{}+ \left[ 2 \, L_2(x) - \ln^2\!x + \frac{2\ln x}{1-x}
    - (3+2i\pi)\ln x - (x\leftrightarrow 1-x) \right] 
\end{eqnarray}
\begin{equation}\label{vvgperp}
   V^\perp_V = \int_0^1\! dx\, \left[ 2 \, L_2(x) - \ln^2\!x - 
   (1+2i\pi)\ln x - (x\leftrightarrow 1-x)\right] \,\Phi^V_{v}(x) 
\end{equation}
where $L_2(x)$ is the dilogarithm
\begin{equation}\label{l2def}
L_2(x)=-\int_0^x\, dt\, \frac{\ln(1-t)}{t}
\end{equation}

The expansion of $\phi^V_{||}$ in Gegenbauer polynomials gives
\begin{equation}\label{vgegenb}
   \int_0^1\!dx\,g(x)\,\phi^V_{||}(x)
   = -\frac12 - 3i\pi + \left( \frac{11}{2} - 3i\pi \right)
   \alpha_1^V - \frac{21}{20}\,\alpha_2^V + \dots 
\end{equation}
Replacing $g(x)$ by $g(1-x)$ leads to a change of sign
in front of the odd Gegenbauer coefficients on the right-hand side. 

\begin{figure}[t]
\begin{center}
\includegraphics[width=\linewidth]{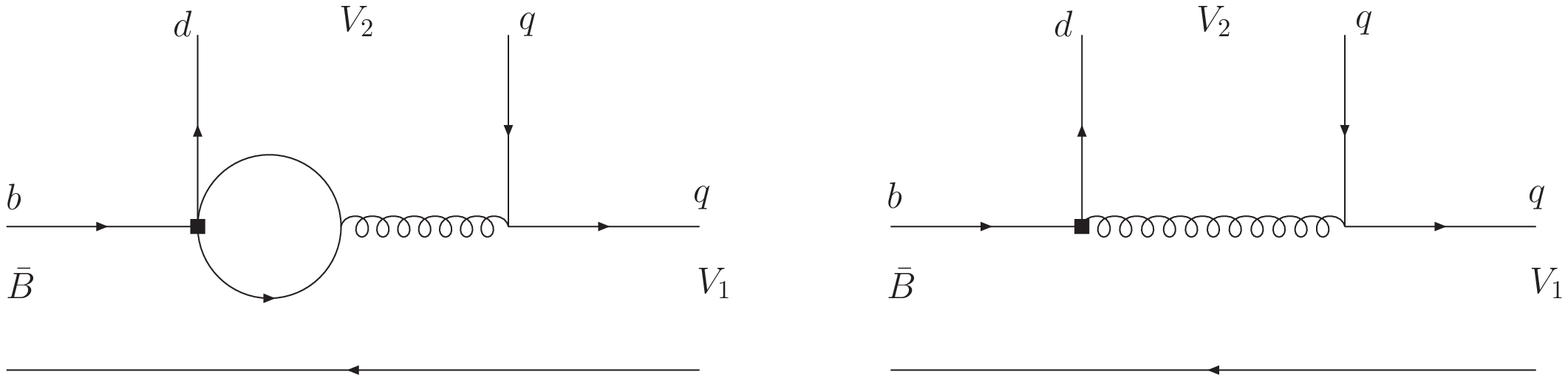}
\end{center}
{\caption{\label{fig:penguin} Penguin diagrams.}}
\end{figure}

Next, the penguin contributions (Fig. \ref{fig:penguin}) are
\begin{eqnarray}\label{pv2}
   P_{V,2}^p &=& C_1 \left[ \frac43\ln\frac{m_b}{\mu}
    + \frac23 - G_V(s_p) \right]
    + C_3 \left[ \frac83\ln\frac{m_b}{\mu} + \frac43
    - G_V(0) - G_V(1) \right] \nonumber\\
   &&\mbox{}+ (C_4+C_6) \left[ \frac{20}{3}\ln\frac{m_b}{\mu}
    - 3 G_V(0) - G_V(s_c) - G_V(1) \right] \nonumber\\
   &&\mbox{}- 2 C_{8g}^{\rm eff} \int_0^1 \frac{dx}{1-x}\,
    \phi^V_{||}(x) \,, \nonumber\\
   P_{V,2}^{p,{\rm EW}} &=& (C_1+N_c C_2) \left[
    \frac43\ln\frac{m_b}{\mu} + \frac23 - G_V(s_p) \right]
    - 3\,C_{7\gamma}^{\rm eff} \int_0^1 \frac{dx}{1-x}\,\phi^V_{||}(x) 
\end{eqnarray}
where $s_u=0$ and $s_c=(m_c/m_b)^2$.
Small contributions from the electroweak coefficients $C_7,\ldots,C_{10}$ 
are consistently neglected in $P^p_{V,2}$ within our approximation
scheme.
Also, the very small corrections from $C_3,\ldots,C_6$
in $P_{V,2}^{p,{\rm EW}}$ (and in $P_{V,3}^{p,{\rm EW}}$,
see (\ref{hatpv3}) below) are omitted for simplicity.

The function $G_V(s)$ is 
\begin{eqnarray}\label{gvs}
   G_V(s) &=& \int_0^1\!dx\,G(s-i\epsilon,1-x)\,\phi^V_{||}(x) \,, \\
   G(s,x) &=& -4\int_0^1\!du\,u(1-u) \ln[s-u(1-u)x] \nonumber\\
   &=& \frac{2(12s+5x-3x\ln s)}{9x}
    - \frac{4\sqrt{4s-x}\,(2s+x)}{3x^{3/2}}
    \arctan\sqrt{\frac{x}{4s-x}}  
\end{eqnarray}
Expanding in Gegenbauer moments one finds 
\begin{eqnarray}
   G_V(s_c) &=& \frac53 - \frac23\ln s_c + \frac{\alpha_1^V}{2}
    + \frac{\alpha_2^V}{5} + \frac43 \left( 8 + 9 \alpha_1^V
    + 9 \alpha_2^V \right) s_c \nonumber\\
   &&\mbox{}+ 2(8 + 63 \alpha_1^V + 214 \alpha_2^V) s_c^2
    - 24 (9 \alpha_1^V + 80 \alpha_2^V) s_c^3
    + 2880 \alpha_2^V s_c^4 \nonumber\\
   &&\mbox{}- \frac23\sqrt{1-4s_c}\,\bigg[ 1 + 2 s_c
    + 6 (4 + 27 \alpha_1^V + 78 \alpha_2^V) s_c^2 \nonumber\\
   &&\quad\mbox{}- 36 (9 \alpha_1^V + 70 \alpha_2^V) s_c^3 
    + 4320 \alpha_2^V s_c^4 \bigg] \left(
    2 \,\mbox{arctanh}\sqrt{1-4 s_c} - i\pi \right) \nonumber\\
   &&\mbox{}+ 12 s_c^2\,\bigg[ 1 + 3 \alpha_1^V + 6 \alpha_2^V
    - \frac43 \left( 1 + 9 \alpha_1^V + 36 \alpha_2^V \right) s_c
    \nonumber\\
   &&\quad\mbox{}+ 18 (\alpha_1^V + 10 \alpha_2^V) s_c^2 
    - 240 \alpha_2^V s_c^3 \bigg] \left(
    2 \,\mbox{arctanh}\sqrt{1-4 s_c} - i\pi \right)^2 + \dots \,,
    \nonumber\\
   G_V(0) &=& \frac53 + \frac{2i\pi}{3} + \frac{\alpha_1^V}{2}
   + \frac{\alpha_2^V}{5} + \dots \,, \nonumber\\
   G_V(1) &=& \frac{85}{3} - 6\sqrt3\,\pi + \frac{4\pi^2}{9}
    - \left( \frac{155}{2} - 36\sqrt3\,\pi+12\pi^2 \right)
    \alpha_1^V \nonumber\\
   &&\mbox{}+ \left( \frac{7001}{5} - 504\sqrt3\,\pi
    + 136\pi^2 \right) \alpha_2^V + \dots 
\end{eqnarray}
The function $G_V(s)$ and its expansion in Gegenbauer moments have the 
same form as $G_K(s)$ in the case of $B\to K\pi$ discussed in 
\cite{Beneke:2001ev}.
Likewise the integrals proportional to 
$C_{7\gamma}^{\rm eff}$ and $C_{8g}^{\rm eff}$ in (\ref{pv2})
are similar to those in $B\to K\pi$. They read
\begin{equation}\label{eq54}
   \int_0^1 \frac{dx}{1-x}\,\phi^V_{||}(x)
   = 3(1+\alpha_1^V + \alpha_2^V + \dots) 
\end{equation}

The twist-3 terms from the penguin diagrams are obtained from the twist-2
terms by the replacement $\phi^V_{||}(x)\to\Phi^V_v(x)$, except for the 
terms proportional to $C_{7\gamma}^{\rm eff}$ and $C_{8g}^{\rm eff}$. 
Here the factor of $(1-x)$ in the denominator of the integral
in (\ref{pv2}) is canceled by the twist-3 projection.
An important difference between the twist-3 penguin contributions in 
$B\to V_LV_L$ and $B\to K\pi$ arises from the different properties of 
the twist-3 wave functions in these two cases.
Since $\int_0^1dx\, \Phi^V_v(x)=0$ it follows that the contributions
from $C_{7\gamma}^{\rm eff}$ and $C_{8g}^{\rm eff}$ vanish in the former 
case. The same
holds for all $x$-independent constants in the hard-scattering
kernel, in particular for the scale and scheme dependent terms.
We then find
\begin{equation}
   P_{V,3}^p = -\left[
    C_1 \, \hat G_V(s_p) + C_3 \, (\hat G_V(0) + \hat G_V(1)) 
   + (C_4+C_6)\, 
     (3 \hat G_V(0) + \hat G_V(s_c) + \hat G_V(1)) \right] \nonumber
\end{equation}
\begin{equation}\label{hatpv3}
   P_{V,3}^{p,{\rm EW}} = -(C_1+N_c C_2) \, \hat G_V(s_p) 
\end{equation}
with 
\begin{equation}\label{penfunction1}
   \hat G_V(s) = \int_0^1\!dx\,G(s-i\epsilon,1-x)\,\Phi_v^V(x) 
\end{equation}
Using the asymptotic form of the wave function
$\Phi^V_v(x)=3(2x-1)$ leads to
\begin{eqnarray}\label{hatgvs}
\hat G_V(s_c) &=& 1 - 36 s_c + 
  12 s_c\sqrt{1-4s_c}\left( 2\,\mbox{arctanh}\sqrt{1-4 s_c} - i\pi \right)
  \nonumber\\ 
 && - 12 s^2_c\left( 2\,\mbox{arctanh}\sqrt{1-4 s_c} - i\pi \right)^2
\,, \nonumber\\
   \hat G_V(0) &=& 1 \,, \qquad\quad
    \hat G_V(1) = \frac{4}{3}\pi^2 + 4\sqrt{3}\pi -35  
\end{eqnarray}

Finally, we give the electromagnetic penguin contributions 
$P_{V,n}^{p,{\rm EW}}$ ($p=u$, $c$).
For intermediate charm, $p=c$, these are calculable in perturbation theory
and read
\begin{equation}\label{pewnc}
P_{V,n}^{c,{\rm EW}}=(C_1+N_c C_2)\left[ \frac{4}{3}\ln\frac{m_b}{\mu} 
 +\frac{2}{3}+\frac{4}{3}\ln\frac{m_c}{m_b}\right] -3 C^{\rm eff}_{7\gamma}
\end{equation}
In the case of the up-quark loop, $p=u$, the amplitude becomes sensitive
to additional long-distance dynamics, which is not strictly
calculable.
Using a suitable hadronic representation of the light-quark loop, we estimate
\begin{eqnarray}\label{pewnu}
P_{V,n}^{u,{\rm EW}} &=& (C_1+N_c C_2)\Bigg[ \frac{4}{3}\ln\frac{m_b}{\mu} 
 -\frac{10}{9}+\frac{4\pi^2}{3}
\sum_{r=\rho,\omega}\frac{f^2_r}{m^2_{V}-m^2_r + i m_r\Gamma_r}
\nonumber\\
&& -\frac{2\pi}{3}\frac{m^2_{V}}{t_c}i+\frac{2}{3}\ln\frac{m^2_{V}}{m^2_b}
+\frac{2}{3}\, \frac{t_c-m^2_{V}}{t_c}\ln\frac{t_c-m^2_{V}}{m^2_{V}} \Bigg] 
 - 3 C^{\rm eff}_{7\gamma}
\end{eqnarray}
where $t_c=4\pi^2(f^2_\rho + f^2_\omega)$.
This point is discussed further in appendix \ref{sec:ldempeng}.

The authors of \cite{Beneke:2003zv} factorize the term (\ref{pewnu})
into a short-distance and a long-distance part, separated by a scale
$\nu$. The short-distance part is equivalent to (\ref{pewnc}) with
$m_c$ replaced by $\nu$. The long-distance part is not considered
explicitly in \cite{Beneke:2003zv}. Our treatment is consistent with
the framework of \cite{Beneke:2003zv}, but supplies a concrete model
representation for the long-distance contribution of the electromagnetic
penguin.

\subsubsection{Hard spectator scattering}

\begin{figure}[t]
\begin{center}
\includegraphics[width=\linewidth]{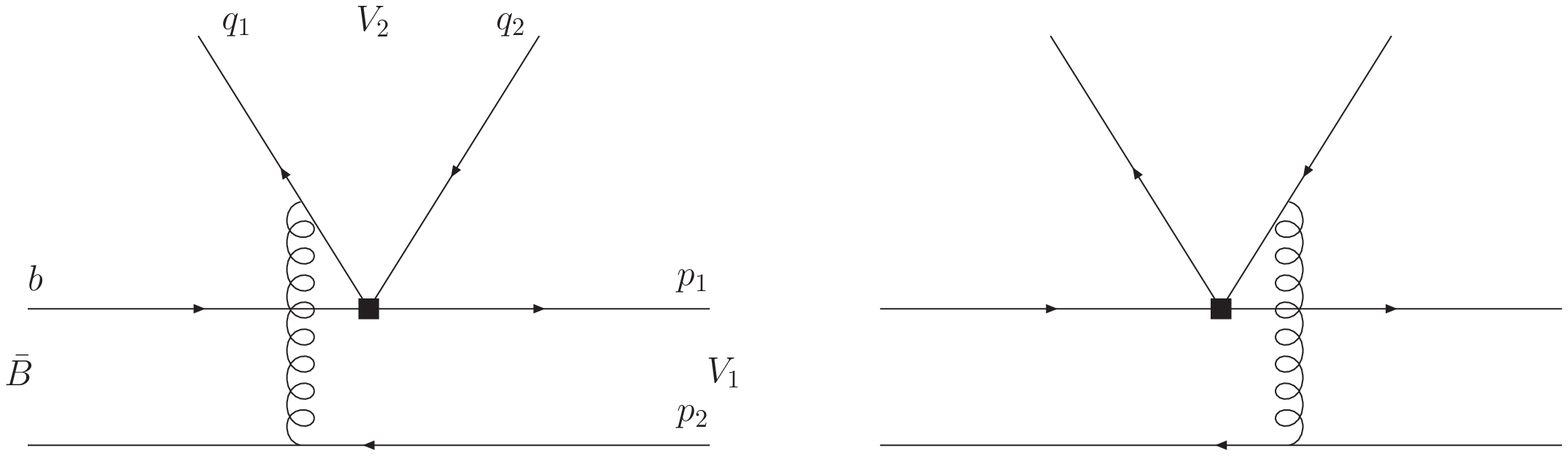}
\end{center}
{\caption{\label{fig:hspec} Hard spectator diagrams.}}
\end{figure}
The hard spectator interactions (Fig. \ref{fig:hspec}) determining the 
coefficients $a_{i,II}$ in (\ref{ai}) are governed by the quantities
\begin{align}
H_{V_1 V_2} &=\frac{f_B f_{V_1}}{m_B^2 A_0^{B\rightarrow V_1}(0)}
\int_0^1\frac{d\xi}{\xi}\Phi_B(\xi)
\int_0^1\frac{dx}{\bar{x}}\phi_{||}^{V_2}(x)\int_0^1\frac{dy}{\bar{y}}
\left[\phi_{||}^{V_1}(y)+r^{V_1}_\perp\frac{\bar{x}}{x}\Phi_v^{V_1}(y)\right]
\label{hv1v2}\\
H'_{V_1 V_2} &=\frac{f_B f_{V_1}}{m_B^2 A_0^{B\rightarrow V_1}(0)}
\int_0^1\frac{d\xi}{\xi}\Phi_B(\xi)\int_0^1\frac{dx}{x}\phi_{||}^{V_2}(x)
\int_0^1\frac{dy}{\bar{y}}
\left[\phi_{||}^{V_1}(y)+r^{V_1}_\perp\frac{x}{\bar{x}}\Phi_v^{V_1}(y)\right]
\label{hpv1v2}
\end{align}
Strictly speaking, only the twist-2 components $\phi_{||}$
of the vector meson distribution amplitudes contribute
at leading power and are consistently
calculable in the present approach. The twist-3 part
described by $\Phi_v$ leads to terms with
logarithmic endpoint singularities, but these terms are
suppressed by one power of $\Lambda/m_b$.
We shall include them in our analysis using a simple model, in order
to estimate the potential impact of power corrections from this source.
Following \cite{Beneke:2001ev}, we parametrize the endpoint
singularity by
\begin{equation}\label{xhdef}
X_H=\int_0^1\frac{dy}{\bar y}= \left( 1+\rho_H \, e^{i\phi_H}\right)
\ln\frac{m_B}{\Lambda_h}
\end{equation}
The logarithm comes from cutting off the 
lower range of integration at $\bar y_{min}=\Lambda_h/m_B$, and
$\rho_H$, $\phi_H$ are real model parameters to allow for
a complex $X_H$ and a deviation from the default value 
$\ln(m_B/\Lambda_h)$.
The integral over $\Phi_v$ in (\ref{hv1v2}), (\ref{hpv1v2})
then becomes
\begin{equation}\label{intphiv}
\int^1_0\frac{dy}{\bar y}\, \Phi_v(y) =3(X_H-2)
\end{equation}

Throughout we use
$\mu_h=\sqrt{\Lambda_h\,\mu}$ with $\Lambda_h=0.5$\,GeV as the scale
in the spectator-scattering contributions.

\subsection{\boldmath $\bar B\to V_{1L}V_{2L}$ decay amplitudes\unboldmath}
\label{subsec:bv1v2amp}

The transition operators ${\cal T}^d$, ${\cal T}^s$ describe
a total of 28 two-body decays of $B^-$, $\bar B_d$ and $\bar B_s$ into the 
charmless vector mesons $K^{*+}$, $K^{*-}$, $K^{*0}$, $\bar K^{*0}$,
$\rho^+$, $\rho^-$, $\rho^0$, $\omega$ and $\phi$. There are 
15 $\Delta S=0$ ($b\to d$) and 13 $\Delta S=1$ ($b\to s$) transitions.
In this section we give the expressions for the amplitudes of these
processes in terms of the factorization coefficients $a_i$.
All light vector mesons are taken to be longitudinally polarized,
that is, $\rho^-\rho^0$ here means $\rho^-_L\rho^0_L$. 
We use the abreviation
\begin{equation}\label{av1v2}
A_{V_1V_2}=i\frac{G_F}{\sqrt{2}} m^2_B\, A^{B\to V_1}_0(m^2_{V_2})\, f_{V_2}
\end{equation}
suppressing the dependence of $A_{V_1V_2}$ on the $B$-meson flavour
in the notation.

The decay amplitudes, up to the factor (\ref{av1v2}), are conveniently 
obtained from the transition operator by the following 'bosonization'
of the bilinear quark currents in (\ref{tpdef}):
\begin{eqnarray}\label{ubdbsb}
(\bar ub)_{V-A} &=& \frac{B^-\rho^0+B^-\omega}{\sqrt{2}} +\bar B_d \rho^+
+\bar B_s K^{*+}\nonumber\\
(\bar db)_{V-A} &=& B^- \rho^- + 
\frac{\bar B_d\omega-\bar B_d\rho^0}{\sqrt{2}} +\bar B_s K^{*0}\nonumber\\
(\bar sb)_{V-A} &=& B^- K^{*-} + \bar B_d \bar K^{*0} +\bar B_s\phi
\end{eqnarray}
\begin{equation}\label{duudsd}
(\bar du)_V = \rho^-, \qquad (\bar ud)_V = \rho^+, \qquad
(\bar sd)_V = \bar K^{*0}
\end{equation}
\begin{equation}\label{dssuus}
(\bar ds)_V = K^{*0}, \qquad\quad (\bar su)_V = K^{*-}, \qquad\quad
(\bar us)_V = K^{*+}
\end{equation}
\begin{equation}\label{uuddss}
(\bar uu)_V = \frac{\rho^0+\omega}{\sqrt{2}}, \qquad\quad 
(\bar dd)_V = \frac{\omega-\rho^0}{\sqrt{2}}, \qquad\quad
(\bar ss)_V = \phi
\end{equation}
In our notation the charge of $B$ mesons (light mesons) corresponds
to that of particles in the initial (final) state.
Note that the axial vector parts of the light-quark currents
don't contribute for final-state vector mesons.  
Insertion of these expressions in ${\cal T}^d$ 
(${\cal T}^s$) from eq. (\ref{tpdef}) generates all
$\Delta S=0$ ($\Delta S=1$) amplitudes. For a specific
process $\bar B\to V_1V_2$ the amplitude is found as the coefficient 
of $(\bar B V_1)V_2$ (and of $(\bar B V_2)V_1$ if $V_1\not= V_2$).  
This procedure automatically keeps track of all sign and Clebsch-Gordan
factors. Note, however, that an extra symmetry factor of 2 has to be
included for amplitudes with two identical particles in the final state.
From the structure of ${\cal T}^{d(s)}$ it follows that the coefficients
$a_3$, $a_5$ and $a^p_7$, $a^p_9$ always appear in the combination
$a_3+a_5$ and $a^p_7 + a^p_9$, respectively.
Representative numerical values for the coefficients $a_i$
can be found in appendix \ref{sec:coeffaibi}.

The $\Delta S=0$ transition amplitudes then read
(a summation over $p=u$, $c$ is understood):
\begin{equation}\label{rmr0}
\sqrt{2}{\cal A}(B^-\to\rho^-\rho^0)=\left[\lambda_u(a_1+a_2)+
\frac{3}{2}\lambda_p(a^p_7+a^p_9+a^p_{10})\right] A_{\rho\rho}
\end{equation}
\begin{eqnarray}
\sqrt{2}{\cal A}(B^-\to\rho^-\omega) &=& 
\left[\lambda_u a_1 +\lambda_p (a^p_4+a^p_{10})\right] A_{\omega\rho}
\nonumber\\
&& +\left[\lambda_u a_2 +\lambda_p \left(a^p_4+2(a_3+a_5)
+\frac{1}{2}(a^p_7+a^p_9-a^p_{10})\right)\right] A_{\rho\omega}
\end{eqnarray} 
\begin{equation}\label{rmph}
{\cal A}(B^-\to\rho^-\phi)=\lambda_p \left[ a_3+a_5-
\frac{1}{2} (a^p_7+a^p_9)\right] A_{\rho\phi}
\end{equation}
\begin{equation}\label{kmk0}
{\cal A}(B^-\to K^{*-}K^{*0})=
\lambda_p\left[ a^p_4-\frac{1}{2} a^p_{10} \right] A_{K^*K^*}
\end{equation}

\begin{equation}\label{r0r0}
{\cal A}(\bar B_d\to\rho^0\rho^0)=\left[-\lambda_u a_2 +
\lambda_p\left( a^p_4-\frac{3}{2}(a^p_7+a^p_9)
-\frac{1}{2} a^p_{10}\right)\right] A_{\rho\rho}
\end{equation}
\begin{eqnarray}
{\cal A}(\bar B_d\to\rho^0\omega) &=& \frac{1}{2}
\left[\lambda_u a_2 +\lambda_p \left(-a^p_4+\frac{3}{2}(a^p_7+a^p_9)+
\frac{1}{2}a^p_{10}\right)\right] A_{\omega\rho}
\nonumber\\
&& -\frac{1}{2}\left[\lambda_u a_2 +\lambda_p \left(a^p_4+2(a_3+a_5)
+\frac{1}{2}(a^p_7+a^p_9-a^p_{10})\right)\right] A_{\rho\omega}
\end{eqnarray}
\begin{equation}\label{omom}
{\cal A}(\bar B_d\to\omega\omega)=\left[\lambda_u a_2 +\lambda_p \left(a^p_4
+2(a_3+a_5)+\frac{1}{2}(a^p_7+a^p_9-a^p_{10})\right)\right] A_{\omega\omega}
\end{equation}
\begin{equation}\label{r0ph}
\sqrt{2}{\cal A}(\bar B_d\to\rho^0\phi)=-\lambda_p \left[ a_3+a_5-
\frac{1}{2} (a^p_7+a^p_9)\right] A_{\rho\phi}
\end{equation}
\begin{equation}\label{omph}
\sqrt{2}{\cal A}(\bar B_d\to\omega\phi)=\lambda_p \left[ a_3+a_5-
\frac{1}{2} (a^p_7+a^p_9)\right] A_{\omega\phi}
\end{equation}
\begin{equation}\label{rprm}
{\cal A}(\bar B_d\to\rho^+\rho^-) =
\left[\lambda_u a_1 +\lambda_p (a^p_4+a^p_{10})\right] A_{\rho\rho}
\end{equation}
\begin{equation}\label{kb0k0}
{\cal A}(\bar B_d\to \bar K^{*0}K^{*0})=
\lambda_p\left[ a^p_4-\frac{1}{2} a^p_{10} \right] A_{K^*K^*}
\end{equation}

\begin{equation}\label{k0r0}
\sqrt{2}{\cal A}(\bar B_s\to K^{*0}\rho^0)=\left[\lambda_u a_2 +
\lambda_p\left( -a^p_4+\frac{3}{2}(a^p_7+a^p_9)
+\frac{1}{2} a^p_{10}\right)\right] A_{K^*\rho}
\end{equation}
\begin{equation}\label{k0om}
\sqrt{2}{\cal A}(\bar B_s\to K^{*0}\omega)=
\left[\lambda_u a_2 +\lambda_p \left(a^p_4
+2(a_3+a_5)+\frac{1}{2}(a^p_7+a^p_9-a^p_{10})\right)\right] A_{K^*\omega}
\end{equation}
\begin{equation}\label{k0ph}
{\cal A}(\bar B_s\to K^{*0}\phi)=
\lambda_p\left[ a^p_4-\frac{1}{2} a^p_{10} \right] A_{\phi K^*}
+\lambda_p \left[ a_3+a_5-
\frac{1}{2} (a^p_7+a^p_9)\right] A_{K^*\phi}
\end{equation}
\begin{equation}\label{kprm}
{\cal A}(\bar B_s\to K^{*+}\rho^-) =
\left[\lambda_u a_1 +\lambda_p (a^p_4+a^p_{10})\right] A_{K^*\rho}
\end{equation}

\vspace*{0.5cm}

The amplitudes for $\Delta S=1$ transitions are found to be:
 
\begin{equation}\label{kmr0}
\sqrt{2}{\cal A}(B^-\to K^{*-}\rho^0) =
\left[\lambda'_u a_1 +\lambda'_p (a^p_4+a^p_{10})\right] A_{\rho K^*}+
\left[\lambda'_u a_2 +\frac{3}{2}\lambda'_p (a^p_7+a^p_9)\right] A_{K^*\rho}
\end{equation}
\begin{eqnarray}
\sqrt{2}{\cal A}(B^-\to K^{*-}\omega) &=&
\left[\lambda'_u a_1 +\lambda'_p (a^p_4+a^p_{10})\right] A_{\omega K^*}
\nonumber\\
&&+\left[\lambda'_u a_2 + \lambda'_p \left( 2(a_3+a_5)+
  \frac{1}{2}(a^p_7+a^p_9)\right)\right] A_{K^*\omega}
\label{kmom}
\end{eqnarray}
\begin{equation}\label{kmph}
{\cal A}(B^-\to K^{*-}\phi) =\lambda'_p
\left[ a^p_4+a_3+a_5-\frac{1}{2}(a^p_7+a^p_9+a^p_{10})\right] A_{K^*\phi}
\end{equation}
\begin{equation}\label{kb0rm}
{\cal A}(B^-\to \bar K^{*0}\rho^-) =
\lambda'_p \left[ a^p_4 - \frac{1}{2} a^p_{10}\right] A_{\rho K^*}
\end{equation}

\begin{equation}\label{kb0r0}
\sqrt{2}{\cal A}(\bar B_d\to \bar K^{*0}\rho^0) =
\left[\lambda'_u a_2 +\frac{3}{2}\lambda'_p (a^p_7+a^p_9)\right] A_{K^*\rho}
-\lambda'_p \left[ a^p_4 - \frac{1}{2} a^p_{10}\right] A_{\rho K^*}
\end{equation}
\begin{eqnarray}
\sqrt{2}{\cal A}(\bar B_d \to \bar K^{*0}\omega) &=&
\left[\lambda'_u a_2 + \lambda'_p \left( 2(a_3+a_5)+
  \frac{1}{2}(a^p_7+a^p_9)\right)\right] A_{K^*\omega}
\nonumber\\
&&+\lambda'_p \left[ a^p_4 - \frac{1}{2} a^p_{10}\right] A_{\omega K^*}
\label{kb0om}
\end{eqnarray}
\begin{equation}\label{kb0ph}
{\cal A}(\bar B_d\to \bar K^{*0}\phi) =\lambda'_p
\left[ a^p_4+a_3+a_5-\frac{1}{2}(a^p_7+a^p_9+a^p_{10})\right] A_{K^*\phi}
\end{equation}
\begin{equation}\label{kmrp}
{\cal A}(\bar B_d\to K^{*-}\rho^+) =
\left[\lambda'_u a_1 +\lambda'_p (a^p_4+a^p_{10})\right] A_{\rho K^*}
\end{equation}

\begin{equation}\label{sr0ph}
\sqrt{2}{\cal A}(\bar B_s\to\rho^0\phi) =
\left[\lambda'_u a_2 +\frac{3}{2}\lambda'_p (a^p_7+a^p_9)\right] A_{\phi\rho}
\end{equation}
\begin{equation}\label{somph}
\sqrt{2}{\cal A}(\bar B_s \to\omega\phi) =
\left[\lambda'_u a_2 + \lambda'_p \left( 2(a_3+a_5)+
  \frac{1}{2}(a^p_7+a^p_9)\right)\right] A_{\phi\omega}
\end{equation}
\begin{equation}\label{phph}
{\cal A}(\bar B_s\to \phi\phi) = 2 \lambda'_p
\left[ a^p_4+a_3+a_5-\frac{1}{2}(a^p_7+a^p_9+a^p_{10})\right] A_{\phi\phi}
\end{equation}
\begin{equation}\label{kpkm}
{\cal A}(\bar B_s\to K^{*+}K^{*-}) =
\left[\lambda'_u a_1 +\lambda'_p (a^p_4+a^p_{10})\right] A_{K^* K^*}
\end{equation}
\begin{equation}\label{skb0k0}
{\cal A}(\bar B_s\to \bar K^{*0} K^{*0}) =
\lambda'_p \left[ a^p_4 - \frac{1}{2} a^p_{10}\right] A_{K^* K^*}
\end{equation}

\subsection{Weak annihilation amplitudes}
\label{subsec:annihilation}

\begin{figure}[t]
\begin{center}
\includegraphics[width=0.7\linewidth]{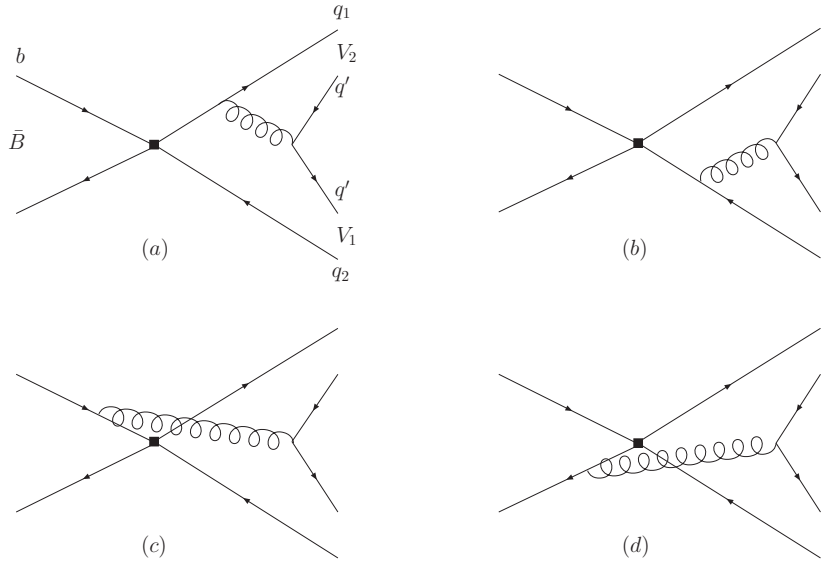}
\end{center}
{\caption{\label{fig:ann} Annihilation diagrams.}}
\end{figure}

The decay mechanism of weak annihilation (Fig. \ref{fig:ann})
gives contributions
to the amplitudes for $B\to V_L V_L$ decays that are supressed
by $\Lambda_{QCD}/m_b$. These power corrections are not calculable
in the usual factorization framework. This is indicated by end-point
singularities from the integrals over light-cone momentum
fractions in a hard-scattering ansatz. We shall use the model of
\cite{Beneke:2001ev}, which is based on this ansatz together with
a cut-off procedure, to estimate 
the impact of annihilation effects on the leading decay amplitudes.  
Following the notation of \cite{Beneke:2001ev} we write
\begin{equation}\label{htpann}
\langle V_{1L} V_{2L}|{\cal H}_{\rm eff}^{\Delta S=0}|\bar B\rangle_{\rm ann}
   = \frac{G_F}{\sqrt 2} \sum_{p=u,c} \lambda_p\,
   \langle V_{1L} V_{2L}|{\cal T}_p^{{\rm ann},d}|\bar B\rangle 
\end{equation}
with $\lambda_p=V_{pb} V^*_{pd}$ and
\begin{eqnarray}
   {\cal T}_p^{{\rm ann},d}
   &=& \delta_{up}\,(\delta_{rd}\,b_1\,\sigma_u^u
    + \delta_{ru}\,b_2\,\sigma_d^u) + b_3\,\sigma_d^r
    + \delta_{rd}\,b_4\,\mbox{tr}(\sigma) \nonumber\\
   &&\mbox{}+ \frac{3}{2}\,b_3^{\rm EW}\,e_r\,\sigma_d^r
    + \frac{3}{2}\,\delta_{rd}\,b_4^{\rm EW}\,\mbox{tr}(Q\,\sigma) 
\label{tpannd}
\end{eqnarray}
Here the index $r=u$, $d$, $s$ denotes the flavour of the spectator quark 
in the $B$ meson and $Q={\rm diag}(2/3,-1/3,-1/3)$. 
The corresponding formulas for $\Delta S=1$ transitions
are obtained by interchanging the labels $d\leftrightarrow s$
in the expressions (\ref{htpann}), (\ref{tpannd}) for $\Delta S=0$. 
In particular,
$\lambda_p$ is then replaced by $\lambda_p'=V_{pb} V^*_{ps}$.

The operators 
\begin{equation}\label{sigmadef}
   \sigma_{q_1}^{q_2} = \sum_{q'=u,d,s}\,(\bar q' q_2)\times(\bar q_1 q') 
\end{equation}
encode the valence quarks of the final state mesons. 
Matrix elements of the product of currents in (\ref{sigmadef})
are defined as
\begin{equation}
   \langle V_{1L} V_{2L}| j_1\times j_2 |\bar B\rangle
   \equiv i c f_{B} f_{V_1} f_{V_2} 
\end{equation}
where $c=0$, $1$, $\pm 1/\sqrt{2}$, etc.,  is the appropriate
Clebsch-Gordan coefficient and symmetry factor, relating the currents 
$j_1$ and $j_2$ to the mesons $V_1$, $V_2$. 
The coefficients $b_i$ are given by
\begin{eqnarray}\label{bidef}
   b_1 &=& \frac{C_F}{N_c^2}\,C_1 A_1^i \,, \qquad
    b_3 = \frac{C_F}{N_c^2} \Big[ C_3 A_1^i + C_5 (A_3^i+A_3^f)
    + N_c C_6 A_3^f \Big] \,, \nonumber\\
   b_2 &=& \frac{C_F}{N_c^2}\,C_2 A_1^i \,, \qquad
    b_4 = \frac{C_F}{N_c^2}\,\Big[ C_4 A_1^i + C_6 A_2^i \Big] \,,
    \nonumber\\
   b_3^{\rm EW} &=& \frac{C_F}{N_c^2} \Big[ C_9 A_1^i
    + C_7 (A_3^i+A_3^f) + N_c C_8 A_3^f \Big] \,, \nonumber\\
   b_4^{\rm EW} &=& \frac{C_F}{N_c^2}\,\Big[ C_{10} A_1^i
    + C_8 A_2^i \Big] 
\end{eqnarray}
and correspond to current--current annihilation ($b_1,b_2$), penguin
annihilation ($b_3,b_4$), and electroweak penguin annihilation
($b_3^{\rm EW},b_4^{\rm EW}$). These coefficients depend on the
final-state mesons, $b_i=b_i(V_1V_2)$, but this dependence will
be left implicit in the following.
Finally, the $A^{i,f}_k$ read 
\begin{align}
{A}^i_1 &=\pi\alpha_s\int_0^1 dx dy\;
\left\{\phi_{||}^{V_1}(y)\phi_{||}^{V_2}(x)\left[\frac{1}{y(1-\bar{y}x)}
+\frac{1}{y\bar{x}^2}\right]-\Phi_{v}^{V_1}(y)\Phi_{v}^{V_2}(x)
    r^{V_1}_\perp r^{V_2}_\perp\frac{2}{y\bar{x}}
\right\} ,\nonumber\\
{A}^f_1 &=0 ,\nonumber\\
{A}^i_2 &=\pi\alpha_s\int_0^1 dx dy\;
\left\{\phi_{||}^{V_1}(y)\phi_{||}^{V_2}(x)\left[\frac{1}{\bar{x}(1-\bar{y}x)}
+\frac{1}{y^2\bar{x}}\right]-\Phi_{v}^{V_1}(y)\Phi_{v}^{V_2}(x)
    r^{V_1}_\perp r^{V_2}_\perp\frac{2}{y\bar{x}}
\right\} , \nonumber\\
{A}^f_2 &=0 ,\nonumber\\
{A}^i_3 &=\pi\alpha_s\int_0^1 dx dy\;
\left\{\phi_{||}^{V_1}(y)\Phi_{v}^{V_2}(x) r^{V_2}_\perp
\frac{2x}{y\bar{x}(1-\bar{y}x)}
 +\phi_{||}^{V_2}(x)\Phi_{v}^{V_1}(y)r^{V_1}_\perp
\frac{2\bar{y}}{y\bar{x}(1-\bar{y}x)}\right\} ,\nonumber\\
{A}^f_3 &=\pi\alpha_s\int_0^1 dx dy\;\left\{-\phi_{||}^{V_1}(y)
   \Phi_{v}^{V_2}(x)r^{V_2}_\perp\frac{2(1+y)}{y^2\bar{x}}
+\phi_{||}^{V_2}(x)
   \Phi_{v}^{V_1}(y) r^{V_1}_\perp\frac{2(1+\bar{x})}{y\bar{x}^2}
\right\}\label{aifk}
\end{align}
The superscript $i$ ($f$) denotes gluon emission from the
initial- (final-)state quarks, as shown in Fig. \ref{fig:ann}
(c) and (d) ((a) and (b)). The subscript $k$ 
indicates the Dirac structure of the four-quark operators,
$\Gamma_1\otimes\Gamma_2=(V-A)\otimes(V-A)$ ($k=1$), 
$(V-A)\otimes(V+A)$ ($k=2$), $(-2)(S-P)\otimes(S+P)$ ($k=3$). 
The various quantities in (\ref{bidef}) will be
evaluated at the scale $\mu_h=\sqrt{\Lambda_h\,\mu}$,
similarly to the spectator-interaction terms.

For the numerical estimate of weak annihilation
the IR-divergent quantities $A^{i,f}_k$ in (\ref{aifk})
will be parametrized by
\begin{equation}\label{xadef}
X_A=\int_0^1\frac{dx}{x}= \left( 1+\rho_A \, e^{i\phi_A}\right)
\ln\frac{m_B}{\Lambda_h}
\end{equation}
The quantity $X_A$ is the cut-off regulated integral 
$\int_{\Lambda_h/m_B}^1 dx/x=\ln(m_B/\Lambda_h)$,
with scale $\Lambda_h=0.5\, {\rm GeV}$, modified by 
a phenomenological magnitude $\rho_A$ and phase $\phi_A$ 
\cite{Beneke:2001ev}. 
Using $SU(3)$ flavour symmetry and the asymptotic forms
of the meson wavefunctions $\phi_{||}$ and $\Phi_v$,
one finds $A^i_1=A^i_2$, $A^i_3=0$ and 
\begin{eqnarray}
A^i_1 &\approx& \pi\alpha_s\left[18\left(X_A-4+\frac{\pi^2}{3}\right)
  +18 (r^V_\perp)^2 \left(X_A-2\right)^2\right]\nonumber\\
A^f_3 &\approx& -36\pi\alpha_s r^V_\perp\, \left(2X^2_A-5X_A+2\right) 
\label{ai1af3}
\end{eqnarray}

We next give the results for the annihilation amplitudes in terms of the 
coefficients $b_i$, where we define 
\begin{equation}\label{bv1v2}
B_{V_1V_2}=i\frac{G_F}{\sqrt{2}}\, f_B f_{V_1} f_{V_2}
\end{equation}
The decay constant $f_B$ depends on the flavour of the
decaying $B$ meson, even though this is not made explicit
in the notation for the $B_{V_1V_2}$.
The following expressions can be efficiently obtained
with a procedure similar to the one described at the beginning of
sec. \ref{subsec:bv1v2amp}.
Typical numerical values for the coefficients $b_i$ are given in
appendix \ref{sec:coeffaibi}.
For the channels with $\Delta S=0$ the annihilation contributions read

\begin{equation}\label{rmr0ann}
{\cal A}_{\rm ann}(B^-\to\rho^-\rho^0) = 0
\end{equation}
\begin{equation}\label{rmomann}
\sqrt{2}{\cal A}_{\rm ann}(B^-\to\rho^-\omega) =
\left[\lambda_u\, 2 b_2 + 
 (\lambda_u +\lambda_c)\, 2(b_3+b^{\rm EW}_3)\right]B_{\rho\omega}
\end{equation} 
\begin{equation}\label{rmphann}
{\cal A}_{\rm ann}(B^-\to\rho^-\phi)=0
\end{equation}
\begin{equation}\label{kmk0ann}
{\cal A}_{\rm ann}(B^-\to K^{*-}K^{*0})=
\left[\lambda_u\, b_2 + 
 (\lambda_u +\lambda_c)\, (b_3+b^{\rm EW}_3)\right]B_{K^*K^*}
\end{equation}

\begin{equation}\label{r0r0ann}
{\cal A}_{\rm ann}(\bar B_d\to\rho^0\rho^0)=
\left[\lambda_u\, b_1 + (\lambda_u +\lambda_c)\, 
\left(b_3+2b_4-\frac{1}{2}b^{\rm EW}_3+\frac{1}{2}b^{\rm EW}_4 \right)\right]
B_{\rho\rho}
\end{equation}
\begin{equation}\label{r0omann}
{\cal A}_{\rm ann}(\bar B_d\to\rho^0\omega) =
\left[\lambda_u\, b_1 + (\lambda_u +\lambda_c)\, 
\left(-b_3+\frac{1}{2}b^{\rm EW}_3+\frac{3}{2}b^{\rm EW}_4 \right)\right]
B_{\rho\omega}
\end{equation}
\begin{equation}\label{omomann}
{\cal A}_{\rm ann}(\bar B_d\to\omega\omega)=
\left[\lambda_u\, b_1 + (\lambda_u +\lambda_c)\, 
\left(b_3+2b_4-\frac{1}{2}b^{\rm EW}_3+\frac{1}{2}b^{\rm EW}_4 \right)\right]
B_{\omega\omega}
\end{equation}
\begin{equation}\label{r0phann}
{\cal A}_{\rm ann}(\bar B_d\to\rho^0\phi)=0
\end{equation}
\begin{equation}\label{omphann}
{\cal A}_{\rm ann}(\bar B_d\to\omega\phi)=0
\end{equation}
\begin{equation}\label{rprmann}
{\cal A}_{\rm ann}(\bar B_d\to\rho^+\rho^-) =
\left[\lambda_u\, b_1 + (\lambda_u +\lambda_c)\, 
\left(b_3+2b_4-\frac{1}{2}b^{\rm EW}_3+\frac{1}{2}b^{\rm EW}_4 \right)\right]
B_{\rho\rho}
\end{equation}
\begin{equation}\label{kb0k0ann}
{\cal A}_{\rm ann}(\bar B_d\to \bar K^{*0}K^{*0})=
(\lambda_u +\lambda_c)\, 
\left(b_3+2b_4-\frac{1}{2}b^{\rm EW}_3 - b^{\rm EW}_4 \right)B_{K^*K^*}
\end{equation}

\begin{equation}\label{k0r0ann}
\sqrt{2}{\cal A}_{\rm ann}(\bar B_s\to K^{*0}\rho^0)=
(\lambda_u +\lambda_c)\, 
\left(-b_3+\frac{1}{2}b^{\rm EW}_3 \right)B_{K^*\rho}
\end{equation}
\begin{equation}\label{k0omann}
\sqrt{2}{\cal A}_{\rm ann}(\bar B_s\to K^{*0}\omega)=
(\lambda_u +\lambda_c)\, 
\left(b_3-\frac{1}{2}b^{\rm EW}_3 \right)B_{K^*\omega}
\end{equation}
\begin{equation}\label{k0phann}
{\cal A}_{\rm ann}(\bar B_s\to K^{*0}\phi)=
(\lambda_u +\lambda_c)\, 
\left(b_3-\frac{1}{2}b^{\rm EW}_3 \right)B_{K^*\phi}
\end{equation}
\begin{equation}\label{kprmann}
{\cal A}_{\rm ann}(\bar B_s\to K^{*+}\rho^-) =
(\lambda_u +\lambda_c)\, 
\left(b_3-\frac{1}{2}b^{\rm EW}_3 \right)B_{K^*\rho}
\end{equation}
In addition there are two $\Delta S=0$ decay modes that proceed
only through annihilation diagrams:
\begin{equation}\label{dkpkmann}
{\cal A}_{\rm ann}(\bar B_d\to K^{*+}K^{*-}) =
\left[\lambda_u\, b_1 + (\lambda_u +\lambda_c)\, 
\left( 2 b_4 +\frac{1}{2}b^{\rm EW}_4 \right)\right]B_{K^*K^*}
\end{equation}
\begin{equation}\label{dphphann}
{\cal A}_{\rm ann}(\bar B_d\to\phi\phi)=
(\lambda_u +\lambda_c)\, 
\left( 2 b_4 - b^{\rm EW}_4 \right)B_{\phi\phi}
\end{equation}

\vspace*{0.5cm}

For the annihilation amplitudes with $\Delta S=1$ we obtain:
 
\begin{equation}\label{kmr0ann}
\sqrt{2}{\cal A}_{\rm ann}(B^-\to K^{*-}\rho^0) =
\left[\lambda'_u\, b_2 + (\lambda'_u +\lambda'_c)\, 
\left(b_3 + b^{\rm EW}_3 \right)\right]B_{K^*\rho}
\end{equation}
\begin{equation}\label{kmomann}
\sqrt{2}{\cal A}_{\rm ann}(B^-\to K^{*-}\omega) =
\left[\lambda'_u\, b_2 + (\lambda'_u +\lambda'_c)\, 
\left(b_3 + b^{\rm EW}_3 \right)\right]B_{K^*\omega}
\end{equation}
\begin{equation}\label{kmphann}
{\cal A}_{\rm ann}(B^-\to K^{*-}\phi) =
\left[\lambda'_u\, b_2 + (\lambda'_u +\lambda'_c)\, 
\left(b_3 + b^{\rm EW}_3 \right)\right]B_{K^*\phi}
\end{equation}
\begin{equation}\label{kb0rmann}
{\cal A}_{\rm ann}(B^-\to \bar K^{*0}\rho^-) =
\left[\lambda'_u\, b_2 + (\lambda'_u +\lambda'_c)\, 
\left(b_3 + b^{\rm EW}_3 \right)\right]B_{K^*\rho}
\end{equation}

\begin{equation}\label{kb0r0ann}
\sqrt{2}{\cal A}_{\rm ann}(\bar B_d\to \bar K^{*0}\rho^0) =
(\lambda'_u +\lambda'_c)\, 
\left(-b_3 + \frac{1}{2} b^{\rm EW}_3 \right) B_{K^*\rho}
\end{equation}
\begin{equation}\label{kb0omann}
\sqrt{2}{\cal A}_{\rm ann}(\bar B_d \to \bar K^{*0}\omega) =
(\lambda'_u +\lambda'_c)\, 
\left(b_3 - \frac{1}{2} b^{\rm EW}_3 \right) B_{K^*\omega}
\end{equation}
\begin{equation}\label{kb0phann}
{\cal A}_{\rm ann}(\bar B_d\to \bar K^{*0}\phi) =
(\lambda'_u +\lambda'_c)\, 
\left(b_3 - \frac{1}{2} b^{\rm EW}_3 \right) B_{K^*\phi}
\end{equation}
\begin{equation}\label{kmrpann}
{\cal A}_{\rm ann}(\bar B_d\to K^{*-}\rho^+) =
(\lambda'_u +\lambda'_c)\, 
\left(b_3 - \frac{1}{2} b^{\rm EW}_3 \right) B_{K^*\rho}
\end{equation}

\begin{equation}\label{sr0phann}
{\cal A}_{\rm ann}(\bar B_s\to\rho^0\phi) = 0
\end{equation}
\begin{equation}\label{somphann}
{\cal A}_{\rm ann}(\bar B_s \to\omega\phi) = 0
\end{equation}
\begin{equation}\label{phphann}
{\cal A}_{\rm ann}(\bar B_s\to \phi\phi) = 
(\lambda'_u +\lambda'_c)\, 
\left(2b_3 + 2b_4 - b^{\rm EW}_3 - b^{\rm EW}_4\right) B_{\phi\phi}
\end{equation}
\begin{equation}\label{kpkmann}
{\cal A}_{\rm ann}(\bar B_s\to K^{*+}K^{*-}) =
\left[\lambda'_u\, b_1 + (\lambda'_u +\lambda'_c)\, 
\left(b_3+ 2 b_4 -\frac{1}{2}b^{\rm EW}_3 +
\frac{1}{2}b^{\rm EW}_4 \right)\right]B_{K^*K^*}
\end{equation}
\begin{equation}\label{skb0k0ann}
{\cal A}_{\rm ann}(\bar B_s\to \bar K^{*0} K^{*0}) =
(\lambda'_u +\lambda'_c)\, 
\left(b_3 + 2b_4 -\frac{1}{2} b^{\rm EW}_3 - b^{\rm EW}_4\right) B_{K^*K^*}
\end{equation}
In the case of $\Delta S=1$ transitions there are four
pure annihilation modes. Their amplitudes have the form:
\begin{equation}\label{srprmann}
{\cal A}_{\rm ann}(\bar B_s\to \rho^{+}\rho^{-}) =
\left[\lambda'_u\, b_1 + (\lambda'_u +\lambda'_c)\, 
\left(2 b_4 +\frac{1}{2}b^{\rm EW}_4 \right)\right]B_{\rho\rho}
\end{equation}
\begin{equation}\label{sr0r0ann}
{\cal A}_{\rm ann}(\bar B_s\to \rho^{0}\rho^{0}) =
\left[\lambda'_u\, b_1 + (\lambda'_u +\lambda'_c)\, 
\left(2 b_4 +\frac{1}{2}b^{\rm EW}_4 \right)\right]B_{\rho\rho}
\end{equation}
\begin{equation}\label{somomann}
{\cal A}_{\rm ann}(\bar B_s\to \omega\omega) =
\left[\lambda'_u\, b_1 + (\lambda'_u +\lambda'_c)\, 
\left(2 b_4 +\frac{1}{2}b^{\rm EW}_4 \right)\right]B_{\omega\omega}
\end{equation}
\begin{equation}\label{sr0omann}
{\cal A}_{\rm ann}(\bar B_s\to \rho^{0}\omega) =
\left[\lambda'_u\, b_1 + (\lambda'_u +\lambda'_c)\, 
\frac{3}{2}b^{\rm EW}_4 \right] B_{\rho\omega}
\end{equation}
Because of similarities in the flavour structure of ${\cal T}^{d(s)}_p$
and ${\cal T}^{ann,d(s)}_p$, in all amplitudes the coefficient $b_3$ 
appears together with the factorization coefficient $a^p_4$ in
the combination $a^p_4 + B_{V_1V_2}/A_{V_1 V_2} b_3$. This has been noted 
before in the context of $PP$ and $PV$ final states \cite{Beneke:2003zv}.

\afterpage{\clearpage}

\section{Experimental results and input parameters}
\label{sec:expinput}

Available data on the decays of $B^-$ and $\bar B_d$ mesons
into a pair of light vector mesons are displayed in Tables
\ref{tab:brexp} and \ref{tab:brexpmore}. The results are from
\cite{Barberio:2007cr} unless indicated otherwise.
CP averaging is understood for branching ratios and $f_L$.
No data are available yet on $\bar B_s\to VV$ decays.
\begin{table}[t]
\centerline{\parbox{14cm}{\caption{\label{tab:brexp}
Experimental results \cite{Barberio:2007cr}
for CP-averaged branching ratios 
and longitudinal polarization fractions $f_L$
of $\bar B\to VV$ decays. Here the $B$ meson is
either a $B^-$ or a $\bar B_d$.
The branching ratios for $\bar B\to V_L V_L$ have been
obtained as $f_L\, B(\bar B\to VV)$.
Also shown are the direct CP asymmetries $A_{CP}\equiv -C\equiv 
(B(\bar B\to\bar f)-B(B\to f))/(B(\bar B\to\bar f)+B(B\to f))$.
The label $(L)$ indicates that $A_{CP}$ refers to vector mesons
with longitudinal polarization only.}}}
\begin{center}
\begin{tabular}{|c|c|c|c|c|}
\hline\hline
$VV$ & $B(\bar B\to VV)/10^{-6}$ & $f_L$ & $B(\bar B\to V_L V_L)/10^{-6}$ 
& $A_{CP}$ \\ 
\hline\hline
$\rho^+\rho^-$ & $24.2^{+3.1}_{-3.2}$ & $0.978^{+0.025}_{-0.022}$ 
& $23.7\pm 3.2$ & $0.06\pm 0.13$ $(L)$ \\
\hline
$\rho^0\rho^0$ & $0.68\pm 0.27$ & $0.70\pm 0.15$ 
& $0.48\pm 0.21$ & $-0.4\pm 0.9$ $(L)$\\
\hline
$\rho^-\rho^0$ & $18.2\pm 3.0$  & $0.912^{+0.044}_{-0.045}$ 
& $16.6\pm 2.9$ & $-0.08\pm 0.13$ \\
\hline
$\rho^-\omega$ & $10.6^{+2.6}_{-2.3}$ & $0.82\pm 0.11$ 
& $8.7\pm 2.4$ & $0.04\pm 0.18$ \\
\hline
$\bar K^{*0}K^{*0}$ & $1.28^{+0.37}_{-0.32}$ \cite{Aubert:2007xc} 
& $0.80^{+0.12}_{-0.13}$ \cite{Aubert:2007xc} & $1.02\pm 0.32$   & --- \\
\hline\hline
$\bar K^{*0}\rho^0$ & $5.6\pm 1.6$& $0.57\pm 0.12$ 
& $3.2\pm 1.1$ & $0.09\pm 0.19$ \\
\hline
$K^{*-}\rho^+$ & $< 12$ & --- & $<12$ & --- \\
\hline
$K^{*-}\rho^0$ & $< 6.1$ & $0.96^{+0.06}_{-0.16}$ 
& $<6.1$ & $0.20^{+0.32}_{-0.29}$ \\
\hline
$\bar K^{*0}\rho^-$ & $9.2\pm 1.5$& $0.48\pm 0.08$ 
& $4.4\pm 1.0$ & $-0.01\pm 0.16$ \\
\hline
$\bar K^{*0}\phi$ & $9.5\pm 0.8$ & $0.484\pm 0.034$ 
& $4.6\pm 0.5$ & $-0.01\pm 0.06$ \\
\hline
$K^{*-}\phi$ & $10.0\pm 1.1$ & $0.50\pm 0.05$ 
& $5.0\pm 0.7$ & $-0.01\pm 0.08$ \\
\hline
$\bar K^{*0}\omega$  & $1.8^{+0.8}_{-0.7}$ \cite{Goldenzweig:2008sz} 
& $0.56^{+0.34}_{-0.30}$ \cite{Goldenzweig:2008sz} & $1.0\pm 0.7$ & --- \\
\hline
\hline
\end{tabular}
\end{center}
\end{table}

\begin{table}[t]
\centerline{\parbox{14cm}{\caption{\label{tab:brexpmore}
Experimental results for further $B\to VV$ decays \cite{Barberio:2007cr}.
Quoted are the CP-averaged branching fractions in units of $10^{-6}$. 
The $B$ meson is either a $B^-$ or a $\bar B_d$.}}}
\begin{center}
\begin{tabular}{|c|c|c|c|c|}
\hline\hline
$K^{*-}\omega$ & $K^{*-}K^{*0}$ & $\rho^-\phi$ \cite{Aubert:2008fq} & 
$\bar K^{*0}\omega$ & $K^{*+}K^{*-}$\\
\hline
$< 3.4$ & $< 71$ & $< 3$ & 
$< 2.7$ & $<141$ \\
\hline\hline 
$\rho^0\omega$ & $\omega\omega$ & $\rho^0\phi$ \cite{Aubert:2008fq} & 
$\omega\phi$ & $\phi\phi$ \cite{Aubert:2008fq}\\
\hline
$<1.5$ & $<4.0$ & $<0.33$ & $<1.2$ & $<0.2$ \\
\hline
\hline
\end{tabular}
\end{center}
\end{table}

Table \ref{tab:input} collects the input parameters used in
our analysis.
\begin{table}[t]
\centerline{\parbox{14cm}{\caption{\label{tab:input}
Input parameters for $B\to V_LV_L$ decays.
Here $B_q$ stands for either $B^-$ or $\bar B_d$.
The values of the scale dependent quantities $f^\perp_V=f^\perp_V(\mu)$ are
given for $\mu=1\,\rm{GeV}$. The scale dependence of $\alpha^V_{1,2}$
is neglected.}}}
\begin{center}
\begin{tabular}{|c||c|c|c|c|c|c|}
\hline\hline
\multicolumn{7}{|c|}{Light vector mesons} \\
\hline
$V$ & $m_V/{\rm MeV}$ & $\Gamma_V/{\rm MeV}$ & $f_V/{\rm MeV}$ & 
 $f^\perp_V/{\rm MeV}$ & $\alpha^V_1$ & $\alpha^V_2$ \\
\hline
$\rho$ & 776 & 149 & 209 & $165\pm 9$ & 0 & $0.1\pm 0.3$ \\
$\omega$ & 783 & 8 & 187 & $151\pm 9$ & 0 & $0.1\pm 0.3$ \\
$K^*$ & 894 & 51 & 218 & $185\pm 10$ & $0.1\pm 0.1$ & $0.1\pm 0.3$ \\
$\phi$ & 1019 & 4 & 221 & $186\pm 9$ & 0 & $0.1\pm 0.3$ \\
\hline\hline
\multicolumn{7}{|c|}{$B$ mesons} \\
\hline
$B$ & \multicolumn{2}{|c|}{$m_B/{\rm GeV}$} & $\tau_B/{\rm ps}$ &
\multicolumn{2}{|c|}{$f_B/{\rm MeV}$} & $\lambda_B/{\rm MeV}$ \\
\hline
$B^-$ & \multicolumn{2}{|c|}{$5.28$} & $1.64$ &
\multicolumn{2}{|c|}{$200\pm 30$} & $350\pm 150$ \\
\hline
$\bar B_d$ & \multicolumn{2}{|c|}{$5.28$} & $1.53$ &
\multicolumn{2}{|c|}{$200\pm 30$} & $350\pm 150$ \\
\hline
$\bar B_s$ & \multicolumn{2}{|c|}{$5.37$} & $1.53$ &
\multicolumn{2}{|c|}{$230\pm 30$} & $350\pm 150$ \\
\hline\hline
\multicolumn{7}{|c|}{Form factors} \\
\hline
\multicolumn{2}{|c|}{$A^{B_q\to\rho}_0(0)$} &
$A^{B_q\to\omega}_0(0)$ &
$A^{B_q\to K^*}_0(0)$ &
$A^{B_s\to K^*}_0(0)$ &
\multicolumn{2}{|c|}{$A^{B_s\to\phi}_0(0)$}  \\
\hline
\multicolumn{2}{|c|}{$0.30\pm 0.04$} &
$0.28\pm 0.05$ &
$0.37\pm 0.05$ &
$0.36\pm 0.05$ &
\multicolumn{2}{|c|}{$0.47\pm 0.06$}  \\
\hline\hline
\multicolumn{7}{|c|}{SM parameters} \\
\hline
\multicolumn{2}{|c|}{$\Lambda^{(5)}_{\overline{\rm MS}}/{\rm MeV}$} &
$m_b(m_b)/{\rm GeV}$ & $m_c(m_b)/{\rm GeV}$ & $m_t(m_t)/{\rm GeV}$ &
\multicolumn{2}{|c|}{$M_W/{\rm GeV}$}  \\
\hline
\multicolumn{2}{|c|}{$225$} &
$4.2$ & $1.3\pm 0.2$ & $167$ &
\multicolumn{2}{|c|}{$80.4$}  \\
\hline\hline
\multicolumn{2}{|c|}{$|V_{us}|$} & $|V_{cb}|$ & 
$|V_{ub}/V_{cb}|$ & $\gamma$ & 
\multicolumn{2}{|c|}{$\sin 2\beta$}  \\
\hline
\multicolumn{2}{|c|}{$0.226$} & $0.0416$ &
$0.09\pm 0.01$ & $(67\pm 12)^\circ$ & 
\multicolumn{2}{|c|}{$0.681\pm 0.025$}  \\
\hline
\hline
\end{tabular}
\end{center}
\end{table}
The values of $m_V$, $\Gamma_V$, $m_B$, $|V_{us}|$ and $|V_{cb}|$
have been taken from \cite{Yao:2006px}. They have only small
uncertainties, which we neglect. Our choice for $|V_{ub}/V_{cb}|$
is compatible with the exclusive determinations quoted in 
\cite{Yao:2006px}. We prefer those over the inclusive values
since we use $V_{ub}$ in exclusive processes where the form factors 
rely on similar theoretical methods (light-cone QCD sum rules, lattice)
as in the exclusive extraction of $|V_{ub}|$.
The lifetimes of $B^-$ and $\bar B_d$ are also from \cite{Yao:2006px}.
On the other hand, the lifetime of $\bar B_s$ is put equal to
$\tau_{\bar B_d}$, which is expected theoretically to hold to very high
accuracy. The value of $\tau_{\bar B_s}$ from \cite{Yao:2006px}
is compatible with this, but is still affected by a larger error.
  
The number for $\sin 2\beta$ is the average of \cite{Barberio:2007cr}
from CP violation in $b\to c\bar cs$ modes. The angle $\gamma$ corresponds 
to the result of global CKM fits \cite{Charles:2004jd,Bona:2005vz}. 
The standard model parameters $\Lambda^{(5)}_{\overline{\rm MS}}$,
$m_b$, $m_c$, $m_t$ and $M_W$ are the same as in \cite{Beneke:2001ev}.
Changes in these values have been small in comparison with
the relevant uncertainties. The quark masses are running 
${\overline{\rm MS}}$-masses.

The decay constants $f_V$ can be determined from data on
$V\to l^+l^-$ and $\tau\to V\nu$. 
We use the values quoted in \cite{Beneke:2003zv}.
The transverse decay constants $f^\perp_V$ need to be computed
theoretically, for instance with QCD sum rules. The results we
use for $f^\perp_V$ have been compiled in \cite{Ball:2006eu}.   
The $B\to V$ form factors are from QCD sum rules on the
light cone \cite{Ball:2004rg}. These results do not yet
incorporate some improvements in the treatment of $SU(3)$
breaking that has been achieved in the meantime (see comments
in sec. 2.3 of \cite{Buchalla:2008jp}). The uncertainties on the 
form factors in Table \ref{tab:input} are taken to be somewhat larger
than reported in \cite{Ball:2004rg}.
The Gegenbauer coefficients $\alpha^V_{1,2}$ are still rather
uncertain. We adopt numbers of the typical size found in
QCD sum rule calculations \cite{Ball:2004rg}, \cite{Ball:2007rt}   
and allow for sizable uncertainties.
The range of numbers for the $B$-meson decay constants is representative
of results from recent unquenched lattice simulations (see sec. 2.4
in \cite{Buchalla:2008jp} for a review and detailed references). 
The parameter $\lambda_B$ is not well known at present. We shall
consider here the generous range already used in \cite{Beneke:2001ev}.
No attempt is made to account for $SU(3)$ breaking in this quantity.
 

\section{Phenomenological analysis}
\label{sec:analysis}

\subsection{\boldmath $B\to V_LV_L$ branching fractions\unboldmath}
\label{subsec:bvvbr}

The branching fraction of a decay $\bar B\to V_{1L}V_{2L}$
is obtained from the corresponding amplitude ${\cal A}$
as
\begin{equation}\label{bramp}
B(\bar B\to V_{1L}V_{2L}) =
S\, \frac{\tau_B}{16\pi m_B}\, |{\cal A}(\bar B\to V_{1L}V_{2L})|^2
\end{equation}
Here $S$ is a symmetry factor with $S=1/2$ if $V_1$ and $V_2$
are identical and $S=1$ otherwise.

Predictions of CP averaged branching ratios are compiled in 
Table \ref{tab:brbd} and Table \ref{tab:brbs} for strangeness-conserving
and strangeness-changing $\bar B\to V_{1L}V_{2L}$ decays, respectively.
\begin{table}[p]
\centerline{\parbox{14cm}{\caption{\label{tab:brbd}
CP-averaged branching fractions for $B\to V_LV_L$ decays with $\Delta S=0$.
The sensitivity to variations in the input parameters according to
Table \ref{tab:input} is displayed where the upper (lower) entry
corresponds to the larger (smaller) value of the parameter.
The renormalization scale $\mu$ is varied between $2m_b$ and $m_b/2$.
The model parameters $X_{A,H}(\rho_{A,H},\phi_{A,H})$ from power corrections
are varied within the range given by $0\leq\rho_{A,H}\leq 1$ and 
$0\leq\phi_{A,H}\leq 2\pi$.
Here upper (lower) entries refer to positive (negative) ${\rm Im}X_{A,H}$.
The appropriate units for each mode are given in square brackets. }}}
\begin{center}
\begin{tabular}{|c|c|c|c|c|c|c|c|c|c|c|}
\hline\hline
mode & central & $A_0$ & $\alpha^V_2$ & $\lambda_B$ & $f_B$ & $\mu$ &
$X_A$ & $X_H$ & $\left|\frac{V_{ub}}{V_{cb}}\right|$ & $\gamma$ \\
\hline
$B^-\to\rho^-\rho^0$ & $17.5[10^{-6}]$ & $^{+4.5}_{-4.0}$ & $^{+1.9}_{-1.4}$ & 
$^{-0.9}_{+2.5}$ & $^{+0.5}_{-0.5}$ & $^{+0.1}_{-0.0}$ & $-$ & 
$^{+1.5}_{-1.4}$ & $^{+4.1}_{-3.6}$ & $^{-0.2}_{+0.2}$ \\
\hline
$B^-\to\rho^-\omega$ & $15.5[10^{-6}]$ & $^{+5.6}_{-4.7}$ & $^{+1.4}_{-0.9}$ &
$^{-0.8}_{+2.0}$ & $^{+0.3}_{-0.3}$ & $^{+0.3}_{-0.3}$ & $^{-1.2}_{+0.9}$ &
$^{+1.2}_{-1.1}$ & $^{+3.4}_{-3.0}$ & $^{-0.8}_{+0.7}$ \\
\hline
$B^-\to\rho^-\phi$ & $6.0[10^{-9}]$ & $^{+0.9}_{-0.8}$ & $^{+3.7}_{-2.2}$ &
$^{-1.5}_{+4.8}$ & $^{+0.9}_{-0.8}$ & $^{-1.1}_{+3.4}$ & $-$ &
$^{+2.8}_{-2.2}$ & $^{+0.0}_{-0.0}$ & $^{+1.1}_{-1.0}$ \\
\hline
$B^-\to K^{*-}K^{*0}$ & $2.7[10^{-7}]$ & $^{+0.9}_{-0.8}$ & $^{-0.8}_{+0.9}$ &
$^{+0.2}_{-0.4}$ & $^{-0.1}_{+0.1}$ & $^{-0.4}_{+0.5}$ & $^{-2.5}_{+3.9}$ &
$^{-0.3}_{+0.3}$ & $^{+0.0}_{-0.0}$ & $^{+0.5}_{-0.5}$ \\
\hline
$\bar B_d\to\rho^0\rho^0$ & $3.3[10^{-7}]$ & $^{+0.3}_{-0.3}$ & $^{+4.3}_{-1.6}$ &
$^{-1.3}_{+6.1}$ & $^{+0.7}_{-0.6}$ & $^{-0.0}_{+0.7}$ & $^{+2.2}_{-1.7}$ &
$^{-2.2}_{+3.7}$ & $^{+0.8}_{-0.7}$ & $^{+0.3}_{-0.3}$ \\
\hline
$\bar B_d\to\rho^0\omega$&$8.0[10^{-8}]$&$^{+3.1}_{-2.5}$ & $^{-3.3}_{+3.9}$ &
$^{+1.1}_{-2.4}$ & $^{-0.4}_{+0.5}$ & $^{-1.4}_{+1.4}$ & $^{-2.7}_{+22.5}$ &
$^{-1.6}_{+1.8}$ & $^{+0.4}_{-0.3}$ & $^{+2.3}_{-2.1}$ \\
\hline
$\bar B_d\to\omega\omega$&$5.0[10^{-7}]$&$^{+0.7}_{-0.6}$ & $^{+3.7}_{-1.7}$ &
$^{-1.4}_{+5.4}$ & $^{+1.0}_{-0.9}$ & $^{-0.2}_{+1.0}$ & $^{-1.3}_{+2.1}$ &
$^{-2.3}_{+3.3}$ & $^{+0.9}_{-0.8}$ & $^{-0.4}_{+0.4}$ \\
\hline
$\bar B_d\to\rho^0\phi$ & $2.8[10^{-9}]$&$^{+0.4}_{-0.4}$ & $^{+1.7}_{-1.0}$ &
$^{-0.7}_{+2.2}$ & $^{+0.4}_{-0.4}$ & $^{-0.5}_{+1.6}$ & $-$ &
$^{-1.0}_{+1.3}$ & $^{+0.0}_{-0.0}$ & $^{+0.5}_{-0.5}$ \\
\hline
$\bar B_d\to\omega\phi$&$2.4[10^{-9}]$ & $^{+0.5}_{-0.4}$ & $^{+1.4}_{-0.9}$ &
$^{-0.6}_{+1.8}$ & $^{+0.3}_{-0.3}$ & $^{-0.4}_{+1.2}$ & $-$ &
$^{-0.9}_{+1.1}$ & $^{+0.0}_{-0.0}$ & $^{+0.4}_{-0.4}$ \\
\hline
$\bar B_d\to\rho^+\rho^-$ & $25.8[10^{-6}]$ & $^{+7.6}_{-6.6}$ & $^{-2.0}_{+1.6}$ &
$^{+1.0}_{-2.4}$ & $^{-0.3}_{+0.3}$ & $^{-0.3}_{+0.1}$ & $^{+2.4}_{-1.8}$ &
$^{-1.5}_{+1.5}$ & $^{+5.8}_{-5.2}$ & $^{-0.9}_{+0.8}$ \\
\hline
$\bar B_d\to\bar K^{*0}K^{*0}$&$3.2[10^{-7}]$&$^{+0.9}_{-0.8}$&$^{-0.9}_{+0.9}$ &
$^{+0.2}_{-0.4}$ & $^{+0.0}_{-0.0}$ & $^{-0.7}_{+0.9}$ & $^{-2.5}_{+3.0}$ &
$^{-0.3}_{+0.4}$ & $^{+0.0}_{-0.0}$ & $^{+0.5}_{-0.4}$ \\
\hline
$\bar B_s\to K^{*0}\rho^0$ & $5.6[10^{-7}]$ & $^{+0.4}_{-0.4}$ & $^{+6.7}_{-2.7}$ &
$^{-2.1}_{+8.8}$ & $^{+1.2}_{-1.0}$ & $^{-0.0}_{+1.1}$ & $^{-0.4}_{+1.5}$ &
$^{-4.2}_{+7.5}$ & $^{+1.3}_{-1.2}$ & $^{+0.4}_{-0.4}$ \\
\hline
$\bar B_s\to K^{*0}\omega$&$6.5[10^{-7}]$&$^{+0.9}_{-0.7}$& $^{+5.5}_{-2.3}$ &
$^{-1.9}_{+7.6}$ & $^{+1.0}_{-0.9}$ & $^{-0.0}_{+0.9}$ & $^{-0.9}_{+1.6}$ &
$^{-3.9}_{+6.6}$ & $^{+1.2}_{-1.1}$ & $^{-0.5}_{+0.4}$ \\
\hline
$\bar B_s\to K^{*0}\phi$&$3.4[10^{-7}]$& $^{+1.1}_{-1.0}$ & $^{-1.3}_{+1.4}$ &
$^{+0.3}_{-0.7}$ & $^{-0.2}_{+0.2}$ & $^{-0.6}_{+0.6}$ & $^{-3.3}_{+5.1}$ &
$^{-0.7}_{+0.8}$ & $^{-0.0}_{+0.0}$ & $^{+0.5}_{-0.5}$ \\
\hline
$\bar B_s\to K^{*+}\rho^-$ & $37.2[10^{-6}]$ & $^{+11.8}_{-10.2}$ & $^{-2.9}_{+2.3}$ &
$^{+1.4}_{-3.3}$ & $^{-0.6}_{+0.6}$ & $^{+0.0}_{-0.3}$ & $^{-0.2}_{+0.4}$ &
$^{-2.6}_{+2.7}$ & $^{+8.4}_{-7.5}$ & $^{-1.2}_{+1.1}$ \\
\hline
$\bar B_d\to K^{*+}K^{*-}$&$2.9[10^{-8}]$&$-$ & $-$ &
$-$ & $^{+0.9}_{-0.8}$ & $^{-1.0}_{+2.2}$ & $^{-2.8}_{+19.1}$ &
$-$ & $^{+0.5}_{-0.4}$ & $^{-0.4}_{+0.3}$ \\
\hline
$\bar B_d\to\phi\phi$ & $2.5[10^{-9}]$ & $-$ & $-$ &
$-$ & $^{+0.8}_{-0.7}$ & $^{-1.3}_{+3.4}$ & $^{+16.7}_{-2.4}$ &
$-$ & $^{+0.0}_{-0.0}$ & $^{+0.5}_{-0.4}$ \\
\hline
\hline
\end{tabular}
\end{center}
\end{table}
\begin{table}[p]
\centerline{\parbox{14cm}{\caption{\label{tab:brbs}
CP-averaged branching fractions for $B\to V_LV_L$ decays with $\Delta S=1$.
The sensitivity to variations in the input parameters according to
Table \ref{tab:input} is displayed where the upper (lower) entry
corresponds to the larger (smaller) value of the parameter.
The renormalization scale $\mu$ is varied between $2m_b$ and $m_b/2$.
The model parameters $X_{A,H}(\rho_{A,H},\phi_{A,H})$ from power corrections
are varied within the range given by $0\leq\rho_{A,H}\leq 1$ and 
$0\leq\phi_{A,H}\leq 2\pi$.
Here upper (lower) entries refer to positive (negative) ${\rm Im}X_{A,H}$.
The appropriate units for each mode are given in square brackets. }}}
\begin{center}
\begin{tabular}{|c|c|c|c|c|c|c|c|c|c|c|}
\hline\hline
mode & central & $A_0$ & $\alpha^V_1$ & $\alpha^V_2$ & $\lambda_B$ & $f_B$ & $\mu$ &
$X_A$ & $X_H$ & $\gamma$ \\
\hline
$B^-\to K^{*-}\rho^0$ & $3.4[10^{-6}]$ & $^{+1.0}_{-0.9}$ & $^{-0.1}_{+0.1}$ & $^{-0.6}_{+0.7}$ & 
$^{+0.1}_{-0.2}$ & $^{-0.1}_{+0.1}$ & $^{-0.2}_{+0.1}$ & $^{-2.5}_{+3.7}$ & 
$^{-0.1}_{+0.1}$ & $^{+0.7}_{-0.7}$ \\
\hline
$B^-\to K^{*-}\omega$ & $1.7[10^{-6}]$ & $^{+0.8}_{-0.6}$ & $^{-0.1}_{+0.1}$ & $^{-0.4}_{+0.6}$ &
$^{+0.2}_{-0.3}$ & $^{-0.1}_{+0.1}$ & $^{-0.1}_{+0.0}$ & $^{-1.0}_{+2.4}$ &
$^{-0.2}_{+0.3}$  & $^{+0.5}_{-0.4}$ \\
\hline
$B^-\to K^{*-}\phi$ & $4.1[10^{-6}]$ & $^{+1.5}_{-1.3}$ & $^{-0.1}_{+0.1}$ & $^{-1.7}_{+2.0}$ &
$^{+0.5}_{-1.2}$ & $^{-0.3}_{+0.3}$ & $^{-0.7}_{+0.5}$ & $^{-4.1}_{+7.8}$ &
$^{-0.9}_{+1.0}$ & $^{-0.0}_{+0.0}$ \\
\hline
$B^-\to\bar K^{*0}\rho^-$ & $3.3[10^{-6}]$ & $^{+1.1}_{-1.0}$ & $^{-0.3}_{+0.3}$ & $^{-1.1}_{+1.2}$ &
$^{+0.3}_{-0.6}$ & $^{-0.2}_{+0.2}$ & $^{-0.4}_{+0.3}$ & $^{-3.3}_{+6.5}$ &
$^{-0.4}_{+0.4}$ & $^{-0.0}_{+0.0}$ \\
\hline
$\bar B_d\to \bar K^{*0}\rho^0$ & $5.0[10^{-7}]$ & $^{+1.7}_{-1.4}$ & $^{-0.6}_{+0.6}$ & $^{-1.7}_{+2.5}$ &
$^{+0.2}_{-0.2}$ & $^{-0.3}_{+0.3}$ & $^{-1.1}_{+0.9}$ & $^{-4.7}_{+24.2}$ &
$^{-0.2}_{+0.3}$ & $^{-0.2}_{+0.2}$\\ 
\hline
$\bar B_d\to \bar K^{*0}\omega$ & $1.4[10^{-6}]$&$^{+0.8}_{-0.6}$ & $^{-0.2}_{+0.2}$ & $^{-0.7}_{+0.8}$ &
$^{+0.3}_{-0.6}$ & $^{-0.2}_{+0.2}$ & $^{-0.1}_{+0.0}$ & $^{-1.3}_{+2.7}$ &
$^{-0.4}_{+0.5}$ & $^{+0.0}_{-0.0}$  \\
\hline
$\bar B_d\to \bar K^{*0}\phi$ & $3.7[10^{-6}]$&$^{+1.4}_{-1.2}$ & $^{-0.1}_{+0.1}$ & $^{-1.6}_{+1.8}$ &
$^{+0.5}_{-1.1}$ & $^{-0.3}_{+0.3}$ & $^{-0.6}_{+0.4}$ & $^{-3.7}_{+7.5}$ &
$^{+0.9}_{-0.8}$ & $^{-0.0}_{+0.0}$  \\
\hline
$\bar B_d\to K^{*-}\rho^+$ & $3.0[10^{-6}]$&$^{+1.0}_{-0.8}$ & $^{-0.2}_{+0.2}$ & $^{-0.7}_{+0.8}$ &
$^{+0.1}_{-0.3}$ & $^{-0.1}_{+0.1}$ & $^{-0.3}_{+0.2}$ & $^{-1.8}_{+5.4}$ &
$^{-0.1}_{+0.1}$ & $^{+0.8}_{-0.7}$ \\
\hline
$\bar B_s\to \rho^0\phi$ & $5.9[10^{-7}]$&$^{+1.8}_{-1.5}$ & $-$ &$^{-0.5}_{+0.5}$ &
$^{+0.4}_{-0.6}$ & $^{-0.1}_{+0.1}$ & $^{+0.1}_{-0.0}$ & $-$ &
$^{-0.6}_{+0.9}$ & $^{+0.5}_{-0.4}$ \\ 
\hline
$\bar B_s\to \omega\phi$ & $4.4[10^{-8}]$ & $^{+1.3}_{-0.7}$ & $-$ & $^{+5.1}_{-0.0}$ &
$^{-0.0}_{+8.4}$ & $^{+0.4}_{-0.0}$ & $^{-0.1}_{+4.0}$ & $-$ &
$^{-1.8}_{+11.3}$ & $^{+0.5}_{-0.4}$\\ 
\hline
$\bar B_s\to \phi\phi$ & $15.5[10^{-6}]$&$^{+5.0}_{-4.3}$ & $-$ & $^{-5.8}_{+6.5}$ &
$^{+1.6}_{-3.6}$ & $^{-0.7}_{+0.7}$ & $^{-3.1}_{+3.3}$ & $^{-14.4}_{+20.2}$ &
$^{-3.2}_{+3.6}$ & $^{-0.1}_{+0.1}$\\ 
\hline
$\bar B_s\to K^{*+}K^{*-}$ & $5.9[10^{-6}]$ & $^{+1.7}_{-1.5}$ & $^{-0.3}_{+0.3}$ & $^{-1.3}_{+1.4}$ &
$^{+0.2}_{-0.4}$ & $^{+0.0}_{-0.0}$ & $^{-0.9}_{+1.1}$ & $^{-3.8}_{+6.5}$ &
$^{-0.3}_{+0.3}$ & $^{+1.5}_{-1.3}$\\ 
\hline
$\bar B_s\to\bar K^{*0}K^{*0}$ & $6.2[10^{-6}]$&$^{+1.9}_{-1.7}$ & $^{-0.3}_{+0.3}$ & $^{-1.9}_{+2.0}$ &
$^{+0.4}_{-1.0}$ & $^{-0.1}_{+0.1}$ & $^{-1.2}_{+1.5}$ & $^{-5.6}_{+7.5}$ &
$^{-0.7}_{+0.7}$ & $^{-0.0}_{+0.0}$  \\
\hline
$\bar B_s\to\rho^+\rho^-$ & $1.0[10^{-7}]$ & $-$ & $-$ & $-$ &
$-$ & $^{+0.3}_{-0.2}$ & $^{-0.5}_{+1.4}$ & $^{+6.7}_{-1.0}$ &
$-$ & $^{+0.0}_{-0.0}$\\ 
\hline
$\bar B_s\to\rho^0\rho^0$ & $5.1[10^{-8}]$ & $-$ & $-$ & $-$ &
$-$ & $^{+1.4}_{-1.2}$ & $^{-2.7}_{+7.0}$ & $^{+33.7}_{-5.0}$ &
$-$ & $^{+0.2}_{-0.2}$ \\ 
\hline
$\bar B_s\to \omega\omega$ & $3.2[10^{-8}]$& $-$ & $-$ & $-$ &
$-$ & $^{+0.9}_{-0.8}$ & $^{-1.7}_{+4.5}$ & $^{-3.2}_{+21.6}$ &
$-$ & $^{+0.1}_{-0.1}$ \\
\hline
$\bar B_s\to\rho^0\omega$ & $1.5[10^{-9}]$ & $-$ & $-$ & $-$ &
$-$ & $^{+0.4}_{-0.4}$ & $^{-0.5}_{+1.1}$ & $^{+9.8}_{-1.4}$ &
$-$ & $^{-0.1}_{+0.1}$ \\
\hline
\hline
\end{tabular}
\end{center}
\end{table}
Absolute branching fractions have in general sizable uncertainties from
hadronic input quantities, for instance from $B\to V_L$ form factors.
Taking ratios or other combinations of suitable branching fractions
can eliminate part of the uncertainties and lead to theoretically
cleaner observables. In spite of this it is still interesting 
to present the theory expectations for the branching fractions, 
which can be directly confronted with experimental data.
In addition, we use Tables \ref{tab:brbd} and \ref{tab:brbs} 
to display in detail the sensitivity of the results on the most
important input parameters. 

\begin{figure}[p]
\begin{center}
\includegraphics[scale=0.5]{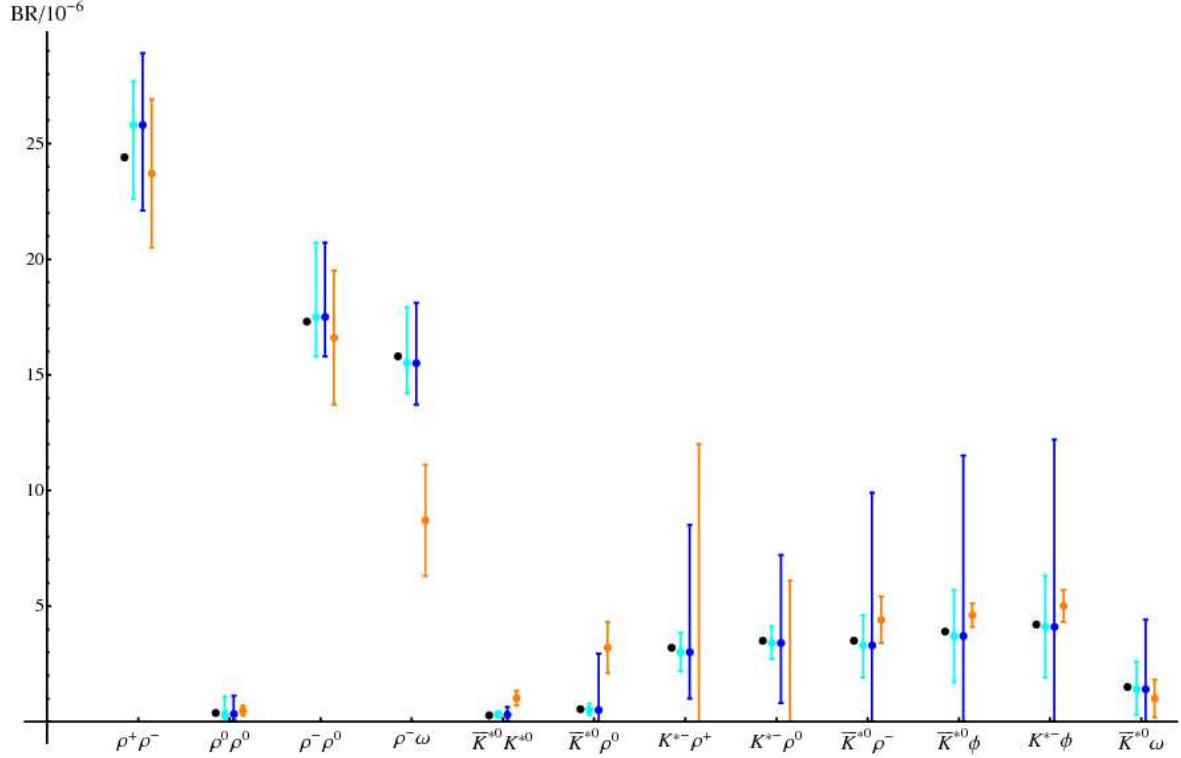}
\end{center}
{\caption{\label{fig:brbd-thexp}
Comparison between theory predictions 
[dots (black), left bar (cyan), middle bar (marine blue)] and 
experimental results [right bar (orange)] for $\bar B_d \to V_L V_L$ 
modes, for which measurements are available.
The theoretical error bars display the hadronic errors without 
[left (cyan)] and with [middle (marine blue)] the model-dependent error 
estimate for annihilation topologies. The form-factor
uncertainties are not included in the error bars. 
The black dots are the central values of the theory predictions
where all power corrections have been omitted. 
From experiment only upper
limits are known for the two $K^{*-}\rho$ channels.}}
\end{figure}

In Fig. \ref{fig:brbd-thexp} we compare theory and experiment
for $\bar B\to V_LV_L$ branching fractions, for which measurements are
available. For this comparison the form factors have been fixed to their
central values.
In the present discussion we will assume that the Standard Model is valid.
Under this assumption the comparison with experimental data will serve as a
test of the theory of QCD interactions in hadronic weak decays.
It should be kept in mind that possible deviations between predictions
and measurements may in principle indicate the existence of New Physics.
In order to disentangle New Physics from QCD effects it is important
to consider observables with very little hadronic uncertainty.
We will discuss several examples for this in the following sections.
For the moment we stay with the discussion of the theoretically less clean
absolute branching fractions for the purpose of testing the method
of QCD factorization, under the (provisional) assumption that physics 
beyond the Standard Model is absent.

Fig. \ref{fig:brbd-thexp} shows good agreement of theory and
experiment within errors. An exception is $B^-\to\rho^-\omega$,
where the measured branching ratio is somewhat low with respect
to the expectation from theory. The reason could be an overestimate
of the $B\to\omega$ form factor, or a statistical fluctuation.
The only other measured decay with an $\omega$ is 
$\bar B_d\to\bar K^{*0}\omega$. Here the theory result is also above the 
central experimental value, but the uncertainties in the latter are
still larger than for $B^-\to\rho^-\omega$ and prevent any firm
conclusion.

We emphasize that theory and experiment agree very well in the three
$\rho$-meson channels $\bar B_d\to\rho^+\rho^-$, $\bar B_d\to\rho^0\rho^0$
and $B^-\to\rho^-\rho^0$, as it has also been found in \cite{Beneke:2006hg}.
The $\rho^0\rho^0$ channel is a colour-suppressed mode and comes with
large uncertainties. Hard spectator scattering plays an important role
and therefore the sensitivity to the poorly known parameter $\lambda_B$ 
is large. Still the experimental result can be accounted for naturally 
with default values of the hadronic parameters.
It is remarkable that the observed pattern of all three $\rho\rho$
branching ratios, which exhibit rather different values, 
is nicely reproduced within QCD factorization.
The fact that this works for the central $B\to\rho$ form factor
$A_0$ supports the numerical value used for this quantity.

The penguin modes $\bar B_d\to\bar K^{*0}K^{*0}$ and 
$\bar B_d\to\bar K^{*0}\rho^0$ tend to have relatively small
predicted branching ratios, which however stretch into the range of 
measured values within errors.
The compatibility is better for $\bar B_d\to\bar K^{*0}K^{*0}$ than
for $\bar B_d\to\bar K^{*0}\rho^0$.
At the same time the latter mode is also seen to be very sensitive
to the annihilation contributions.

It is interesting to note that the central values of the experimental and
theoretical results are particularly close for the penguin decays
$B^-\to\bar K^{*0}\rho^-$, $\bar B_d\to\bar K^{*0}\phi$, $B^-\to K^{*-}\phi$.
On the other hand, the dependence on weak annihilation is very strong.
The huge variations from these effects shown in Fig. \ref{fig:brbd-thexp}
suggest that, at least for these channels, the annihilation model
used by us is likely to overestimate the related uncertainty. 

Further branching ratio predictions and information on the various
error sources for all 34 $\bar B\to V_LV_L$ decays can be obtained
from Tables \ref{tab:brbd} and \ref{tab:brbs}.

Our results include estimates of some effects that are 
suppressed by a factor of $\Lambda_{QCD}/m_b$. These corrections
are weak annihilation and the effects proportional to $r^V_\perp$
(see eq. (\ref{rperpdef})). Terms at this order are not calculable
in QCD factorization. They have still been included as model estimates
in order to permit us to assess the sensitivity of factorization
predictions on potentially important power corrections. Weak annihilation
is the most prominent example.
For the default choice of input parameters the impact of power corrections
on the predicted branching ratios is in general small.
This can already be seen from Fig. \ref{fig:brbd-thexp}, where central
results with all power corrections omitted are indicated by the black dots.
They differ very little from the central theory predictions that include
such effects.  
To make these statements more quantitative, we list the differences 
between the central values for all $\Delta S=0$ branching ratios without 
and including power corrections, 
$BR({\rm no\ power\ corr.})/BR({\rm default})-1$, 
in the order of appearance in Table \ref{tab:brbd}, in $\%$:
\begin{equation}\label{delbrbd}
\begin{array}{ccccccccccccccccc}
-1, & +2, & -6, & +6, & +15, & +7, & -27, & -6, & -6, & -5, & 
-16, & -12, & -9, & +6, & +1, & -100, & -100 \\
\end{array}
\end{equation}
The same information for the $\Delta S=1$ decays of Table \ref{tab:brbs}
reads  
\begin{equation}\label{delbrbs}
\begin{array}{ccccccccccccccccc}
+5, & +5, & +3, & +7, & +9, & +12, & +6, & +7, & +2, & +0, &
-8, & -17, & -19, & -100, & -100, & -100, & -100 \\
\end{array}
\end{equation}
The deviation is $-100\%$ for the six pure annihilation decays,
which have no leading-power contribution.
In all other cases the impact of the default power corrections
is rather moderate or indeed very small, notably for the dominant
decay channels.

We finally comment on the impact of the long-distance
electromagnetic penguin correction defined in (\ref{pewnu})
and discussed in appendix \ref{sec:ldempeng}.
This contribution affects only decays with the emission of
$\rho^0$, $\omega$ or $\phi$, where it enters through the coefficient
$a^u_7+a^u_9$. The long-distance effects are sizable, on the scale of this
coefficient, for $\rho^0$ and $\omega$, but much less in the case of $\phi$.
Since the long-distance terms are of order $\alpha=1/129$ their overall
contribution is in general very small. This is particularly true for
the $\Delta S=1$ decays where the up-quark sector is also
CKM suppressed. For the $\Delta S=0$ transitions the absence of the 
term in (\ref{pewnu}) would change branching ratios at the level of a few
percent at most and below the size of most of the other uncertainties.
The situation is similar for the direct CP asymmetries in the
$\Delta S=0$ modes with the exception of $\rho^-\rho^0$, $\rho^-\phi$,
$\rho^0\omega$, $\rho^0\phi$ and $\omega\phi$, where the impact
is relatively large. However, in any case, the direct CP asymmetry is 
very uncertain for $\rho^0\omega$ and it is very small for the 
remaining channels.
 

\subsection{\boldmath Direct CP violation in $B\to V_LV_L$\unboldmath}
\label{subsec:cpvbvv}

Direct CP asymmetries require the presence of a strong
as well as a weak phase difference between two interfering
amplitudes. In the heavy-quark limit this phase difference
arises at order $\alpha_s$. It is therefore parametrically suppressed
and at the same time sensitive to uncalculable power corrections.
This makes it difficult to obtain accurate predictions for direct
CP violation. At present the most precisely measured direct
CP asymmetry in $B$ decays is $A_{CP}(B\to K^+\pi^-)=-0.097\pm 0.012$
\cite{Barberio:2007cr}. The corresponding strong phase difference
is small ($\sim 15^\circ$) \cite{Buchalla:2008tg}, but has a sign
opposite to the ${\cal O}(\alpha_s)$ result in the heavy-quark limit.
This may indicate the importance of $\Lambda/m_b$ corrections.

In Tables \ref{tab:brbdcpa} and \ref{tab:brbscpa} we show
estimates of direct CP asymmetries for the decays under discussion.
\begin{table}[ht]
\centerline{\parbox{14cm}{\caption{\label{tab:brbdcpa}
CP asymmetries for $B\to V_LV_L$ decays with $\Delta S=0$,
defined as $A_{CP}\equiv 
(B(\bar B\to\bar f)-B(B\to f))/(B(\bar B\to\bar f)+B(B\to f))$.
The sensitivity to variations in the input parameters according to
Table \ref{tab:input} is displayed where the upper (lower) entry
corresponds to the larger (smaller) value of the parameter.
The renormalization scale $\mu$ is varied between $2m_b$ and $m_b/2$.
The model parameters $X_{A,H}(\rho_{A,H},\phi_{A,H})$ from power corrections
are varied within the range given by $0\leq\rho_{A,H}\leq 1$ and 
$0\leq\phi_{A,H}\leq 2\pi$. 
The appropriate units for each mode are given in square brackets.
We refrain from quoting estimates of CP asymmetries for pure annihilation 
modes.}}}
\begin{center}
\begin{tabular}{|c|c|c|c|c|c|c|c|c|c|c|c|}
\hline\hline
mode & central & $m_c$ 
& $\alpha^V_2$ & $\lambda_B$ & $f_B$ & $\mu$ & $X_A$ & $X_H$ & $\left|\frac{V_{ub}}{V_{cb}}\right|$ & $\gamma$\\ 
\hline
$B^-\to\rho^-\rho^0$ & $-2.6[10^{-4}]$ & $^{+0.8}_{-0.7}$ 
& $^{+0.2}_{-0.3}$ & $^{+0.1}_{-0.2}$ & $^{-0.0}_{+0.0}$ & $^{-5.6}_{+7.4}$ & $-$ & $^{+3.7}_{-3.7}$ & $^{+0.3}_{-0.3}$ & $^{-0.2}_{+0.3}$ \\
\hline
$B^-\to\rho^-\omega$ & $-9.3[10^{-2}]$ & $^{+2.4}_{-2.0}$ 
& $^{+1.0}_{-1.0}$ & $^{-0.4}_{+0.8}$ & $^{+0.2}_{-0.2}$ & $^{+1.4}_{-2.2}$ & $^{+22.4}_{-21.6}$ & $^{+3.1}_{-3.1}$ & $^{+0.8}_{-1.0}$ & $^{-1.1}_{+1.4}$ \\
\hline
$B^-\to\rho^-\phi$ & $-1.2[10^{-2}]$ & $^{-0.0}_{+0.0}$ 
& $^{+0.2}_{-0.3}$ & $^{-0.2}_{+0.3}$ & $^{+0.1}_{-0.1}$ & $^{-1.1}_{+1.1}$ & $-$ & $^{+0.2}_{-0.4}$ & $^{-0.1}_{+0.1}$ & $^{+0.1}_{-0.1}$ \\
\hline
$B^-\to K^{*-}K^{*0}$ & $-1.0[10^{-1}]$ & $^{-1.0}_{+0.9}$ 
& $^{-0.0}_{+0.1}$ & $^{+0.0}_{-0.1}$ & $^{+0.1}_{-0.1}$ & $^{+0.1}_{-0.2}$ &
$^{+8.9}_{-7.9}$ & $^{+0.1}_{-0.1}$ & $^{-0.1}_{+0.1}$ & $^{+0.1}_{-0.1}$ \\
\hline
$\bar B_d\to\rho^0\rho^0$ & $+5.3[10^{-1}]$ & $^{-0.7}_{+0.4}$ 
& $^{-2.7}_{+3.8}$ & $^{+2.2}_{-2.8}$ & $^{-0.7}_{+0.8}$ & $^{-1.0}_{+0.7}$ &
$^{+4.3}_{-5.8}$ & $^{+3.6}_{-2.8}$ & $^{-0.5}_{+0.6}$ & $^{-0.2}_{+0.0}$ \\
\hline
$\bar B_d\to\rho^0\omega$ & $+7.8[10^{-2}]$ & $^{-17.4}_{+14.9}$ 
& $^{+13.9}_{-5.3}$ & $^{-2.0}_{+7.4}$ & $^{+6.7}_{-6.1}$ & $^{-5.3}_{+10.8}$ &
$^{+92.2}_{-107.8}$ & $^{+10.2}_{-10.2}$ & $^{+0.5}_{-0.6}$ 
& $^{-1.3}_{+1.6}$\\
\hline
$\bar B_d\to\omega\omega$ & $-4.5[10^{-1}]$ & $^{+0.5}_{-0.4}$ 
& $^{+1.9}_{-2.4}$ & $^{-1.5}_{+2.1}$ & $^{+0.6}_{-0.8}$ & $^{+0.7}_{-0.5}$ &
$^{+5.0}_{-2.7}$ & $^{+2.7}_{-2.8}$ & $^{+0.3}_{-0.3}$ & $^{-0.7}_{+0.8}$ \\
\hline
$\bar B_d\to\rho^0\phi$ & $-1.2[10^{-2}]$ & $^{-0.0}_{+0.0}$ 
& $^{+0.2}_{-0.3}$ & $^{-0.2}_{+0.3}$ & $^{+0.1}_{-0.1}$ & $^{-1.1}_{+1.1}$ & $-$ & $^{+0.2}_{-0.4}$ & $^{-0.1}_{+0.1}$ & $^{+0.1}_{-0.1}$ \\
\hline
$\bar B_d\to\omega\phi$ & $-1.2[10^{-2}]$ & $^{-0.0}_{+0.0}$ 
& $^{+0.2}_{-0.3}$ & $^{-0.2}_{+0.3}$ & $^{+0.1}_{-0.1}$ & $^{-1.1}_{+1.1}$ & $-$ & $^{+0.3}_{-0.4}$ & $^{-0.1}_{+0.1}$ & $^{+0.1}_{-0.1}$ \\
\hline
$\bar B_d\to\rho^+\rho^-$ & $-3.7[10^{-2}]$ & $^{+1.4}_{-1.2}$ 
& $^{+0.0}_{-0.0}$ & $^{+0.1}_{-0.2}$ & $^{-0.0}_{+0.0}$ & $^{+0.3}_{-0.3}$ & $^{+10.7}_{-10.5}$ &
$^{+0.2}_{-0.2}$ & $^{+0.3}_{-0.4}$ & $^{-0.4}_{+0.5}$ \\
\hline
$\bar B_d\to\bar K^{*0}K^{*0}$ & $-1.5[10^{-1}]$ & $^{-0.7}_{+0.6}$ 
& $^{-0.2}_{+0.2}$ & $^{+0.0}_{-0.1}$ & $^{-0.0}_{+0.0}$ & $^{+0.2}_{-0.4}$ & $^{+0.9}_{-2.2}$ &
$^{+0.1}_{-0.1}$ & $^{-0.2}_{+0.2}$ & $^{+0.1}_{-0.0}$ \\
\hline
$\bar B_s\to K^{*0}\rho^0$ & $+4.3[10^{-1}]$ & $^{-0.7}_{+0.4}$ 
& $^{-2.1}_{+3.4}$ & $^{+1.7}_{-2.1}$ & $^{-0.6}_{+0.7}$ & $^{-0.7}_{+0.3}$ &
$^{+4.5}_{-8.2}$ & $^{+4.2}_{-2.5}$ & $^{-0.4}_{+0.5}$ & $^{+0.0}_{-0.2}$ \\
\hline
$\bar B_s\to K^{*0}\omega$ & $-5.2[10^{-1}]$ & $^{+0.5}_{-0.4}$ 
& $^{+2.4}_{-2.9}$ & $^{-1.7}_{+2.5}$ & $^{+0.6}_{-0.7}$ & $^{+0.9}_{-0.5}$ &
$^{+6.7}_{-3.6}$ & $^{+3.9}_{-3.7}$ & $^{+0.3}_{-0.3}$ & $^{-0.8}_{+0.9}$ \\
\hline
$\bar B_s\to K^{*0}\phi$ & $-2.0[10^{-1}]$ & $^{-0.8}_{+0.7}$ 
& $^{-0.5}_{+0.3}$ & $^{+0.1}_{-0.3}$ & $^{-0.1}_{+0.1}$ & $^{+0.3}_{-0.7}$ &
$^{+2.3}_{-7.6}$ & $^{+0.3}_{-0.3}$ & $^{-0.2}_{+0.2}$ & $^{+0.1}_{-0.1}$ \\
\hline
$\bar B_s\to K^{*+}\rho^-$ & $-3.9[10^{-2}]$ & $^{+1.5}_{-1.2}$ 
& $^{+0.0}_{-0.0}$ & $^{+0.1}_{-0.2}$ & $^{-0.0}_{+0.0}$ & $^{+0.3}_{-0.4}$ &
$^{+13.6}_{-13.5}$ & $^{+0.2}_{-0.2}$ & $^{+0.4}_{-0.4}$ & $^{-0.4}_{+0.5}$ \\
\hline
\hline
\end{tabular}
\end{center}
\end{table}
\begin{table}[ht]
\centerline{\parbox{14cm}{\caption{\label{tab:brbscpa}
CP asymmetries for $B\to V_LV_L$ decays with $\Delta S=1$
(see caption of Table \ref{tab:brbdcpa} for details).}}}
\begin{center}
\begin{tabular}{|c|c|c|c|c|c|c|c|c|c|c|}
\hline\hline
mode & central & $m_c$ & $\alpha^V_1$ & $\alpha^V_2$ & $\lambda_B$ & $\mu$ &
$X_A$ & $X_H$ & $\left|\frac{V_{ub}}{V_{cb}}\right|$ & $\gamma$ \\
\hline
$B^-\to K^{*-}\rho^0$ & $2.9[10^{-1}]$ & $^{-0.7}_{+0.6}$ & $^{-0.1}_{+0.1}$ & $^{+0.5}_{-0.4}$ &
$^{-0.1}_{+0.3}$ & $^{-0.1}_{+0.3}$ & $^{+7.0}_{-9.0}$ & $^{+0.8}_{-0.9}$ &
$^{+0.3}_{-0.3}$ & $^{-0.4}_{+0.3}$ \\
\hline
$B^-\to K^{*-}\omega$ & $4.9[10^{-1}]$ & $^{-1.1}_{+0.9}$ & $^{-0.0}_{+0.0}$ & $^{+1.4}_{-1.1}$ &
$^{-0.5}_{+1.1}$ & $^{-0.5}_{+1.4}$ & $^{+5.1}_{-13.3}$ & $^{+1.9}_{-2.2}$ &
$^{+0.2}_{-0.3}$  & $^{-0.8}_{+0.9}$ \\
\hline
$B^-\to K^{*-}\phi$ & $5.4[10^{-3}]$ & $^{+5.8}_{-5.0}$ & $^{+0.1}_{-0.1}$ & $^{-0.5}_{+0.0}$ &
$^{-0.2}_{+0.5}$ & $^{-0.5}_{+0.7}$ & $^{+994.6}_{-1005.4}$ & $^{+0.9}_{-0.8}$ &$^{+0.6}_{-0.6}$ & $^{+0.4}_{-0.6}$ \\
\hline
$B^-\to\bar K^{*0}\rho^-$ & $6.0[10^{-3}]$ & $^{+5.2}_{-4.5}$ & $^{+1.3}_{-1.1}$ & $^{+0.2}_{-0.5}$ &
$^{-0.1}_{+0.4}$ & $^{-0.6}_{+1.1}$ & $^{+994.0}_{-1006.0}$ & $^{+0.5}_{-0.5}$ &$^{+0.7}_{-0.7}$ & $^{+0.5}_{-0.7}$ \\
\hline
$\bar B_d\to \bar K^{*0}\rho^0$ & $-3.7[10^{-1}]$ & $^{+0.3}_{-0.2}$ & $^{-0.3}_{+0.3}$ & $^{-2.6}_{+1.4}$ &
$^{+0.7}_{-1.6}$ & $^{-0.3}_{+0.0}$ & $^{+13.7}_{-6.3}$ & $^{+2.1}_{-1.8}$ &
$^{-0.3}_{+0.3}$ & $^{-0.4}_{+0.5}$\\ 
\hline
$\bar B_d\to \bar K^{*0}\omega$ & $2.1[10^{-1}]$&$^{-0.1}_{+0.1}$ & $^{+0.2}_{-0.2}$ & $^{+2.0}_{-0.8}$ &
$^{-0.5}_{+2.0}$ & $^{-0.1}_{+0.6}$ & $^{+7.9}_{-11.0}$ & $^{+1.5}_{-1.5}$ &
$^{+0.2}_{-0.2}$ & $^{+0.1}_{-0.2}$  \\
\hline
$\bar B_d\to \bar K^{*0}\phi$ & $1.1[10^{-2}]$&$^{+0.4}_{-0.4}$ & $^{+0.0}_{-0.0}$ & $^{+0.3}_{-0.2}$ &
$^{-0.1}_{+0.3}$ & $^{-0.2}_{+0.4}$ & $^{+98.9}_{-101.1}$ & $^{+0.2}_{-0.1}$ &
$^{+0.1}_{-0.1}$ & $^{+0.1}_{-0.1}$  \\
\hline
$\bar B_d\to K^{*-}\rho^+$ & $3.3[10^{-1}]$&$^{-1.2}_{+1.1}$ & $^{-0.2}_{+0.2}$
& $^{+0.7}_{-0.5}$ &
$^{-0.0}_{+0.1}$ & $^{-0.0}_{+0.1}$ & $^{+6.7}_{-13.2}$ & $^{+0.1}_{-0.1}$ &
$^{+0.2}_{-0.3}$ & $^{-0.5}_{+0.6}$ \\
\hline
$\bar B_s\to \rho^0\phi$ & $3.0[10^{-1}]$&$^{+0.0}_{-0.0}$ & $-$ &$^{+0.2}_{-0.2}$ &
$^{-0.2}_{+0.3}$ & $^{-0.4}_{+0.8}$ & $-$ & $^{+2.5}_{-2.8}$ &
$^{+0.3}_{-0.3}$ & $^{+0.0}_{-0.1}$ \\ 
\hline
$\bar B_s\to \omega\phi$ & $9.0[10^{-1}]$ & $A_0$:$^{+0.8}_{-1.9}$ & $-$ & $^{-7.0}_{+0.8}$ &
$^{+0.9}_{-8.0}$ & $^{+0.0}_{-3.4}$ & $-$ & $^{+1.0}_{-18.0}$ &
$^{-0.5}_{+0.4}$ & $^{-0.3}_{+0.0}$\\ 
\hline
$\bar B_s\to \phi\phi$ & $9.7[10^{-3}]$&$^{+4.0}_{-3.4}$ & $-$ & $^{+2.4}_{-1.6}$ &
$^{-0.6}_{+1.8}$ & $^{-1.5}_{+3.1}$ & $^{+30.8}_{-6.8}$ & $^{+1.5}_{-1.2}$ &
$^{+1.1}_{-1.1}$ & $^{+0.7}_{-1.1}$\\ 
\hline
$\bar B_s\to K^{*+}K^{*-}$ & $2.6[10^{-1}]$ & $^{-1.0}_{+0.9}$ & $^{-0.2}_{+0.1}$ & $^{+0.5}_{-0.3}$ &
$^{-0.0}_{+0.1}$ & $^{+0.2}_{-0.3}$ & $^{+7.0}_{-9.0}$ & $^{+0.1}_{-0.1}$ &
$^{+0.2}_{-0.2}$ & $^{-0.4}_{+0.4}$\\ 
\hline
$\bar B_s\to\bar K^{*0}K^{*0}$ & $8.7[10^{-3}]$&$^{+3.3}_{-2.9}$ & $^{+0.9}_{-0.9}$ & $^{+1.6}_{-1.2}$ &
$^{-0.3}_{+0.9}$ & $^{-1.2}_{+2.3}$ & $^{+19.7}_{-4.9}$ & $^{+0.6}_{-0.6}$ &
$^{+1.0}_{-1.0}$ & $^{+0.6}_{-1.0}$  \\
\hline
\hline
\end{tabular}
\end{center}
\end{table}
The values have large uncertainties, as anticipated.
Most of the asymmetries are small or moderate, but there can be
exceptions. Large asymmetries may occur when the interfering
amplitudes have comparable magnitude and a substantial strong relative
phase. Examples are the $\Delta S=0$ decays with a colour suppressed
tree contribution ($\sim a_2$), as $\bar B_d\to\rho^0\rho^0$, $\omega\omega$
or $\bar B_s\to K^{*0}\rho^0$, $K^{*0}\omega$. Despite the $\alpha_s$
factor the strong phase difference can here be naturally more sizeable.
Generically, a decay amplitude of the form
\begin{equation}\label{ampgen}
A(\bar B\to M_1M_2)\sim e^{-i\gamma}-p e^{i\phi}
\end{equation}
with $p$, $\phi$, $\gamma$ real, leads to the direct CP asymmetry
\begin{equation}\label{acpgen}
A_{CP}=\frac{2 p\sin\phi\, \sin\gamma}{1+p^2-2p\cos\phi\, \cos\gamma}
\end{equation}
For $\sin\gamma\approx 0.92$, a value $p={\cal O}(1)$ and a
substantial phase $\phi$ give a large asymmetry.
In the case of $\bar B_d\to\rho^0_L\rho^0_L$ the central values
$p=0.36$, $\phi=49^\circ$ give $A_{CP}=53\%$.


\subsection{\boldmath Sensitivity to $\omega$-$\phi$ mixing\unboldmath}

In the other sections of this paper the vector mesons
$\phi$ and $\omega$ are always implemented as pure $s\bar s$ and 
$(u\bar u+d\bar d)/\sqrt{2}$-states, respectively.
Here we investigate the sensitivity of our results to the deviation
from this case of ideal mixing. We assume that other effects with
Zweig-rule suppression are negligibly small. We neglect, for example,
Zweig-rule forbidden matrix elements of the type
$\langle\phi(s\bar s)|(\bar ub)_{V-A}|B^-\rangle$.

Mixing can be introduced by the following parametrization:
\begin{align}
\phi(1020)=& 
s\bar s \cos{\theta} + \frac{u\bar u+d\bar d}{\sqrt{2}}\sin{\theta}\\
\omega(782)=& 
\frac{u\bar u+d\bar d}{\sqrt{2}}\cos{\theta}- s\bar s\sin{\theta}.
\end{align}
The ideal mixing angle in this parametrization is $\theta=0$. According to 
sum-rules quadratic in meson masses \cite{Yao:2006px}, the mixing angle can 
be estimated to be $\theta=3.4^\circ$. 
The results of varying the mixing angle up to $\theta=6.8^\circ$ are shown 
in Table \ref{tab:erromphi}.
\begin{table}[ht]
\centerline{\parbox{14cm}{\caption{\label{tab:erromphi}
Dependence of $B\to V_LV_L$ branching fractions
on $\omega$-$\phi$ mixing. The variation of the branching fractions is
given for two values of the mixing angle $\theta$.
The upper (lower) value corresponds to 
$\theta =6.8^\circ$ ($\theta =3.4^\circ$).}}}
\begin{center}
\begin{tabular}{|c|c|c||c|c|c|}
\hline
\hline
mode & default value & deviation & mode & default value & deviation\\
\hline
$B^-\to\rho^-\omega$ & $15.5[10^{-6}]$ & $^{-0.2}_{-0.0}$ &
$B^-\to K^{*-}\omega$ & $1.7[10^{-6}]$ & $^{-0.4}_{-0.2}$\\
\hline
$B^-\to\rho^-\phi$ & $6.0[10^{-9}]$ & $^{+207.5}_{+49.8}$ &
$B^-\to K^{*-}\phi$ & $4.1[10^{-6}]$ & $^{+0.4}_{+0.2}$\\
\hline
$\bar B_d\to\rho^0\omega$ & $8.0[10^{-8}]$ & $^{+0.2}_{+0.1}$ &
$\bar B_d\to \bar K^{*0}\omega$ & $1.4[10^{-6}]$ & $^{-0.5}_{-0.3}$\\
\hline
$\bar B_d\to\omega\omega$ & $5.0[10^{-7}]$ & $^{-0.1}_{-0.0}$ &
$\bar B_d\to \bar K^{*0}\phi$ & $3.7[10^{-6}]$ & $^{+0.5}_{+0.3}$\\
\hline
$\bar B_d\to\rho^0\phi$ & $2.8[10^{-9}]$ & $^{-1.9}_{-1.2}$ &
$\bar B_s\to \rho^0\phi$ & $5.9[10^{-7}]$ & $^{-0.1}_{-0.0}$\\
\hline
$\bar B_d\to\omega\phi$ & $2.4[10^{-9}]$ & $^{+10.0}_{+1.7}$ &
$\bar B_s\to \omega\phi$ & $4.4[10^{-8}]$ & $^{+28.7}_{+4.6}$\\
\hline
$\bar B_s\to K^{*0}\omega$ & $6.5[10^{-7}]$ & $^{-0.6}_{-0.3}$ &
$\bar B_s\to \phi\phi$ & $15.5[10^{-6}]$ & $^{-0.3}_{-0.0}$\\
\hline
$\bar B_s\to K^{*0}\phi$ & $3.4[10^{-7}]$ & $^{+0.6}_{+0.3}$ &
$\bar B_s\to \omega\omega$ & $3.2[10^{-8}]$ & $^{+1.6}_{+0.3}$\\
\hline
$\bar B_d\to\phi\phi$ & $2.5[10^{-9}]$ & $^{-0.4}_{-0.3}$ &
$\bar B_s\to \rho^0\omega$ & $1.5[10^{-9}]$ & $^{+11.1}_{+3.5}$\\
\hline
\hline
\end{tabular}
\end{center}
\end{table} 
For most branching fractions the effect of a nonvanishing mixing angle
$\theta\approx 3.4^\circ$ is very small, in particular for the important
modes $B^-\to\rho^-\omega$, $B^-\to K^{*-}\phi$, $\bar B_s\to\phi\phi$.
On the other hand, the modes $\bar B_d\to\rho^0\phi$, $\bar B_d\to\omega\phi$,
$\bar B_s\to\omega\phi$, $\bar B_s\to\rho^0\omega$ have a significant
dependence on deviations from ideal mixing.
The largest effect is observed for $B^-\to\rho^-\phi$. In this case 
$B^-\to\rho^-\omega$ feeds into the former channel through mixing
with a more than three orders of magnitude higher branching ratio
compared to $B^-\to\rho^-\phi(s\bar s)$, which overcompensates the 
small mixing angle:
\begin{equation}\label{brhophimix}
B(B^-\to\rho^-\phi)_{\rm mix}\approx \sin^2\theta\, B(B^-\to\rho^-\omega) 
\end{equation} 
A recent discussion of hadronic $B$ decays, mostly with
charm in the final state, for which $\omega$-$\phi$ mixing has a large
impact can be found in \cite{Gronau:2008kk}. Their estimate of
$B^-\to\rho^-\phi$ is compatible with ours.


\subsection{\boldmath Unitarity triangle from CP violation 
in $B_d\to\rho^+_L\rho^-_L$\unboldmath}
\label{subsec:cpvbrhrh}

\subsubsection{\boldmath Determination of $\bar\rho$, $\bar\eta$, 
$\gamma$ and $\alpha$\unboldmath}
\label{subsubsec:detrega}

The time dependent CP asymmetry in $B_d\to\rho^+_L\rho^-_L$ is
given by
\begin{equation}\label{acpscdef}
{\cal A}_{CP,\rho}(t)=\frac{\Gamma(\bar B_d(t)\to\rho^+_L\rho^-_L)-
\Gamma(B_d(t)\to\rho^+_L\rho^-_L)}{
\Gamma(\bar B_d(t)\to\rho^+_L\rho^-_L)+
\Gamma(B_d(t)\to\rho^+_L\rho^-_L)}=
S_\rho\, \sin(\Delta m_d\, t) - C_\rho\, \cos(\Delta m_d\, t) 
\end{equation} 
The parameters $S_\rho$ and $C_\rho$ have been measured to be
\begin{equation}\label{scrhoexp}
S_\rho=-0.05\pm 0.17\qquad\quad C_\rho=-0.06\pm 0.13
\end{equation}
as quoted by \cite{Barberio:2007cr}, based on results of 
BaBar \cite{Aubert:2007nua} and Belle \cite{Abe:2007ez}.
Together with the experimentally well determined quantity
$\sin 2\beta$ from CP violation in $B\to\psi K^0$ decays,
the parameter $S_\rho$ can be used to fix the CKM unitarity
triangle. The value of $\sin 2\beta$ from Table \ref{tab:input}
implies $\beta=(21.5\pm 1.0)^\circ$ or
\begin{equation}\label{tauctb}
\tau\equiv\cot\beta=2.54\pm 0.13
\end{equation}
In terms of the improved Wolfenstein parameters $\bar\rho$ and
$\bar\eta$ \cite{Buras:1994ec} the unitarity triangle is then
determined by
\begin{equation}\label{rhotaueta}
\bar\rho =1 - \tau \bar\eta
\end{equation}
\begin{eqnarray}\label{etabar}
\bar\eta &=& \frac{1}{(1+\tau^2)S_\rho} 
  \Bigg[(1+\tau S_\rho)(1+r_\rho \cos\phi_\rho)  \nonumber \\
 &&  \qquad -\sqrt{(1-S^2_\rho)(1+r_\rho\cos\phi_\rho)^2-
  S_\rho(1+\tau^2)(S_\rho+\sin 2\beta) r^2_\rho \sin^2\phi_\rho}\Bigg]
\end{eqnarray}
These formulas have been derived in \cite{Buchalla:2003jr,Buchalla:2004tw}
for $B\to\pi^+\pi^-$, but they apply to the case of $B\to\rho^+_L\rho^-_L$
as well. The parameters $r_\rho$ and $\phi_\rho$ are hadronic quantities.
They are defined here through
\begin{equation}\label{rphidef}
r_\rho e^{i\phi_\rho}=-
\frac{a^c_4+a^c_{10}+r^\rho_A\left(b_3+2 b_4-\frac{1}{2}b^{\rm EW}_3
+\frac{1}{2}b^{\rm EW}_4\right)}{a_1+a^u_4+a^u_{10}+r^\rho_A
\left(b_1+b_3+2 b_4-\frac{1}{2}b^{\rm EW}_3 +\frac{1}{2}b^{\rm EW}_4\right)}
\end{equation}
where all coefficients $a_i$, $b_i$ refer to the $\rho^+_L\rho^-_L$ 
final state and
\begin{equation}\label{rann}
r^\rho_A\equiv\frac{B_{\rho\rho}}{A_{\rho\rho}}=
\frac{f_B f_\rho}{m^2_B A^{B\to\rho}_0(0)}\approx 5\cdot 10^{-3}
\end{equation}
The real quantities $r_\rho$ and $\phi_\rho$ are the magnitude and
phase of the penguin-to-tree amplitude ratio in $\bar B\to\rho^+_L\rho^-_L$.
They are independent of CKM parameters.
Numerically we find
\begin{eqnarray}
r_\rho &=& 0.038\pm 0.005\, (\mu,\, \alpha^\rho_2)
\quad ^{+0.019}_{-0.026}\, (\rho_A,\, \phi_A)\label{rrhonum}\\
\phi_\rho &=& 0.23\pm 0.09\, (m_c,\, \alpha^\rho_2)
\quad ^{+0.74}_{-0.73}\, (\rho_A,\, \phi_A)\label{phirhonum}\\
r_\rho\cos\phi_\rho &=& 0.037\pm 0.005\, (\mu,\, \alpha^\rho_2)
\quad ^{+0.018}_{-0.026}\, (\rho_A,\, \phi_A)\label{rerrhonum}
\end{eqnarray}
The first error is from the uncertainties in the input parameters
$A_0$, $\alpha^\rho_2$, $f^\perp_\rho$, $\lambda_B$, $f_B$,
$m_c$ and a variation of the renormalization scale $\mu$ between
$m_b/2$ and $2m_b$ around its default value $\mu=m_b$.
The dominant sources of uncertainty are indicated in brackets.
The second error reflects the sensitivity to the parameters
$\rho_A$, $\phi_A$, $\rho_H$ and $\phi_H$ used to model
power corrections from weak annihilation ($A$) and in
the spectator scattering amplitude ($H$).
We have used $0\leq\rho_{A,H}\leq 1$, $0\leq\phi_{A,H}\leq 2\pi$.
The second error is entirely determined by weak annihilation.

The phase $\phi_\rho$ is parametrically suppressed since it arises
only at order $\alpha_s$ or $\Lambda_{QCD}/m_b$. Its precise value is
rather uncertain, in particular due to the model dependence of
power corrections, which may compete numerically with the calculable
${\cal O}(\alpha_s)$ term. Fortunately the dependence of
$\bar\eta$ in (\ref{etabar}) on $\phi_\rho$ is very weak
\cite{Buchalla:2003jr,Buchalla:2004tw}.
In addition, $r_\rho$ is a small parameter, even smaller than the
corresponding quantity $r_\pi$ in $\bar B_d\to\pi^+\pi^-$.
The smaller size of the penguin contribution in the case
of vector mesons as compared to pseudoscalars has been pointed out
before in the context of QCD factorization 
\cite{Kagan:2004uw,Beneke:2003zv}.
The formulation in (\ref{etabar}) makes it particularly transparent
to analyze the impact of a small penguin correction on the 
determination of the unitarity triangle. To linear order in
$r_\rho$, eq. (\ref{etabar}) implies the simple relation
\begin{equation}\label{etab1}
\bar\eta=\frac{1+\tau S_\rho-\sqrt{1-S^2_\rho}}{(1+\tau^2)S_\rho}
\, (1+r_\rho\cos\phi_\rho)
\end{equation}
In this approximation $\bar\eta$ and $\bar\rho$ depend only
on the real part of the penguin-to-tree ratio.
As can be seen from (\ref{etabar}), second order corrections in $r_\rho$ 
are further suppressed by $\sin^2\phi_\rho$.

With $\bar\eta$ and $\bar\rho$ also the CKM angles $\gamma$ and $\alpha$
can be computed:
\begin{equation}\label{gametalpha}
\gamma=\arctan\frac{\bar\eta}{1-\tau\bar\eta}, \qquad
\alpha=\pi-\beta-\gamma
\end{equation}
It is instructive to write down the expressions for small values
of $S_\rho$, which are suggested by the data in (\ref{scrhoexp}).
To first order in both $S_\rho$ and $r_\rho$ we find
\begin{eqnarray}
\gamma &=& \arctan\tau+\frac{S_\rho}{2} + \tau\, r_\rho\cos\phi_\rho
\label{gamsr} \\
\alpha &=& \frac{\pi}{2} -\frac{S_\rho}{2} - \tau\, r_\rho\cos\phi_\rho
\label{alpsr}
\end{eqnarray}
For $S_\rho=0$ and in the absence of a penguin contribution one has 
$\alpha=90^\circ$ and $\gamma=(68.5\pm 1.0)^\circ$.
Non-zero values of the observable $S_\rho$ and the theoretical quantity
$r_\rho\cos\phi_\rho$ then compete in shifting $\gamma$ and $\alpha$
away from these lowest-order approximations.

Evaluation of the exact formulas (\ref{etabar}) and (\ref{gametalpha})
gives
\begin{equation}\label{etanum}
\bar\eta=0.350\pm 0.013\, (\tau)\, \pm 0.012\, (S_\rho)\,
\pm 0.008\, (r_\rho\cos\phi_\rho)
\end{equation}
\begin{equation}\label{gammanum}
\gamma=72.4^\circ\pm 1.3^\circ\, (\tau)\, \pm 5.1^\circ\, (S_\rho)\,
\pm 3.2^\circ\, (r_\rho\cos\phi_\rho)
\end{equation}
Nearly identical results are obtained for $\gamma$ when the
first order expression (\ref{gamsr}) is employed.
The approximations (\ref{gamsr}) and (\ref{alpsr}) 
work to very good accuracy in the relevant range of $S_\rho$ and $r_\rho$.
This greatly facilitates the determination of $\gamma$ and $\alpha$ and
the analysis of errors, which can simply be read off from
(\ref{gamsr}) and (\ref{alpsr}).

The calculation of $\gamma$ in \cite{Beneke:2006hg} using the longitudinal 
part of the time 
dependent CP-asymmetry in the $\rho^+\rho^-$-system and $\beta$ as input 
yields a similar result for the hadronic error of $\pm 3^\circ$. 

The determination of $\gamma$ in (\ref{gammanum}) is considerably
more precise at present than measurements using $B\to DK$ tree-level decays.
Belle has found \cite{Poluektov:2006ia}
\begin{equation}\label{gammabelle}
\gamma=(53^{+15}_{-18}(stat) \pm 3(sys) \pm 9(model))^\circ
\end{equation}
and a recent analysis from BaBar gives \cite{Aubert:2008bd}
\begin{equation}\label{gammababar}
\gamma=(76\pm 22(stat) \pm 5(sys) \pm 5(model))^\circ
\end{equation}
Within the errors both are well compatible with (\ref{gammanum}).

\subsubsection{Bounds on UT parameters}
\label{subsubsec:utbounds}

Useful information on the angle $\gamma$ can also be obtained in the
form of a lower bound, which is even less sensitive to theory
input than the result in (\ref{gammanum}).
It relies only on the conservative condition that
$r\cos\phi\geq 0$, which holds in the heavy-quark limit.
This bound has been derived in \cite{Buchalla:2003jr,Buchalla:2004tw}.
Further discussions may also be found in 
\cite{Botella:2003xp,Lavoura:2004rs}.
The bound is valid as long as $S > -\sin 2\beta$ and reads
\begin{equation}\label{gammabound}
\gamma > \frac{\pi}{2}-
\arctan\frac{S-\tau(1-\sqrt{1-S^2})}{\tau S+1-\sqrt{1-S^2}}
\end{equation}
The constraint (\ref{gammabound}) can be evaluated using CP violation
in $B\to\pi^+\pi^-$ ($S=S_\pi$) or in $B\to\rho^+_L\rho^-_L$ ($S=S_\rho$).
The derivation of (\ref{gammabound}) is identical for both cases.
In fact, since the (positive) penguin correction $r_\rho\cos\phi_\rho$ is
smaller than $r_\pi\cos\phi_\pi$, the bound is expected to be
more stringent using $S_\rho$ instead of $S_\pi$.
This expectation is indeed bourne out by the experimental
result $S_\rho > S_\pi=-0.61\pm 0.08$ (see sec. \ref{subsec:rbqkkpp}),
which implies that $S_\rho$ gives the better constraint.
Qualitatively, these features can also be understood from
the approximate relation (\ref{gamsr}).   

To linear order in $S$ the bound (\ref{gammabound}) becomes
\begin{equation}\label{gammabound1}
\gamma > \arctan\tau + \frac{S}{2}
\end{equation}
in agreement with (\ref{gamsr}).

Using $S=S_\rho=-0.05$ (central),
$-0.22$ ($1\sigma$), $-0.39$ ($2\sigma$), we obtain from
(\ref{gammabound}), respectively  
\begin{equation}\label{gammaboundnum}
\gamma > 67^\circ\, ,\quad 62^\circ\, ,\quad 57^\circ
\end{equation}
The linear approximation (\ref{gammabound1}) gives practically identical
results. We remark that the relevant values of $S_\rho$
fulfill the condition $S_\rho > -\sin 2\beta$, under which the
bound can be applied.

The penguin correction is expected to shift the numbers
in (\ref{gammaboundnum}) by approximately $+6^\circ$ to
yield the actual value of $\gamma$.
The bound is therefore quite stringent.
Within a Standard Model interpretation it eliminates already a sizable
fraction of the allowed range from the direct measurements in 
(\ref{gammabelle}) and (\ref{gammababar}). 

Bounds similar to the one for $\gamma$ can also be derived for 
$\bar\eta$ and $\bar\rho$ \cite{Buchalla:2003jr,Buchalla:2004tw}.
The lower bound for $\bar\eta$ is given by the right-hand side
of (\ref{etab1}) with $r_\rho$ put to zero, the upper bound
on $\bar\rho$ then follows from $\bar\rho=1-\tau\bar\eta$. 
With the same input for $S_\rho$ as in (\ref{gammaboundnum})
we find
\begin{equation}\label{etaboundnum}
\bar\eta  > 0.338 \, ,\quad  0.326   \, ,\quad 0.314 
\end{equation}
and
\begin{equation}\label{rhoboundnum}
\bar\rho  < 0.143 \, ,\quad  0.172   \, ,\quad 0.203
\end{equation}

\subsubsection{\boldmath Precision determination of $|V_{ub}|$ from 
$\sin 2\beta$ and $S_\rho$\unboldmath}
\label{subsubsec:vubbetas}

The preceding analysis has a further interesting application 
regarding the determination of $|V_{ub}|$ from $\sin 2\beta$ 
and $S_\rho$.
The value of $|V_{ub}|$ determined in this way may be affected
by New Physics entering CP violation in $B_d\to \psi K_S$ and
$B_d\to\rho^+_L\rho^-_L$. The presence of non-standard contributions
can be revealed by comparing the extracted value of $|V_{ub}|$
with the result for $|V_{ub}|$ from an independent method.
An important example is the direct determination of $|V_{ub}|$ from
semileptonic, exclusive or inclusive, $b\to u l\nu$ decays, which
are most likely independent of physics beyond the Standard Model.
It is clear that the usefulness of such a New Physics test will  
depend on how precisely $|V_{ub}|$ can be determined.
We will show that $\sin 2\beta$ and $S_\rho$ offer 
a particularly clean and accurate determination of $|V_{ub}|$.

The magnitude of $|V_{ub}|$ is proportional to 
$R_b\equiv\sqrt{\bar\rho^2+\bar\eta^2}$.
Using the exact formulas in (\ref{rhotaueta}) and (\ref{etabar}), we expand
$R^2_b$ in $S_\rho$ and $r_\rho$. This is motivated by the
smallness of the theoretical parameter $r_\rho$ and the empirical observation
that also $S_\rho$ is small, as we have discussed in section 
\ref{subsubsec:detrega}. Treating $S_\rho$ and $r_\rho$ as small
quantities of the same order we find
\begin{eqnarray}
R_b &=&\sqrt{\bar\rho^2+\bar\eta^2} = 
\frac{1}{\sqrt{1+\tau^2}}   \label{rb2eb2}\\
&\cdot& \left[ 1+   
\frac{1}{2}\left(\frac{S_\rho}{2}+\tau r_\rho\cos\phi_\rho\right)^2
+\frac{r_\rho}{2}\left(\frac{S_\rho}{2}+
\tau r_\rho\cos\phi_\rho\right)
\left(S_\rho\cos\phi_\rho + 2\tau r_\rho\sin^2\phi_\rho\right)\right]
\nonumber
\end{eqnarray}
where we have neglected terms of the fourth order.
Through terms of third order in $S_\rho$ and $r_\rho$
($S^3_\rho$, $S^2_\rho r_\rho$, $S_\rho r^2_\rho$, $r^3_\rho$)
eq. (\ref{rb2eb2}) is exact.

The basic features of (\ref{rb2eb2}) are easy to understand from the
geometry of the unitarity triangle. If $S_\rho=r_\rho=0$ then
$\alpha=\pi/2$. In this case 
$R_b=\sin\beta\equiv 1/\sqrt{1+\tau^2}$,
which gives the leading term in (\ref{rb2eb2}). Because $\sin\beta$
is the minimum value that $R_b$ can take for
fixed $\beta$, first order corrections in $r_\rho$ and $S_\rho$ are
absent and the second-order term is strictly positive.
The protection of (\ref{rb2eb2}) from first-order corrections
in $S_\rho$ and $r_\rho$ is the basis for a precise determination
of $V_{ub}$.

The quantity $S_\rho/2+\tau r_\rho\cos\phi_\rho$ appeared already
in (\ref{gamsr}), (\ref{alpsr}).
For $S_\rho < 0$ there is a further cancellation in this 
term with the penguin shift $\tau r_\rho\cos\phi_\rho$.
Taking $S_\rho=-0.05\pm 0.17$ (\ref{scrhoexp}), 
$\tau=2.54\pm 0.13$ (\ref{tauctb}) and the conservative range
$r_\rho\cos\phi_\rho = 0.04\pm 0.03$ we have
\begin{equation}\label{corrstr}
\frac{S_\rho}{2}+\tau r_\rho\cos\phi_\rho = 0.077\pm 0.114
\end{equation}
The range of $r_\rho\cos\phi_\rho$ covers the result obtained
from the QCD analysis in sec. \ref{subsubsec:detrega}.
As we will show in sec. \ref{subsec:rbqkk}, $r_\rho\cos\phi_\rho$
can also be determined by independent experimental information
on the penguin mode $\bar B_d\to\bar K^{*0}_L K^{*0}_L$, which
confirms the values employed here.

Through second order in $S_\rho$ and $r_\rho$ the correction factor
relative to the lowest-order result in (\ref{rb2eb2}) reads,
using (\ref{corrstr}),
\begin{equation}\label{corrrb}
1+\frac{1}{2}\left(\frac{S_\rho}{2}+\tau r_\rho\cos\phi_\rho\right)^2 = 
1.003^{+0.015}_{-0.003}
\end{equation}
We remark that the lower limit of $1$ for this factor is an
absolute bound.
The third-order term in (\ref{rb2eb2}) is less than about
$0.2 r_\rho S_\rho/2 \lsim 0.2\cdot 0.04\cdot 0.1\lsim 0.001$
and thus completely negligible.

Using $\sin\beta=0.366\pm 0.016$ from Table \ref{tab:input} we obtain  
\begin{equation}\label{rbres}
R_b=\sin\beta \left[1+
\frac{1}{2}\left(\frac{S_\rho}{2}+\tau r_\rho\cos\phi_\rho\right)^2\right]
=0.367\pm 0.016 ^{+0.005}_{-0.002}
\end{equation}
From \cite{Yao:2006px} we have
\begin{equation}\label{vusvcb}
\lambda=|V_{us}|=0.226\pm 0.002\qquad\quad  
|V_{cb}|=0.0416\pm 0.0006
\end{equation}
This implies
\begin{equation}\label{vubbsnum}
|V_{ub}|\equiv\left|\frac{V_{cb} V_{cd}}{V_{ud}}\right| R_b=
\frac{\lambda}{1-\frac{\lambda^2}{2}}\, R_b\, |V_{cb}|=
(3.54^{+0.16}_{-0.15}(R_b)\, \pm 0.05 (V_{cb})\, \pm 0.03(V_{us}))
\cdot 10^{-3} 
\end{equation}
The uncertainty is dominated by the error in 
$\beta$ ($(\pm 0.15)\cdot 10^{-3}$), 
followed by the error in the correction from $S_\rho$, $r_\rho$ 
($(^{+0.05}_{-0.02})\cdot 10^{-3}$) and the error in $V_{cb}$.
Adding errors in quadrature the final result reads
\begin{equation}\label{vubnum2}
|V_{ub}|=(3.54\pm 0.17)\cdot 10^{-3}
\end{equation}
It corresponds to a ratio $|V_{ub}/V_{cb}|=0.085\pm 0.004$, in agreement
with Table \ref{tab:input}.
The value in (\ref{vubnum2}) should be compared with the direct measurements
of $|V_{ub}|$ in $b\to ul\nu$ transitions.
A recent analysis of inclusive decays gives \cite{Neubert:2008cp}
\begin{equation}\label{vubincl}
|V_{ub}|=(3.70\pm 0.32)\cdot 10^{-3}
\end{equation}
The exclusive determination from $B\to\pi l\nu$ decays
has been investigated in \cite{Bourrely:2008za} with the result
\begin{equation}\label{vubexcl}
|V_{ub}|=(3.36\pm 0.23)\cdot 10^{-3}
\end{equation}
Related discussions, in the context of QCD sum rules, 
can be found for instance in \cite{Ball:2006jz} and \cite{Duplancic:2008ix}.
Using an average of data from lattice QCD, \cite{Lubicz:2008am}
quotes for the determination from exclusive decays 
\begin{equation}\label{vublat}
|V_{ub}|=(3.54\pm 0.40)\cdot 10^{-3}
\end{equation}

The results in (\ref{vubnum2}), (\ref{vubincl}), (\ref{vubexcl})
and (\ref{vublat})
are in very good agreement with each other. They provide us with 
a test of Standard Model CP violation in $B\to\psi K_S$ and
$B\to\rho^+\rho^-$ (\ref{vubnum2}) against the $|V_{ub}|$ determination
from tree-level, semileptonic $b\to u l\nu$ decays (\ref{vubincl}),
(\ref{vubexcl}) and (\ref{vublat}).

Numbers for $|V_{ub}|$ very similar to (\ref{vubnum2}) have been
obtained from global fits of the unitarity triangle performed by the
CKMfitter \cite{Charles:2004jd} and UTfit \cite{Bona:2005vz}
collaborations, which quote 
\begin{eqnarray}
|V_{ub}| &=& (3.57\pm 0.17)\cdot 10^{-3}\qquad ({\rm CKMfitter})
\label{vubckmf}\\
|V_{ub}| &=& (3.55\pm 0.15)\cdot 10^{-3}\qquad ({\rm UTfit})
\label{vubutf}
\end{eqnarray}
While such global fit results summarize our current overall
knowledge of quark-mixing parameters, they do not exhibit explicitly
the individual pieces of information that determine this knowledge.
We emphasize here that the precise result in (\ref{vubnum2})
can be obtained from $\sin 2\beta$ and $S_\rho$ alone, with only
very moderate requirements on the accuracy of the penguin
contribution $\sim r_\rho$ from theory.
The representation proposed in (\ref{rbres}) makes this
statement particularly transparent.

The result in (\ref{vubnum2}) is currently the most precise 
determination of $|V_{ub}|$. Since the error is dominated by
the uncertainty in $\sin\beta$, an even
higher precision will be achieved by a more accurate
measurement of $\sin\beta$ as it is expected at the
upcoming LHC experiments. For instance, with a determination
of $\sin\beta$ to $1\%$, the error in (\ref{vubnum2}) would
shrink to $\pm 0.08\cdot 10^{-3}$, corresponding to
a precision of $2\%$ for $|V_{ub}|$.

\subsubsection*{\boldmath Constraint on New Physics phase in 
$B_d-\bar B_d$ mixing\unboldmath}

The preceding analyses rely on the assumption of a Standard Model phase 
in $B_d-\bar B_d$ mixing. We would like to examine the effect of a small 
New Physics phase entering only in $B_d-\bar B_d$-meson 
mixing \cite{Grossman:1997dd,Fleischer:2003xx}. In this 
scenario the New Physics phase shall not violate unitarity of the Standard 
Model CKM matrix. The modified mixing phase $\beta+\Omega$, with the 
New Physics contribution $\Omega$, enters the analysis in the determination 
of $\tau=\cot(\beta+\Omega)$ from $\bar B_d\to J/\psi K_s$ and through 
mixing-induced CP violation in $\bar B_d\to \rho^+\rho^-$. 
The relation (\ref{rhotaueta}) for $\bar\rho$ will 
depend now on $\Omega$:
\begin{equation}\label{rhotauetaNP}
\bar\rho=\frac{(1-\tau\bar\eta)-(\tau+\bar\eta)\tan{\Omega}}
        {1-\tau \tan{\Omega}}
\end{equation}
The measurement of $S_\rho$ determines $\bar\eta$ up to the mixing phase 
$\Omega$. The new relation for $\bar\eta$ in an expansion in $r_\rho$ and 
$S_\rho$ reads
\begin{equation}\label{etaNPexp}
\bar\eta=
\frac{\tau(1-\tau \tan{\Omega})}{1+\tau^2}+\frac{(1-\tau\tan{\Omega})
\left(\tau r_\rho\cos{\phi_\rho}+\frac{S_\rho}{2}(1-\tau\tan{\Omega})\right)}{
1+\tau^2}+\mathcal{O}(r_\rho^2,r_\rho S_\rho,S_\rho^2)
\end{equation}
We note that $\Omega$ enters the leading term in the expansion with an 
enhancement of $\tau\approx 2.5$. The first order term is suppressed relative 
to the leading order term by a factor of $\sim 0.1$. 

Using the exact relations for $\bar\rho$ and $\bar\eta$ one finds the 
following expansion for $R_b$:
\begin{equation}\label{RbNPexp}
R_b = \frac{|1-\tau \tan{\Omega}|}{\sqrt{1+\tau^2}}
\left[1-\tan{\Omega}
\left(\frac{\tau r_\rho\cos{\phi_\rho}}{1-\tau \tan{\Omega}}+\frac{S_\rho}{2}
\right)\right]+\mathcal{O}(r_\rho^2,r_\rho S_\rho,S_\rho^2)
\end{equation}
Again we have an enhancement of the dependence on $\Omega$ by $\tau$. 
For $R_b$ the error from $S_\rho$
and $r_\rho\cos{\phi_\rho}$ is much less important than the error from 
$\tau$ (\ref{tauctb}), in contrast to the case of $\bar\eta$ (\ref{etanum}).

Relating $R_b$ to $|V_{ub}|$ as in (\ref{vubbsnum}), taking 
$|V_{ub}|$, $S_\rho$, $\tau$ from 
experiment and $r_\rho$ and $\phi_\rho$ from QCD factorization, 
the angle $\Omega$ can be extracted.
In general the solution is not unique.
Here we assume that the new phase $\Omega$ is small,
neglecting discrete ambiguities.
Such ambiguities may be eliminated with additional measurements.
In particular, a second solution with large $\Omega$ would imply
a negative sign of $\cos(2(\beta+\Omega))$, which is disfavoured
by experiment \cite{Aubert:2007rp}.
A more general discussion on the New Physics aspects of this analysis
can be found in \cite{Fleischer:2003xx}. 

If one disregards solutions with 
$|\Omega|>\arctan{\left(1/\tau\right)}\approx 20^\circ$
and uses the exclusive determination (\ref{vublat}) of $|V_{ub}|$, then 
$\Omega$ can be determined with an accuracy of few degrees.
Using the exact expression for $R_b(\tau,S_\rho,r_\rho,\phi_\rho,\Omega)$
we find
\begin{equation}\label{omeganum}
\Omega=
\left(0.0^{-0.9}_{+1.0}(\tau)\, ^{+0.2}_{-0.1}(S_\rho)\, ^{+0.2}_{-0.1}
(r_\rho)\, ^{+0.3}_{-0.3}(V_{cb})\,
^{-2.5}_{+2.5}(V_{ub})\right)^\circ
\end{equation}
One may note the very small impact of $S_\rho=-0.05\pm 0.17$
and $r_\rho=0.04\pm 0.03$. A different value for $|V_{ub}|$
from other direct determinations such as (\ref{vubincl}) or (\ref{vubexcl})
leads to very similar results. Combining the errors in (\ref{omeganum})
in quadrature one finds $\Omega=(0.0\pm 2.7)^\circ$.


\subsection{\boldmath Extracting $r_\rho$ from 
$B^-\to\bar K^{*0}_L\rho^-_L$\unboldmath}
\label{subsec:rbqkr}

The precision of CKM angles extracted from CP violation in 
$B\to\rho^+_L\rho^-_L$ is ultimately limited by our knowledge
of the penguin parameters $r_\rho$ and, to a lesser extent, $\phi_\rho$.
Since $r_\rho$ is small, a very moderate accuracy in this quantity
is sufficient to obtain a small theoretical error for CKM parameters.
In \cite{Beneke:2006rb} it has been proposed to constrain the
penguin parameter $r_\rho$ using the penguin dominated
decay $B^-\to\bar K^{*0}_L\rho^-_L$. 
We will discuss this method in the context of our analysis,
comment on the benefits and limitations, present an updated numerical
evaluation, and compare with the theory results of sec. \ref{subsec:cpvbrhrh}. 

The main idea of \cite{Beneke:2006rb} is to determine the penguin amplitude
from the pure-penguin process $B^-\to\bar K^{*0}_L\rho^-_L$ through
\begin{equation}\label{bqkrac}
B(B^-\to\bar K^{*0}_L \rho^-_L) =
\frac{\tau_{B_u}G^2_F\, |\lambda'_c|^2}{32 \pi m_{B}}\, |a_c(\rho K^*)|^2 
\end{equation}

Here we defined $a_p(\rho K^*)\equiv a_p(B^-\to\bar K^{*0}_L \rho^-_L)$
as the coefficient of $(i G_F/\sqrt{2})\lambda_p$ in the
amplitude for $B^-\to\bar K^{*0}_L \rho^-_L$, eqs. 
(\ref{kb0rm}) and (\ref{kb0rmann}). 
They correspond to the charm- and up-quark penguin amplitudes
for this process. Since $a_u(\rho K^*)$ and $a_c(\rho K^*)$ are of
comparable size, and the up-quark amplitude is strongly
CKM suppressed, the charm penguin completely dominates the
branching ratio (\ref{bqkrac}).
The penguin amplitude $a_c(\rho K^*)$ can be related to the (similarly 
normalized) penguin amplitude $a_c(\rho)$ in $\bar B_d\to\rho^+\rho^-$
by introducing the factor
\begin{equation}\label{kapdef}
|\kappa|=\left|\frac{a_c(\rho K^*)}{a_c(\rho)}\right|\approx 0.84 
\end{equation}
To lowest order (in $\alpha_s$ and $\Lambda/m_b$) this factor
would be given by $\kappa=f_K^*/f_\rho=1.04$. Including QCD corrections
this value is reduced to $|\kappa|=1.01$, and further to $|\kappa|=0.97$
by the effects of electroweak penguins.
The estimate in (\ref{kapdef}) includes also the weak annihilation
terms with default model parameters. Annihilation contributions
are thus seen to be potentially important. These observations
agree with the discussion in \cite{Beneke:2006rb}.
In that paper the ratio of the penguin amplitudes
in $B^-\to\bar K^{*0}_L \rho^-_L$ and $\bar B_d\to\rho^+\rho^-$
has been parametrized in terms of a factor $F$, which is 
related to $\kappa$ through $|\kappa|=\sqrt{F} f_{K^*}/f_\rho$.
In \cite{Beneke:2006rb} a rather wide range for $F$ is assumed,
$0.3 < F < 1.5$. We will use the same range, which corresponds to
$|\kappa|=0.93\pm 0.36$.     

For a given value of $|\kappa|$ the penguin parameters $r_\rho$,
$\phi_\rho$ are then constrained by the ratio 
\begin{equation}\label{brkbrh}
\frac{B(B^-\to\bar K^{*0}_L \rho^-_L)}{B(\bar B\to\rho^+_L\rho^-_L)}=
\frac{\tau_{B_u}}{\tau_{B_d}}\left|\frac{V_{cs}}{V_{cd}}\right|^2\,
\frac{|\kappa|^2\, r^2_\rho}{\bar\rho^2+
\bar\eta^2+r^2_\rho + 2\bar\rho\, r_\rho\cos\phi_\rho}
\end{equation}
where CP averaged rates are understood.
Using $\tau=\cot\beta$, $S_\rho$, $C_\rho$ and the ratio
of branching fractions in (\ref{brkbrh}) as experimental inputs,
the four quantities $\bar\rho$, $\bar\eta$, $r_\rho$ and $\phi_\rho$
can be determined as functions of $|\kappa|$.
A discrete ambiguity in the sign of $\cos\phi_\rho$ can be
resolved using the heavy-quark limit. The suppression of
$\phi_\rho$ in this limit singles out the solution with $\cos\phi_\rho >0$.
A similar use of the qualitative result $\cos\phi_\rho >0$ from
factorization has been made in \cite{Buchalla:2003jr,Buchalla:2004tw}.
Further details on the extraction of 
$\bar\rho$, $\bar\eta$, $r_\rho$, $\phi_\rho$ are discussed in
\ref{subsec:rbqkk} in the context of a similar analysis.  
The results for the present method are collected in
Table \ref{tab:ckmkrrr}.
\begin{table}[ht]
\centerline{\parbox{14cm}{\caption{\label{tab:ckmkrrr}
CKM and penguin parameters extracted from $\tau=\cot\beta=2.54\pm 0.13$,
$S_\rho=-0.05\pm 0.17$, $C_\rho=-0.06\pm 0.13$ and
$b=B(B^-\to\bar K^{*0}_L\rho^-_L)/B(\bar B\to\rho^+_L\rho^-_L)
=0.186\pm 0.049$. The penguin correction factor is taken
to be $|\kappa|=0.93\pm 0.36$.}}}
\begin{center}
\begin{tabular}{|c|c|c|c|c|c|c|}
\hline\hline
 & central & $\tau$ & $S_\rho$ & $C_\rho$ & $b$ & $|\kappa|$ \\
\hline
$\bar\rho$ & $0.110$ & $^{-0.010}_{+0.012}$ & $^{-0.030}_{+0.031}$ &
$^{-0.002}_{+0.028}$ & $^{-0.005}_{+0.005}$ & $^{+0.010}_{-0.024}$ \\
\hline
$\bar\eta$ & $0.350$ & $^{-0.013}_{+0.014}$ & $^{+0.012}_{-0.012}$ &
$^{+0.001}_{-0.011}$ & $^{+0.002}_{-0.002}$ & $^{-0.004}_{+0.009}$ \\
\hline
$\gamma$[deg] & $72.5$ & $^{+1.0}_{-1.1}$ & $^{+5.0}_{-5.1}$ &
$^{+0.3}_{-4.8}$ & $^{+0.8}_{-0.9}$ & $^{-1.7}_{+3.9}$ \\
\hline
$r_\rho$ & $0.040$ & $^{-0.002}_{+0.002}$ & $^{+0.000}_{-0.000}$ &
$^{+0.000}_{-0.001}$ & $^{+0.005}_{-0.006}$ & $^{-0.011}_{+0.027}$ \\
\hline
$\phi_\rho$ & $-0.32$ & $^{+0.00}_{-0.00}$ & $^{+0.01}_{-0.01}$ &
$^{+0.69}_{-1.17}$ & $^{+0.04}_{-0.06}$ & $^{-0.13}_{+0.13}$ \\
\hline
\hline
\end{tabular}
\end{center}
\end{table}
Combining and symmetrizing errors we obtain from Table \ref{tab:ckmkrrr} 
\begin{equation}\label{gamalkrrr}
\gamma=(72.5\pm 6.9)^\circ \qquad\quad
\alpha=\pi-\beta-\gamma=(86.0\pm 7.0)^\circ
\end{equation}
where $S_\rho$ is the largest source of uncertainty.
The results agree very well with those in (\ref{gammanum}).
Eq. (\ref{gamalkrrr}) is an update of the results quoted
in \cite{Beneke:2006rb}. We have checked that we obtain
the numbers given in that paper if we use the same input.

A disadvantage of the method just described is that the
charm-penguin amplitudes in $B^-\to\bar K^{*0}_L\rho^-_L$
and in $\bar B_d\to\rho^+_L\rho^-_L$ are not related in full
QCD by $SU(3)$ flavour symmetry alone. The $SU(3)$ argument relating
$a_c(\rho K^*)$ and $a_c(\rho)$ strictly holds only to leading
order in the heavy-quark limit. At the level of power corrections
from weak annihilation these penguin amplitudes are not related
by $SU(3)$. This can be seen from eqs. (\ref{rprmann}) and
(\ref{kb0rmann}), which show that the QCD annihilation penguins
are determined by the coefficient $b_3$ for $a_c(\rho K^*)$, 
but by $b_3+2b_4$ instead for $a_c(\rho)$.
This difference has been discussed in \cite{Beneke:2006rb}.
In order to account for the corresponding $SU(3)$ breaking, a rather
generous correction factor $\kappa$ (\ref{kapdef}) has been allowed for.
While this is certainly a valid procedure, it is 
somewhat against the spirit of using experimental
data to constrain the penguin in $\bar B_d\to\rho^+_L\rho^-_L$.
An unexpectedly large penguin annihilation effect in this
channel, beyond the available model estimates, would not necessarily be
indicated by the $B^-\to\bar K^{*0}_L\rho^-_L$ rate, not even
in the $SU(3)$ limit. In this respect, the method of \cite{Beneke:2006rb}
amounts to the standard analysis of CP violation in 
$\bar B_d\to\rho^+_L\rho^-_L$ with input on $r_\rho$ from factorization
calculations, which are validated by comparing similar theory results
on the penguin mode $B^-\to\bar K^{*0}_L\rho^-_L$ with data.
Indeed, QCD factorization works very well for 
$B^-\to\bar K^{*0}_L\rho^-_L$ with little room for sizable
power corrections. Correspondingly, the values for $r_\rho$ and
the angle $\gamma$ determined in Table \ref{tab:ckmkrrr} are very close
to the values found in the factorization analysis, 
eqs. (\ref{rrhonum}) and (\ref{gammanum}). 
Nevertheless, an independent control of penguin annihilation
corrections in $\bar B_d\to\rho^+_L\rho^-_L$, which is not
guaranteed by $B^-\to\bar K^{*0}_L\rho^-_L$, would be
very desirable.
A variant of the method in \cite{Beneke:2006rb} that can
provide this control will be discussed in the following section.


\subsection{\boldmath Extracting $r_\rho$ from 
$B_d\to\bar K^{*0}_L K^{*0}_L$\unboldmath}
\label{subsec:rbqkk}

In this section we propose a method to constrain the penguin
parameter $r_\rho$ in $\bar B_d\to\rho^+_L\rho^-_L$ (\ref{rphidef}) 
using $SU(3)$ flavour symmetry and data on the penguin decay
$\bar B_d\to\bar K^{*0}_L K^{*0}_L$.
This approach shares the basic idea with the method discussed
in section \ref{subsec:rbqkr}. An important difference is
that now, unlike the case of section \ref{subsec:rbqkr}, 
the penguin process exhibits an exact $SU(3)$ relation to 
the penguin amplitude of $\bar B_d\to\rho^+_L\rho^-_L$. 
Because this relation extends beyond the heavy-quark limit, the
method offers an independent control of all power corrections, in
particular those from weak annihilation topologies.
We show that a precise determination of the unitarity triangle is possible, 
already with present data on $\bar B_d\to\bar K^{*0}_L K^{*0}_L$. 
Since the penguin decay $\bar B_d\to\bar K^{*0}_L K^{*0}_L$ is a
$\Delta S=0$ transition, the up-quark sector of the amplitude does
not have the same CKM suppression as for the $\Delta S=1$ process 
$B^-\to\bar K^{*0}_L\rho^-_L$. We will find that it is still
sufficiently well constrained.

The $SU(3)$ relation between the relevant penguin amplitudes
can be demonstrated as follows.
The penguin contribution for $\bar B_d\to\rho^+_L\rho^-_L$
is given by the component of the amplitude proportional to
$\lambda_c=V_{cb}V^*_{cd}$. The corresponding part of the
effective Hamiltonian (\ref{heff}) has the form
\begin{equation}\label{heffcharm}
   {\cal H}_{\rm QCDP,c} = \frac{G_F}{\sqrt2}
    \bigg( C_1\,Q_1^c + C_2\,Q_2^c
   + \!\sum_{i=3,\dots, 6}\! C_i\,Q_i
   + C_{8g}\,Q_{8g} \bigg) + \mbox{h.c.}
\end{equation}
where we have neglected higher-order electroweak effects.
The operators $Q_i$ are defined in (\ref{qqi}). 
The Hamiltonian in (\ref{heffcharm}) gives rise to the
QCD penguin amplitude in the charm sector of both 
$\bar B_d\to\rho^+_L\rho^-_L$ and $\bar B_d\to\bar K^{*0}_L K^{*0}_L$.
To prove the symmetry relation we note that all operators entering
(\ref{heffcharm}) are invariant under $SU(2)$ rotations
of the doublet $(u,s)$ of quark flavours, the V-spin subgroup
of flavour $SU(3)$. The initial state, a $\bar B_d$ meson in both
cases, is likewise a V-spin singlet. The final states $\rho^+\rho^-$
and $\bar K^{*0} K^{*0}$ are transformed into each other by
interchanging $u$ and $s$ quarks, which represents a particular
V-spin rotation.
In the V-spin symmetry limit, therefore, the relation
\begin{equation}\label{vspinrel}
\langle \bar K^{*0}_L K^{*0}_L|{\cal H}_{\rm QCDP,c}|\bar B_d\rangle =
\langle \rho^+_L\rho^-_L|{\cal H}_{\rm QCDP,c}|\bar B_d\rangle 
\end{equation}
holds as an identity in QCD. As a consequence, 
the QCD penguin amplitudes proportional to $\lambda_c$
in $\bar B_d\to\rho^+_L\rho^-_L$ and $\bar B_d\to\bar K^{*0}_L K^{*0}_L$
have the same form, including the weak annihilation contributions. 
This can be seen from (\ref{rprm}), (\ref{kb0k0}) and 
(\ref{rprmann}), (\ref{kb0k0ann}).

In practice V-spin is broken because the masses of up and strange quarks 
are not the same. This source of V-spin breaking can be expected to be of 
the typical size of flavour $SU(3)$ breaking effects, roughly $20$-$30\%$. 
It is possible to estimate the required correction to the V-spin limit 
using factorization. We will give a more quantitative treatment below.  
Electroweak effects also violate V-spin symmetry. They are similar to
isospin breaking and likely to be much smaller than the $SU(3)$-breaking 
effects due to the strange-quark mass. 
For example, the relative importance of (standard) electroweak penguins 
is governed by the ratio $a^c_{10}/a^c_4\approx 0.03$. This is 
safely negligible in comparison with the dominant V-spin breaking effects.
Contributions from electroweak penguin annihilation are very small
and can also be neglected.

We next turn to the phenomenological implications of the
flavour symmetry relation (\ref{vspinrel}).
Denoting by $a_p(K^*)\equiv a_p(\bar B_d\to\bar K^{*0}_L K^{*0}_L)$
the coefficient of $(i G_F/\sqrt{2})\lambda_p$ in the
amplitude for $\bar B_d\to\bar K^{*0}_L K^{*0}_L$, eqs. 
(\ref{kb0k0}) and (\ref{kb0k0ann}), the CP averaged
branching fraction may be written as
\begin{equation}\label{bqkkac}
B(\bar B_d\to\bar K^{*0}_L K^{*0}_L) =
\frac{\tau_{B_d}G^2_F\, |\lambda_c|^2}{32 \pi m_{B}}
\bigg[ f_0 |a_c(K^*)|^2 + 2 f_1\, {\rm Re}\, a^*_c(K^*)\, \Delta(K^*) 
+ f_2 |\Delta(K^*)|^2
\bigg]
\end{equation} 
\begin{equation}\label{delacuks}
\Delta(K^*)=a_c(K^*)-a_u(K^*) 
\end{equation}
where the functions $f_i$ depend only on CKM parameters.
Expressed in terms of $\tau=\cot\beta$ from (\ref{tauctb}) and
\begin{equation}\label{sigctg}
\sigma\equiv\cot\gamma
\end{equation}
they read
\begin{eqnarray}
f_0(\sigma,\tau) &=& \frac{1+\tau^2}{(\sigma+\tau)^2}  
=\frac{|\lambda_t|^2}{|\lambda_c|^2} \label{f0st}\\
f_1(\sigma,\tau) &=& \frac{\sigma\tau-1}{(\sigma+\tau)^2}
=\frac{{\rm Re}\,\lambda^*_t\lambda_u}{|\lambda_c|^2}\label{f1st}\\
f_2(\sigma,\tau) &=& \frac{1+\sigma^2}{(\sigma+\tau)^2}
=\frac{|\lambda_u|^2}{|\lambda_c|^2}\label{f2st}
\end{eqnarray}
In the region of interest for a Standard Model test,
\begin{equation}\label{sigtaunum}
\sigma=0.447\pm 0.253, \qquad\quad \tau=2.54\pm 0.13
\end{equation}
there is a clear hierarchy among the CKM factors
\begin{equation}\label{f012num}
f_0=0.835^{+0.162}_{-0.125},\qquad
f_1/f_0=0.018\pm 0.086,\qquad
f_2/f_0=0.161^{+0.042}_{-0.026}
\end{equation}
implying $|f_1|\ll f_0$ and $f_2\ll f_0$.
The second inequality is a consequence of the fact that
numerically $|V_{ub}/V_{td}|^2\ll 1$. The first inequality
arises because $f_1\sim\cos\alpha$ and the angle $\alpha$ is close
to $90^\circ$. A similar feature holds for the decay $B\to\rho\gamma$,
where it leads to a suppression of hadronic uncertainties
\cite{Bosch:2004nd,Beneke:2004dp}.
The dominance of the $f_0$ term in (\ref{bqkkac}) is re-inforced
by the hadronic factors since the difference $|\Delta(K^*)|$ is
systematically smaller than $|a_c(K^*)|$.
This difference is a next-to-leading order effect in QCD factorization,
whereas $a_c$ is present at leading order.
In addition, several terms cancel in the difference $a_c-a_u$.
First, the NLO hard spectator corrections are identical in the $c$-
and $u$-sector and drop out, which eliminates the uncertainty
due to $\lambda_B$. Spectator effects in $a_c-a_u$ can only come
from penguin diagrams at NNLO (${\cal O}(\alpha^2_s)$), 
which are very small \cite{Beneke:2006mk}.
Second, also weak annihilation effects cancel in general, in particular
those that are taken into account in our model estimate (coefficients
$b_i$). The only exception would be more complicated 
$b\bar d\to s\bar d d\bar s$ annihilation topologies involving
charm and up-quark loops. These are both power and Zweig rule
suppressed and not expected to give a significant contribution.
Numerically we find  
\begin{equation}\label{acuksdiff}
|a_c(K^*)-a_u(K^*)|/{\rm GeV}^3 =0.021 ^{+0.003}_{-0.003}\, (A_0)\quad
^{-0.003}_{+0.004}\, (\alpha^V_2)\quad
^{+0.006}_{-0.006}\, (m_c)\quad
^{-0.004}_{+0.007}\, (\mu)
\end{equation}
Normalized to the central value of $|a_c(K^*)|=0.084$ we then have
\begin{equation}\label{acuacks}
\frac{|a_c(K^*)-a_u(K^*)|}{|a_c(K^*)|}=0.25^{+0.12}_{-0.10}
\end{equation} 
Together with the CKM factors from (\ref{f012num})
we estimate a relative suppression of the third term
in (\ref{bqkkac}) by $0.010\, (+0.012,-0.007)$ with respect to the
first term. For the second term we estimate a relative size of at most 
$0.009\pm 0.043$, neglecting the phase between $a_c$ and $\Delta$.
In this case the potential magnitude of the correction
depends strongly on the CKM suppression due to $f_1$, which
can be checked after the CKM factors have been determined 
at the end of the analysis. 

Because of the smallness of the $f_1$, $f_2$ terms,
the first term in (\ref{bqkkac}) determines
the branching fraction to very good approximation.
In the $SU(3)$ limit, and up to negligible corrections from
electroweak penguins, $a_c(K^*)$ is equal to the penguin
amplitude $a_c(\rho)$ in $\bar B\to\rho^+_L\rho^-_L$ (in a corresponding
normalization). Introducing the $SU(3)$ factor
\begin{equation}\label{xsu3def}
|\xi|=\left|\frac{a_c(K^*)}{a_c(\rho)}\right|\approx 1.28
\end{equation}
we obtain the ratio of CP averaged branching fractions
\begin{equation}\label{bksbrh}
\frac{B(\bar B_d\to\bar K^{*0}_L K^{*0}_L)}{B(\bar B\to\rho^+_L\rho^-_L)}=
\frac{((1-\bar\rho)^2+\bar\eta^2)\, |\xi|^2\, r^2_\rho}{\bar\rho^2+
\bar\eta^2+r^2_\rho + 2\bar\rho\, r_\rho\cos\phi_\rho}
\end{equation}
This result constrains the penguin parameter $r_\rho$
in $\bar B_d\to\rho^+_L\rho^-_L$.
The four variables $\bar\rho$, $\bar\eta$, $r_\rho$ and $\phi_\rho$  
may now be determined from the four measurements of
$\tau=(1-\bar\rho)/\bar\eta$, $S_\rho(\bar\rho,\bar\eta,r_\rho,\phi_\rho)$,
$C_\rho(\bar\rho,\bar\eta,r_\rho,\phi_\rho)$ and
$b\equiv B(\bar B_d\to\bar K^{*0}_L K^{*0}_L)/B(\bar B\to\rho^+_L\rho^-_L)$.
This analysis then depends on a single theoretical parameter,
the $SU(3)$ factor $|\xi|=1.28\pm 0.14$, where we adopt
the estimate in (\ref{xsu3def}) and assign a $50\%$ error
on the magnitude of $SU(3)$ breaking.
The quantity $\xi=a_c(K^*)/a_c(\rho)$ is real to very good approximation, 
$\xi\approx |\xi|$. 

The expressions for $S$ and $C$ in terms of $\bar\rho$,
$\bar\eta$, $r$ and $\phi$ are identical to the case of 
$\bar B_d\to\pi^+\pi^-$ discussed in 
\cite{Buchalla:2003jr,Buchalla:2004tw}. They read
\begin{equation}\label{srhoeta}
S=\frac{2\bar\eta [\bar\rho^2+\bar\eta^2-r^2-\bar\rho(1-r^2)+
       (\bar\rho^2 +\bar\eta^2-1)r \cos\phi]}{((1-\bar\rho)^2+\bar\eta^2)
         (\bar\rho^2+\bar\eta^2+r^2 +2 r\bar\rho \cos\phi)}
\end{equation}
\begin{equation}\label{crhoeta}
C=\frac{2 r\bar\eta\, \sin\phi}{
   \bar\rho^2+\bar\eta^2+r^2 +2 r\bar\rho \cos\phi}
\end{equation}
We remark that discrete ambiguities in the determination
of $\bar\rho$, $\bar\eta$, $r$ and $\phi$ can be avoided
using other constraints on the unitarity triangle, which
exclude $\bar\rho$, $\bar\eta$ far outside the region
allowed in the Standard Model \cite{Buchalla:2003jr,Buchalla:2004tw}.
A discrete ambiguity in the sign of $\cos\phi_\rho$ can be
resolved by the heavy-quark limit, which favours the solution 
with $\cos\phi_\rho >0$. As pointed out in a similar context
in \cite{Beneke:2006rb}, the discrete choice is still less restrictive
in practice, because the second solution has $\cos\phi_\rho < -0.8$,
which is in fact far smaller than zero.

The result of this analysis is given in Table \ref{tab:ckmkkrr}, 
where we have also summarized the experimental input
from Table \ref{tab:brexp} and section \ref{subsec:cpvbrhrh}. 
\begin{table}[ht]
\centerline{\parbox{14cm}{\caption{\label{tab:ckmkkrr}
CKM and penguin parameters extracted from $\tau=\cot\beta=2.54\pm 0.13$,
$S_\rho=-0.05\pm 0.17$, $C_\rho=-0.06\pm 0.13$ and 
$b=B(\bar B_d\to\bar K^{*0}_L K^{*0}_L)/B(\bar B\to\rho^+_L\rho^-_L)
=0.043\pm 0.015$. The $SU(3)$ breaking parameter is taken 
to be $|\xi|=1.28\pm 0.14$.}}}
\begin{center}
\begin{tabular}{|c|c|c|c|c|c|c|}
\hline\hline
 & central & $\tau$ & $S_\rho$ & $C_\rho$ & $b$ & $|\xi|$ \\
\hline
$\bar\rho$ & $0.088$ & $^{-0.010}_{+0.011}$ & $^{-0.029}_{+0.029}$ & 
$^{-0.000}_{+0.010}$ & $^{-0.009}_{+0.011}$ & $^{+0.006}_{-0.007}$ \\
\hline
$\bar\eta$ & $0.359$ & $^{-0.014}_{+0.015}$ & $^{+0.011}_{-0.011}$ &
$^{+0.000}_{-0.004}$ & $^{+0.004}_{-0.004}$ & $^{-0.002}_{+0.003}$ \\
\hline
$\gamma$[deg] & $76.2$ & $^{+1.0}_{-1.0}$ & $^{+4.7}_{-4.9}$ &
$^{+0.0}_{-1.6}$ & $^{+1.5}_{-1.8}$ & $^{-0.9}_{+1.1}$ \\
\hline
$r_\rho$ & $0.064$ & $^{-0.003}_{+0.004}$ & $^{-0.002}_{+0.003}$ &
$^{+0.000}_{-0.000}$ & $^{+0.010}_{-0.012}$ & $^{-0.006}_{+0.008}$ \\
\hline
$\phi_\rho$ & $-0.20$ & $^{+0.00}_{-0.00}$ & $^{+0.00}_{-0.00}$ &
$^{+0.43}_{-0.47}$ & $^{+0.03}_{-0.05}$ & $^{-0.02}_{+0.02}$ \\
\hline
\hline
\end{tabular}
\end{center}
\end{table}
The output values $\bar\rho$, $\bar\eta$, $r_\rho$ and $\phi_\rho$
are shown together with their sensitivity to the relevant
input quantities. 
From Table \ref{tab:ckmkkrr} we draw the following conclusions:
\begin{description}
\item{a)}
The errors on the CKM quantities $\bar\rho$, $\bar\eta$ and $\gamma$
are rather small. They are dominated by the uncertainty in $S_\rho$
($\bar\eta$ is sensitive also to $\tau=\cot\beta$).
\item{b)}
The error from the $SU(3)$ factor $|\xi|$ is smaller than the 
errors from the experimental quantities $\tau$, $S_\rho$, $b$,
which may still be improved by future measurements.
\item{c)}
The penguin parameter is obtained as $r_\rho=0.064\pm 0.014$.
The central value is somewhat larger than the theoretical number
in (\ref{rrhonum}), but both results are compatible within errors.
This confirms the expected smallness of $r_\rho$, which is the basis
for a precise extraction of CKM quantities.   
\item{d)} 
The phase $\phi_\rho$ is seen to be strongly dependent on $C_\rho$,
but essentially uncorrelated with the remaining parameters and input
quantities. In particular $\bar\rho$, $\bar\eta$ and $\gamma$ are almost
unaffected by the value of $C_\rho$ within the measured range.
This behaviour is in agreement with the general expectation discussed
in \ref{subsec:cpvbrhrh}. The error on $\phi_\rho$ is completely 
dominated by the error on $C_\rho$. The sign of $\phi_\rho$ is opposite
to the central standard model value from factorization in (\ref{phirhonum}).
If higher-order perturbative corrections cannot account for this change
in sign, and we assume it is not due to New Physics, this would mean
that power corrections give an important contribution to the
strong phase. 
A similar situation is known to occur for the direct CP asymmetries in
$\bar B_d\to\pi^+\pi^-$ and $\bar B_d\to\pi^+K^-$.
However, within uncertainties the numbers for $\phi_\rho$ in
(\ref{phirhonum}) and Table \ref{tab:ckmkkrr} are fully consistent
with each other. The result for $\phi_\rho$ in Table \ref{tab:ckmkkrr}
confirms the prediction of a suppressed phase in the heavy quark limit. 
\end{description}
Combining the errors in Table \ref{tab:ckmkkrr} we find for the CKM
angles
\begin{equation}\label{gamalkkrr}
\gamma=(76.2\pm 5.3)^\circ \qquad\quad
\alpha=\pi-\beta-\gamma=(82.3\pm 5.4)^\circ
\end{equation}
where the uncertainty is dominated by the experimental error in $S_\rho$.
The result is in very good agreement with (\ref{gammanum}). It is already
rather accurate at present. From Table \ref{tab:ckmkkrr} we see
that a precision of $\pm 1^\circ$ for $\gamma$ and $\alpha$ from
this method should be possible.

Finally, we remark that the approximations leading to
(\ref{bksbrh}) may be cross-checked using the extracted value
of $\gamma$ or $\sigma=\cot\gamma=0.25\pm 0.10$, and 
$\tau=\cot\beta=2.54\pm 0.13$.
Varying also the hadronic input parameters,
the relative importance of the correction terms in (\ref{bqkkac})
is then smaller than $\pm 3\%$. (For the default parameter
set and $\sigma=0.25$, $\tau=2.54$, the correction is $-0.8\%$.)
The corresponding change in (\ref{bksbrh}) could be absorbed in a 
modification of $|\xi|$ by less than $\pm 1.5\%$, which is entirely
negligible. 


\subsection{\boldmath Unitarity triangle from $B_d\to\pi^+\pi^-$
and $B_d\to\bar K^{0} K^{0}$\unboldmath}
\label{subsec:rbqkkpp}

The analysis of section \ref{subsec:rbqkk} made use of
CP violation in $\bar B_d\to\rho^+_L\rho^-_L$, a measurement
of $\sin 2\beta$, and the $\bar B_d\to\bar K^{*0}_L K^{*0}_L$
branching fraction to obtain an accurate determination
of the unitarity triangle. The decay $\bar B_d\to\bar K^{*0}_L K^{*0}_L$
served to fix the penguin-to-tree ratio in $\bar B_d\to\rho^+_L\rho^-_L$
based on $SU(3)$ flavour symmetry.

The same analysis may also be performed with the $VV$-modes replaced  
by their pseudoscalar counterparts, that is, using CP violation
in $\bar B_d\to\pi^+\pi^-$ and constraining the penguin
parameter $r_\pi$ with $\bar B_d\to\bar K^{0} K^{0}$ and $SU(3)$ symmetry.
The formulas of section \ref{subsec:rbqkk} apply with obvious
substitutions. Related discussions can be found in
\cite{Buchalla:2003jr,Buchalla:2004tw}.

Using form factor estimates based on \cite{Ball:2004ye}
\begin{equation}\label{ffkpi}
f^{B\to\pi}_+(0)=0.258\pm 0.031\qquad\quad 
f^{B\to K}_+(0)=0.304\pm 0.042
\end{equation}
we find from QCD factorization \cite{Beneke:2001ev,Beneke:2003zv} 
\begin{equation}\label{zsu3def}
|\zeta|\equiv\left|\frac{a_c(K)}{a_c(\pi)}\right| = 1.46\pm 0.23
\end{equation}
Again we have assigned a generous $50\%$ uncertainty on the
total amount of $SU(3)$ breaking.
With experimental input from \cite{Barberio:2007cr} we obtain the
results displayed in Table \ref{tab:ckmkkpp}.

\begin{table}[ht]
\centerline{\parbox{14cm}{\caption{\label{tab:ckmkkpp}
CKM and penguin parameters extracted from $\tau=\cot\beta=2.54\pm 0.13$,
$S_\pi=-0.61\pm 0.08$, $C_\pi=-0.38\pm 0.07$ and 
$b\equiv B(\bar B_d\to\bar K^{0} K^{0})/B(\bar B\to\pi^+\pi^-)
=0.186\pm 0.040$. The $SU(3)$ breaking parameter is taken 
to be $|\zeta|=1.46\pm 0.23$.}}}
\begin{center}
\begin{tabular}{|c|c|c|c|c|c|c|}
\hline\hline
 & central & $\tau$ & $S_\pi$ & $C_\pi$ & $b$ & $|\zeta|$ \\
\hline
$\bar\rho$ & $0.179$ & $^{-0.013}_{+0.014}$ & $^{-0.020}_{+0.023}$ & 
$^{-0.014}_{+0.021}$ & $^{-0.015}_{+0.019}$ & $^{+0.023}_{-0.027}$ \\
\hline
$\bar\eta$ & $0.323$ & $^{-0.011}_{+0.011}$ & $^{+0.008}_{-0.009}$ &
$^{+0.005}_{-0.008}$ & $^{+0.006}_{-0.007}$ & $^{-0.009}_{+0.011}$ \\
\hline
$\gamma$[deg] & $61.1$ & $^{+1.0}_{-1.0}$ & $^{+3.3}_{-3.7}$ &
$^{+2.3}_{-3.4}$ & $^{+2.5}_{-3.1}$ & $^{-3.7}_{+4.4}$ \\
\hline
$r_\pi$ & $0.146$ & $^{-0.008}_{+0.009}$ & $^{-0.006}_{+0.008}$ &
$^{-0.002}_{+0.002}$ & $^{+0.015}_{-0.017}$ & $^{-0.020}_{+0.027}$ \\
\hline
$\phi_\pi$ & $-0.88$ & $^{+0.01}_{-0.01}$ & $^{+0.02}_{-0.03}$ &
$^{+0.19}_{-0.24}$ & $^{+0.09}_{-0.13}$ & $^{-0.17}_{+0.14}$ \\
\hline
\hline
\end{tabular}
\end{center}
\end{table}

Combining errors we obtain from Table \ref{tab:ckmkkpp}
\begin{equation}\label{gamkkpp}
\gamma=(61.1\pm 6.8)^\circ 
\end{equation}
This value is lower than the result in (\ref{gamalkkrr})
but it remains consistent at the level of roughly $2\sigma$.
One possible source of this discrepancy is the rather large
value of $C_\pi=-0.38\pm 0.07$, representing the simple average
of the BaBar \cite{Aubert:2007mj} and Belle \cite{Ishino:2006if} results
\begin{equation}\label{cpiexp}
C_\pi=-0.21\pm 0.09 \quad {\rm (BaBar)}\qquad\quad
C_\pi=-0.55\pm 0.09 \quad {\rm (Belle)}
\end{equation}
These results are not in very
good agreement. With a smaller $|C_\pi|$, favoured by QCD factorization
and the BaBar measurement, the extracted value for $\gamma$ would
increase somewhat. For example, with $C_\pi=-0.1$ we obtain
$\gamma=66.6^\circ$, keeping all other inputs fixed.
Particularly important for the resulting $\gamma$ is the value of $S_\pi$.
If it were $2\sigma$ lower in absolute magnitude, at $S_\pi=-0.45$,
the central value of $\gamma$ would shift to $\gamma=67.4^\circ$.
The uncertainties in $b$ and $|\zeta|$ also have a relatively
large impact. This is because of the larger size of the penguin
contribution $r_\pi$ in comparison with $r_\rho$.
Note that the error on $\gamma$ from the uncertainty in $b$
is almost twice as large in Table \ref{tab:ckmkkpp} than in
Table \ref{tab:ckmkkrr}, even though $b$ is known with an accuracy of
$22\%$ in the former case and only to $35\%$ in the latter. 
Thus, because of the larger size of the penguin amplitude, 
and also because of the experimental situation
of $C_\pi$, which is still not entirely resolved,
the determination of the unitarity triangle from
$\bar B\to\pi^+\pi^-$ and $\bar B\to\bar K^0 K^0$ appears
to be somewhat less precise than the determination from
$\bar B\to\rho^+_L\rho^-_L$ and $\bar B\to\bar K^{*0}_L K^{*0}_L$.


\subsection{\boldmath CP violation in $B_s\to\phi_L\phi_L$\unboldmath}
\label{subsec:cpvbsphph}

The decay $\bar B_s\to\phi\phi$ is a pure $b\to s$ penguin transition
and thus of considerable interest as a New Physics probe. Possible hints
of deviations from the Standard Model in CP violation in the
$b\to s$ penguin process $\bar B_d\to\phi K_S$, and similar modes,
have so far remained inconclusive. A detailed experimental study of
$\bar B_s\to\phi\phi$ will become possible with the LHC 
\cite{Buchalla:2008jp}.
In the Standard Model CP violation in $\bar B_s\to\phi\phi$ is small.
Any nonzero effect in excess of the Standard Model contribution will
signal the presence of New Physics.
Based on our next-to-leading order results
we shall investigate the size and uncertainty of CP violation
in the Standard Model, which ultimately limits
the sensitivity to New Physics. 

The time dependent CP asymmetry in $\bar B_s\to\phi_L\phi_L$ decays
is defined by
\begin{equation}\label{scphidef}
{\cal A}_{CP,\phi}(t)=\frac{\Gamma(\bar B_s(t)\to\phi_L\phi_L)-
\Gamma(B_s(t)\to\phi_L\phi_L)}{
\Gamma(\bar B_s(t)\to\phi_L\phi_L)+
\Gamma(B_s(t)\to\phi_L\phi_L)}=
S_\phi\, \sin(\Delta m_s\, t) - C_\phi\, \cos(\Delta m_s\, t) 
\end{equation} 
Here we have neglected the effects of a nonzero width difference
$\Delta\Gamma_{B_s}$, which would modify the time dependence
of the CP asymmetry. This can be taken into account in extracting
$S_\phi$ and $C_\phi$, but would not change the following 
discussion of these parameters. 

For a generic $B$ decay into a CP self-conjugate final state $f$
one has
\begin{equation}\label{scxi}
S=\frac{2\,{\rm Im}\xi}{1+|\xi|^2},\qquad\quad
C=\frac{1-|\xi|^2}{1+|\xi|^2},\qquad\quad
\xi=-\frac{M^*_{12}}{|M_{12}|}\frac{A(\bar B\to f)}{A(B\to f)}
\end{equation}
where $M_{12}=\langle B|{\cal H}_{\Delta B=2}|\bar B\rangle$ is the 
$B$--$\bar B$ mixing amplitude. We use the phase convention 
$CP|\bar B\rangle = -|B\rangle$.
The CP violation parameters $S_\phi$ and $C_\phi$ then become
\begin{equation}\label{scphia}
S_\phi = 2\lambda^2\eta\, {\rm Re}\frac{a_c(\phi)-a_u(\phi)}{a_c(\phi)}\, , 
\qquad\quad
C_\phi = 2\lambda^2\eta\, {\rm Im}\frac{a_c(\phi)-a_u(\phi)}{a_c(\phi)}
\end{equation}
where $a_p(\phi)$, $p=u$, $c$, is the coefficient of 
$(iG_F/\sqrt{2})\lambda'_p$ in the
$\bar B_s\to\phi_L\phi_L$ amplitude (\ref{phph}) and (\ref{phphann}).
It can be seen from (\ref{scphia}) that $S_\phi$ and $C_\phi$ depend on
the same CKM quantity but on different hadronic parameters.
A measurement of $C_\phi$ is therefore only of limited use in controlling
hadronic uncertainties in $S_\phi$.

The hadronic parameters in (\ref{scphia}) depend on the difference
between the penguin amplitudes from the charm and the up-quark sector.
This difference is calculable in factorization. It has the further
advantage that the leading annihilation corrections related
to the parameters $b_i$ (\ref{phphann}) cancel in $a_c(\phi)-a_u(\phi)$. 
A similar cancellation occurs for the hard-spectator scattering
contributions in the NLO approximation.
We then find
\begin{equation}\label{acuphdiff}
|a_c(\phi)-a_u(\phi)|/{\rm GeV}^3 =0.057 ^{+0.007}_{-0.007}\, (A_0)\quad
^{-0.008}_{+0.010}\, (\alpha^V_2)\quad 
^{+0.016}_{-0.015}\, (m_c)\quad
^{-0.012}_{+0.021}\, (\mu)
\end{equation}
where we show the dominant parametric uncertainties and
their origin (in brackets).
In contrast to the difference $a_c(\phi)-a_u(\phi)$, the absolute value
of $a_c(\phi)$ depends on the annihilation contributions $b_i$.
Rather than aiming for an accurate theoretical prediction,  
it therefore appears more reliable to extract
$|a_c(\phi)|$ from experiment. Neglecting the very small
up-quark contribution, we may write
\begin{equation}\label{bbsphphac}
B(\bar B_s\to\phi_L\phi_L) =
\frac{\tau_{B_s}G^2_F\, |\lambda'_c|^2}{64\pi m_{B_s}}\, |a_c(\phi)|^2
\end{equation}
This gives
\begin{equation}\label{acbr}
|a_c(\phi)| = 0.177\, {\rm GeV}^3\,
\left[\frac{B(\bar B_s\to\phi_L\phi_L)}{15\cdot 10^{-6}}\right]^{1/2}\,
\left[\frac{1.53\,{\rm ps}}{\tau_{B_s}}\right]^{1/2}
\end{equation}
First evidence for the decay $\bar B_s\to\phi\phi$ has been reported
by the CDF collaboration, which quotes \cite{Acosta:2005eu}
\begin{equation}\label{bsppexp}
B(\bar B_s\to\phi\phi) = (14^{+6}_{-5} (stat) \pm 6 (syst))\cdot 10^{-6}
\end{equation}
This is not yet the longitudinal branching fraction needed in
(\ref{acbr}), and the error is still large. 
By the time CP violation in $\bar B_s\to\phi\phi$ will be studied
at the LHC, the branching fraction will be known with good
precision and the number in (\ref{acbr}) can be easily updated.

The quantity $S_\phi$ is predicted to be small and positive
in the Standard Model. With our default parameter set
and $\eta=0.36$ we obtain 
\begin{equation}\label{sphidefault}
S_\phi({\rm default}) \approx 0.01
\end{equation}  
From the discussion above we conclude
that the Standard Model upper limit can be written as
\begin{equation}\label{sphilim}
S_\phi\lsim 2 \lambda^2\eta\, \frac{|a_c(\phi)-a_u(\phi)|}{|a_c(\phi)|}
\lsim  \lambda^2 \eta\,
\left[\frac{B(\bar B_s\to\phi_L\phi_L)}{15\cdot 10^{-6}}\right]^{-1/2}
\end{equation}
A similar limit holds for the absolute value of $C_\phi$.
For $\eta\lsim 0.4$ we have
\begin{equation}\label{scphlim}
S_\phi \lsim 0.02 \qquad\quad   |C_\phi| \lsim 0.02
\end{equation}
A rescaling for the actual value of the branching fraction,
which might deviate from the assumed default value of $15\cdot 10^{-6}$,
can be done using (\ref{sphilim}).
Measurements in excess of these Standard Model limits
would constitute evidence for New Physics.
The expected sensitivity of LHCb after five years of data taking
is $\sigma(S_\phi)\approx 0.05$ \cite{Amato:2007jc}. 
Improvements to values of $0.01$ or $0.02$
with the anticipated LHCb upgrade appear possible \cite{Buchalla:2008jp}.
This should allow us to exploit the full New Physics potential
of $S_\phi$ and to detect non-standard effects in $b\to s$
penguins at the few percent level.

\afterpage{\clearpage}

\section{Comparison with the literature}
\label{sec:comparison}

In this section we comment briefly on the existing literature related 
to the subject of the present paper \citer{Kagan:2004uw,Cheng:2008gx}.

We re-emphasize that factorization calculations for
charmless two-body $B$ decays, in particular
$B\to V_LV_L$, are useful for flavour physics analyses such as 
the determination of CKM parameters. This has also been stressed
in \cite{Beneke:2006hg} and earlier in 
\cite{Beneke:2001ev,Buchalla:2003jr,Buchalla:2004tw,Beneke:2003zv}.
The most comprehensive study of $B\to VV$ decays has been presented
in \cite{Beneke:2006hg} with an emphasis on total branching
fractions and polarization observables, for instance the longitudinal
polarization fractions $f_L$. 
More recently these processes were considered in \cite{Cheng:2008gx}
in an extended study, following the analysis of \cite{Beneke:2006hg}.
We do not discuss transverse polarization
here but rather concentrate on the decays with
longitudinal vector mesons $B\to V_LV_L$. These are calculable in QCD
in the heavy-quark limit and thus of special interest for phenomenological
applications in flavour physics. We list the detailed results for
the $B\to V_LV_L$ amplitudes in explicit terms. The corresponding results
of \cite{Beneke:2006hg} can be reconstructed from similar formulas
given for $B\to PV$ decays in \cite{Beneke:2003zv}.
Our main results are consistent with \cite{Beneke:2006hg}.
A minor difference with (the original version of) \cite{Beneke:2006hg}
are the expressions for penguin annihilation $A^i_3\approx 0$ and $A^f_3$
(\ref{ai1af3}). The final expressions in \cite{Beneke:2006hg} give
incorrectly, due to a relative sign change,
a nonvanishing $A^i_3$ and $A^f_3\approx 0$, even though
the basic formulas agree with (\ref{aifk}). The difference leads to
a reduced sensitivity to penguin annihilation in penguin-dominated
$B\to V_LV_L$ decays.
It does not play a role for 
$\bar B\to\rho^+\rho^-$, because the corresponding decay amplitude is tree 
dominated and color allowed, so no deviations from modelling of 
power-suppressed contributions are expected.
This point has previously been noted in \cite{Cheng:2008gx}. 
There is consensus on the expressions (\ref{aifk}), (\ref{ai1af3}) within 
the annihilation model used \cite{Kagan:2004uw,Cheng:2008gx,Beneke:pc} .
Another difference with \cite{Beneke:2006hg,Beneke:2003zv} is the 
treatment of electromagnetic penguin matrix elements contributing
to $a^u_{7,9}$, where we have proposed an explicit model for the
long-distance contributions to these ${\cal O}(\alpha)$ terms. 

The article \cite{Kagan:2004uw} concentrates on transverse polarization
and therefore does not report the complete expressions for the amplitudes
with longitudinal vector mesons.
Where a comparison is possible we agree with the results of
\cite{Kagan:2004uw}, except for two minor discrepancies.
One is the detailed form of the integrand of the annihilation parameter 
$A^f_3$. However, the final result for $A^f_3$ coincides with ours.
Another difference is the annihilation part of $B^-\to K^{*-}\phi$,
which should read $b_3+b^{\rm EW}_3$ instead of $b_3-b^{\rm EW}_3/2$.
Both issues are inconsequential.  

$B\to VV$ decays have been studied within QCD factorization
also in \citer{Cheng:2001aa,Cheng:2008gx}. 
These papers address various $VV$ channels, especially penguin
dominated modes such as $\phi K^*$. Some of them investigate the impact of 
New Physics scenarios \cite{Yang:2004pm,Das:2004hq,Huang:2005qb},
\cite{Cheng:2008gx} extends the analysis to $VA$ and $AA$ modes as well.
In comparison, the present paper, while concentrating on $B\to V_LV_L$,
gives complete NLO results for all channels, a detailed analysis
of uncertainties and applications for precision tests of flavour physics.   
The authors of \cite{Cheng:2008gx} employ $m_c(m_b)=0.91$ GeV,
smaller than the value used here and in \cite{Beneke:2006hg}.
We find that the error due to $m_c$ for the longitudinal amplitude is small 
compared to other experimental input, also for penguin dominated decays.

\afterpage{\clearpage}

\section{Conclusions}
\label{sec:conclusion}

In this paper we presented a systematic analysis of $B$-meson decays
into a pair of longitudinal vector mesons. The main results can be
summarized as follows:
\begin{itemize}
\item
Explicit formulas are given for the complete set of $\Delta S=0$
and $\Delta S=1$ decay amplitudes of $\bar B\to V_LV_L$ at NLO
in QCD factorization. Estimates of power corrections from weak
annihilation are included to study the sensitivity to effects of 
this kind in phenomenological applications.
The set of decays considered comprises 17 $\Delta S=0$ 
and 17 $\Delta S=1$ channels, including 2 and 4 pure annihilation
modes, respectively.

\item
The agreement with the available measured branching ratios of  
$\bar B_d$, $B^-$ decays into $\rho^+\rho^-$, $\rho^0\rho^0$, $\rho^-\rho^0$, 
$\bar K^{*0}K^{*0}$, $\bar K^{*0}\rho^0$, $K^{*-}\rho^+$, $K^{*-}\rho^0$, 
$\bar K^{*0}\rho^-$, $\bar K^{*0}\phi$, $K^{*-}\phi$ and $\bar K^{*0}\omega$
is very good, within current uncertainties and with the central
values used for $\bar B\to V_L$ form factors.
The decay to $\rho^-\omega$ agrees at the level of about $2\sigma$.
We note that QCD factorization works well in particular for the penguin
modes and for the three
$\rho\rho$ channels and their characteristic hierarchy of
branching fractions. 
Our hadronic input is based on the available literature. No tuning of
parameters has been done to improve the fit with data.

\item
Long-distance electromagnetic penguins in $B$ decays with
neutral vector mesons $\rho^0$, $\omega$ or $\phi$
are taken into account using a model description. 
They are of some conceptual interest but their numerical
impact is generally small.

\item
The deviation from ideal mixing in the $\omega$-$\phi$ system is
found to have a small effect on most decay modes with these particles
in the final state. The impact is very large for $B^-\to\rho^-\phi$.

\item
We advocate the direct use of mixing-induced CP violation in 
$\bar B_d\to \rho_L^+\rho_L^-$, measured by $S_\rho$, to extract 
the parameters of the unitarity triangle.
Together with $\sin 2\beta$ the current measurements of $S_\rho$
imply (\ref{gammanum})
\begin{equation}
\gamma=(72.4\pm 6.2)^\circ
\end{equation}
where the error is dominated by $S_\rho$. This analysis 
benefits from the small penguin-to-tree ratio for vector modes 
$r_\rho=0.038\pm 0.024$ (\ref{rrhonum}), which leads to a residual 
theory uncertainty in $\gamma$ of $\pm 3^\circ$.

\item
We propose a method to relate the penguin contribution in
$\bar B_d\to \rho_L^+\rho_L^-$ to the decay 
$\bar B_d\to \bar K_L^{*0}K_L^{*0}$ based on the $V$-spin
subgroup of flavour $SU(3)$. This makes it possible to constrain the 
uncertainties due to penguin power corrections, especially from annihilation 
topologies, and provides us with a check on the penguin-to-tree ratio 
calculated in QCD fatorization. The absolute value determined by the $V$-spin
method, $r_\rho=0.064\pm 0.014$, is consistent with the calculation 
in QCD factorization. The resulting angle 
\begin{equation}
\gamma=(76.2\pm 5.3)^\circ 
\end{equation}
has a residual theory error of $\pm 1^\circ$. 

\item
A comparison of the analyses mentioned in the previous two items 
to the corresponding ones with pseudoscalar decay modes suggests that the 
hadronic uncertainties are better under control in the case
of vector modes. 

\item
We point out that within the SM $\sin 2\beta$ and the CP violation 
parameter $S_\rho$ in $\bar B\to\rho^+_L\rho^-_L$ determine
\begin{equation}
|V_{ub}|=(3.54\pm 0.17)\cdot 10^{-3}
\end{equation} 
where the error is at present still entirely dominated by $\sin 2\beta$.
Hadronic uncertainties enter only at second order in $S_\rho$ and 
the penguin parameter $r_\rho$ and are below $2\%$.
Possible New Physics affecting the $B_d-\bar B_d$ mixing phase
can be constrained by comparing the above value of $|V_{ub}|$
with direct determinations from exclusive or inclusive
$b\to u l\nu$ decays.

\item
In future measurements $\bar B_s\to\phi\phi$ will provide tests 
for New Physics.
We present a bound on $S_\phi$ and $C_\phi$, which will further improve when 
more data are available.
\end{itemize}

The phenomenology of $\bar B\to V_LV_L$ decays is rich and
promising. QCD factorization provides a solid theoretical basis
for these processes. With the measurement of additional channels
and improved precision for the ones already observed, many more
applications in flavour physics may be foreseen. 

\appendix

\section{Long-distance electromagnetic penguins}
\label{sec:ldempeng}

For $B$ decays with neutral vector mesons $\rho^0$, $\omega$ or $\phi$,
the operators $Q^p_{1,2}$ have electromagnetic penguin-type
matrix elements where the photon from the $p\bar p$ loop 
is transformed into one of these mesons. The photon virtuality
$k^2=m^2_V$ is then small. While the penguin loop may still be
considered short-distance dominated for $p=c$ due to the charm-quark mass,
the matrix element becomes sensitive to long-distance hadronic
physics for $p=u$. This can be seen from the perturbative result
for the case of the charm quark (\ref{pewnc}), which diverges in the
limit $m_c\to 0$. 
The up-quark contribution is thus not strictly calculable.
Since this situation arises only in a small electromagnetic correction,
it is not a serious problem for practical purposes. In fact,
additional dependence on long-distance hadronic physics is to be
expected when electromagnetic radiative corrections to hadronic
$B$ decays are considered. 
Still the penguin matrix element under discussion contributes
within our approximation scheme of including leading electroweak
effects. We shall therefore give an estimate of its size using
available information on the long-distance dynamics of the up-quark loop.  
Apart from obtaining a numerical evaluation of the effect,
the long-distance electromagnetic penguin is also interesting
for conceptual reasons.
 
The up-quark loop is closely related to the vacuum polarization
function $\Pi(k^2)$, where the UV subtraction is given by the
standard renormalization prescription of the weak hamiltonian.
We thus write the penguin matrix element, needed at
low photon virtuality $k^2=m^2_V$, as the matrix element evaluated at
$k^2=m^2_b$ plus a remainder proportional to the difference
$\Pi(k^2)-\Pi(m^2_b)$. The contribution of the electromagnetic
up-quark penguin to the coefficients $a^u_{7,9}$ then takes
the form 
\begin{equation}\label{dau79ew}
\Delta a^u_{7,9}=\frac{\alpha}{9\pi}
(C_1+N_c C_2)\left[ \frac{4}{3}\ln\frac{m_b}{\mu}
 -\frac{4}{9}-\frac{2\pi}{3}i-\frac{8\pi^2}{N_c}(\Pi(m^2_V)-\Pi(m^2_b))\right]
\end{equation}
The correlator $\Pi(k^2)$ is defined through
\begin{equation}\label{pidef}
\Pi_{\mu\nu}(k)=i\, \int d^4x\, e^{i k\cdot x}\,
\langle 0|T j_\mu(x) j_\nu(0) | 0 \rangle
\equiv (k_\mu k_\nu -k^2 g_{\mu\nu})\Pi(k^2)
\end{equation}
where $j_\mu=\bar u\gamma_\mu u$.
The first terms in the square brackets of (\ref{dau79ew}) come from the 
perturbative evaluation of the matrix element at $k^2=m^2_b$ and carry
the appropriate scale and scheme dependence.
The remainder depending on $\Pi$ may be computed to lowest
(one-loop) order, which reproduces the perturbative
result for the matrix element.
We shall treat $\Pi(k^2)-\Pi(m^2_b)$ as the full hadronic
correlator, which includes the nonperturbative hadronic
physics relevant at low $k^2$. This procedure assumes
a factorization of the soft hadronic correlator form the
remaining parts of the diagram, which is not strictly
justified. We adopt this additional assumption to obtain 
a rough estimate of the long-distance sensitive penguin contributon. 
A similar method has been proposed and applied in the context of 
$b\to s(d)e^+e^-$ decays in \cite{Kruger:1996cv}.

The function $\Pi(k^2)$ obeys the dispersion relation
\begin{equation}\label{disprel}
\Pi(k^2)=\frac{1}{\pi}\int_0^\infty dt\, 
\frac{{\rm Im}\Pi(t)}{t-k^2-i\epsilon}
\end{equation}
In this form the dispersion relation needs one subtraction,
but the subtraction constant cancels in $\Pi(k^2)-\Pi(m^2_b)$.
In principle ${\rm Im}\Pi(t)$ could be determined experimentally.
Instead, for simplicity, we choose a convenient ansatz that
should capture the essential features of the true hadronic
quantity ${\rm Im}\Pi(t)$.
We write ${\rm Im}\Pi$ as the sum of a resonance and a continuum
contribution
\begin{equation}\label{impirc}
{\rm Im}\Pi = {\rm Im}\Pi_r + {\rm Im}\Pi_c
\end{equation}
where
\begin{eqnarray}
{\rm Im}\Pi_r(t) &=& \sum_{r=\rho,\omega}\frac{1}{2}
\frac{f^2_r m_r \Gamma_r}{(t-m^2_r)^2 + m^2_r \Gamma^2_r} \label{impir}\\
{\rm Im}\Pi_c(t) &=& \frac{t}{4\pi t_c} \Theta(t_c-t)
+\frac{1}{4\pi} \Theta(t-t_c), \qquad\quad t_c=4\pi^2(f^2_\rho+f^2_\omega),
\qquad t > 0 \label{impic}
\end{eqnarray}
The asymptotic QCD result fixes ${\rm Im}\Pi_c$ to
$N_c/(12\pi)=1/(4\pi)$ at large $t$. Imposing quark-hadron
duality for the integral of ${\rm Im}\Pi(t)$ up to (at least)
$t=t_c$ determines the value of 
$t_c=4\pi^2(f^2_\rho+f^2_\omega)\approx 3.1\,{\rm GeV}^2$.
The factor $1/2$ in (\ref{impir}) is an isospin factor
coming from the overlap of $\rho^0$ and $\omega$ with the
$\bar u\gamma_\mu u$ current.
Determining $\Pi$ in (\ref{dau79ew}) with the help of (\ref{disprel})
and (\ref{impirc}), treating the resonances as narrow and
taking the heavy-quark limit $t_c\ll m_b$, we finally obtain (\ref{pewnu}). 
Concerning the factor in square brackets in (\ref{pewnu}),
two limiting cases are worth noting. If $k^2=m^2_V\to 0$,
we recover an expression similar to (\ref{pewnc}) where
the light-quark mass under the logarithm is replaced by
the hadronic scale $\sqrt{t_c}$. In the limit $k^2=m^2_V\to t_c$
the same terms appear, and in addition the perturbative
imaginary part $-2\pi i/3$.

\section{\boldmath Coefficients $a_i$, $b_i$\unboldmath}
\label{sec:coeffaibi}

In the following Table we quote the
central values of the coefficients $a_i$ as defined in (\ref{tpdef}) for 
two final-state $\rho$-mesons (\ref{tab:input}). 
The default value used for the model of power-suppressed 
hard-spectator contributions is $X_H=\ln{\frac{m_B}{\Lambda_h}}$ and
for the renormalization scale it is $\mu=4.2$GeV.

\begin{center}
\begin{tabular}{*{5}{|c}|}
\hline\hline
$a_1$ & $a_2$ & $a_3+a_5$ & $a^u_4$ & $a^c_4$\\
\hline 
$0.991+ 0.020 i$ & $0.177-0.084 i$ & $0.002-0.001 i$ & $-0.025-0.016 i$ & 
$-0.033-0.009 i$\\
\hline
\hline
$(a^u_7+a^u_9)/\alpha$ & $(a^u_7-a^u_9)/\alpha$ & $(a^c_7+a^c_9)/\alpha$ & 
$a^u_{10}/\alpha$ & $a^c_{10}/\alpha$\\
\hline
$-1.84-0.54 i$ & $1.15+0.02 i$ & $-1.10-0.02 i$ & $-0.17+0.09 i$ & 
$-0.17+0.09 i$\\
\hline
\hline
\end{tabular}
\end{center}

The central values of the coefficients $b_i$ as defined in (\ref{tpannd}) for 
two final-state $\rho$-mesons (\ref{tab:input}) are given below.
The default value used for the model of power-suppressed 
annihilation contributions is $X_A=\ln{\frac{m_B}{\Lambda_h}}$ and
for the renormalization scale it is $\mu=4.2$GeV.
Here $r_A=B_{\rho\rho}/A_{\rho\rho}$.

\begin{center}
\begin{tabular}{*{6}{|c}|}
\hline\hline
$r_A b_1$ & $r_A b_2$ & $r_A b_3$ & $r_A b_4$ & $r_A b^{EW}_3/\alpha$ & 
$r_A b^{EW}_4/\alpha$\\
\hline
$0.029$ & $-0.011$ & $0.003$ & $-0.003$ & $-0.035$ & $0.013$\\
\hline
\hline
\end{tabular}
\end{center}

\section*{Acknowledgements}
We thank Martin Beneke and Matthias Neubert for useful discussions.
This work was supported in part by the DFG cluster of excellence 
`Origin and Structure of the Universe' and by the DFG
Graduiertenkolleg GK 1054.


\end{document}